\documentclass{aastex62}
\usepackage{natbib,textcomp,gensymb,wasysym}
\submitjournal{ApJ}
\shorttitle{Spherical orbits in Kerr spacetime}
\shortauthors{O. Kop\'{a}\v{c}ek \& V. Karas}

\begin{document}

\newcommand{\rff}[1]{Fig.~\ref{#1}}
\newcommand{\req}[1]{Eq.~(\ref{#1})}
\newcommand{\rffs}[1]{Figs.~\ref{#1}}
\newcommand{\reqs}[1]{Eqs.~(\ref{#1})}
\newcommand{\refsec}[1]{Section~\ref{#1}}
\newcommand{\newstuff}[1]{{\color{red}#1}}

\title{On Innermost Stable Spherical Orbits near a rotating black hole:\\[0.2em]
{A numerical study of the particle motion near the plunging region}}
\correspondingauthor{Ond\v{r}ej Kop\'{a}\v{c}ek}
\email{kopacek@asu.cas.cz, vladimir.karas@asu.cas.cz}
\author[0000-0002-6489-4010]{Ond\v{r}ej Kop\'{a}\v{c}ek}
\affiliation{Astronomical Institute, Czech Academy of Sciences, Bo\v{c}n\'{i} II 1401, CZ-141\,00~Prague, Czech~Republic}
\affiliation{Faculty of Science, Humanities and Education, Technical University of Liberec, Studentsk\'{a} 1402/2, CZ-461\,17~Liberec, Czech Republic}
\author[0000-0002-5760-0459]{Vladim\'{i}r Karas$^1$}

\begin{abstract}
According to General Relativity, astrophysical black holes are described by a small number of parameters. Apart from the mass of the black hole ($M$), among the most interesting characteristics is the spin ($a$), which determines the degree of rotation, i.e.\ the angular momentum of the black hole. The latter is observationally constrained by the spectral and timing properties of the radiation signal emerging from an accretion disk of matter orbiting near the event horizon. In the case of planar (standard, equatorial) accretion disk, it is the location of the Innermost Stable Circular Orbit (ISCO) that determines the observable radiation characteristics, and this way allows us to measure the spin. In this paper, we discuss a more general case of the Innermost Stable Spherical Orbits (ISSO) extending above and below the equatorial plane. To this end, we study the non-equatorial geodesic motion of particles following inclined, spherical, relativistically precessing trajectories with the aim of exploring the boundary between the regions of stable (energetically bound) and escaping (energetically unbound) motion. The concept of the ISSO radius should play a role in determining the inner rim of a tilted or geometrically thick accretion flow. We demonstrate that the region of inclined bound orbits has a complicated structure due to enhanced precession near the inner rim. We also explore the fate of particles launched below the radius of the Marginally Bound Spherical Orbit (MBSO): these may either plunge into the event horizon or escape to radial infinity.
\end{abstract}

\keywords{black hole physics; accretion, accretion disks; spherical orbits, geodesics; methods: numerical}

\section{Introduction}
\label{intro}
A growing number of studies have been published during the recent decade indicating that black holes are ubiquitous in different types of cosmic objects, ranging from stellar-mass black holes in compact binary systems to supermassive black holes that reside in cores of galaxies \citep{2012bhae.book.....M}. The processes of accretion and ejection of matter onto black holes turn out to be essential for our ability to identify cosmic black holes and to explore their interaction with the surrounding gaseous environment and the radiation field \citep[e.g.]{2002apa..book.....F,2018ASSL..454.....S}. Strong-gravity effects operate in the regions close to the black hole event horizon, and so the General Relativity framework has to be employed in order to describe the motion of particles and fluids \citep{1998mtbh.book.....C,2008bhad.book.....K,mtw}.

The gravitational field near black holes can be described in terms of Kerr metric \citep{kerr63} with mass and spin being the only two astrophysically relevant parameters. The motion of test particles around classical black holes is regular, i.e., without signatures of chaos \citep{carter68}. The special importance has been attributed to the innermost stable circular orbit of the equatorial motion \citep[ISCO][]{bardeen72,2022PhRvL.129p1101M}, when test particles transit from the phase-space region of energetically stable circulation into plunging trajectories. Even in the planar case, complete taxonomy is very rich, as can be seen in a systematic classification by \citet{2008PhRvD..77j3005L}.

Pressure and turbulence must affect the corresponding structure of gaseous flows, nevertheless, ISCO plays an important role in the standard accretion disc scenario, and it may also influence gaseous flows to a certain extent. This raises a long-standing question about a boundary between stable (energetically bound), plunging, and escaping (unbound) trajectories for non-equatorial motion; despite its apparent relevance for geometrically thick and/or inclined accretion flows, the topology of that boundary has been so far explored only indirectly, and the results about a potential analogy of ISCO outside the equatorial plane have been presented only in an implicit form \citep{2012CQGra..29u7001W,rana19,teo21,2021PhRvD.104l4059T}.

In this context, we revisit the issue and give a detailed account of the problem of stability and energetic binding of spherical orbits including a numerical analysis of an unstable case occurring in an inner region of an accreting black hole system. The paper is organized as follows. In \refsec{spec} we review the equations of motion and we introduce a two-dimensional effective potential for the geodesic motion of free test particles in Kerr spacetime. In \refsec{spherical} we review the properties of spherical orbits using two alternative parameterizations (in terms of Carter constant or inclination angle) and we study the locations of the innermost stable spherical orbits and marginally bound spherical orbits with respect to parameters of the system. The results are numerically verified in \refsec{numerical} with the technique of escape-boundary plots. Furthermore, we focus here on the unstable orbits below ISSO and employ this method to demonstrate, how the parameters (namely, spin and Carter constant) affect the instability and what effect they may have on the probability of the plunge or escape. Results are summarized and briefly discussed in \refsec{conclusions}. Details on numerical methods employed for the analysis are provided in \hyperref[appa]{Appendix}.

\section{Equations of motion and effective potential}
\label{spec}
Kerr metric describes the geometry of the spacetime around the rotating black hole \citep{kerr63}. It can be expressed in Boyer-Lindquist coordinates $x^{\mu}= (t,\:r, \:\theta,\:\varphi)$ as follows \citep{mtw}:
\begin{equation}
\label{metric}
ds^2=-\frac{\Delta}{\Sigma}\:[dt-a\sin{\theta}\,d\varphi]^2+\frac{\sin^2{\theta}}{\Sigma}\:[(r^2+a^2)d\varphi-a\,dt]^2+\frac{\Sigma}{\Delta}\;dr^2+\Sigma d\theta^2,
\end{equation}
where
\begin{equation}
{\Delta}\equiv{}r^2-2Mr+a^2,\;\;\;
\Sigma\equiv{}r^2+a^2\cos^2\theta.
\end{equation}
A coordinate singularity at $\Delta(r)=0$ corresponds to the position of the outer and the inner event horizons of the black hole, respectively: $r_\pm=M\pm\sqrt{M^2-a^2}$. The rotation of the black hole is measured by the spin parameter $a\in\left<-M,M\right>$; we can assume $a\geq0$ without any loss of generality.

We note that geometrized units will be used throughout the paper, making the gravitational constant and the speed of light equal to one; $G=c=1$. For the rest of the paper, we switch to dimensionless units scaled by the mass of the black hole, i.e., $M=1$ is set in all formulas.

The Hamiltonian $\mathcal{H}$ of a particle with rest mass $m$ in the spacetime with contravariant metric components $g^{\mu\nu}$ may be defined as \citep{mtw}:
\begin{equation}
\label{hamiltonian}
\mathcal{H}=\textstyle{\frac{1}{2}}g^{\mu\nu}p_{\mu} p_{\nu},
\end{equation}
where $p_{\mu}$ is the generalized (canonical) momentum which in this case corresponds to the kinematical momentum. The equations of motion are expressed as:
\begin{equation}
\label{hameq}
\frac{{\rm d}x^{\mu}}{{\rm d}\lambda}\equiv p^{\mu}=
\frac{\partial \mathcal{H}}{\partial p_{\mu}},
\quad 
\frac{{\rm d} p_{\mu}}{d\lambda}=-\frac{\partial\mathcal{H}}{\partial x^{\mu}},
\end{equation}
where $\lambda\equiv\tau/m$ is a dimensionless affine parameter ($\tau$ denotes the proper time). The conserved value of the Hamiltonian is given as $\mathcal{H}=-m^2/2$. The system is stationary and the time component of canonical momentum $\pi_t$ is an integral of motion, which equals the (negatively taken) energy of the test particle, $p_t\equiv-E$. Moreover, the system is axially symmetric which assures the conservation of the axial component of the angular momentum $L\equiv p_{\varphi}$. Analysis of the geodesics in Kerr spacetime using the Hamilton-Jacobi formalism reveals \citep{carter68} that besides the above obvious integrals $E$, $L$ corresponding to relevant Killing vectors, there exists another independent integral of motion, namely, the Carter constant $Q$, related to the {\em hidden} symmetry of the system corresponding to the existence of the Killing tensor. Carter constant may be expressed as:

\begin{equation}
\label{carter}
Q\equiv p^2_{\theta}+\cos^2{\theta}\left( a^2(m^2-E^2)+\left(\frac{L}{\sin{\theta}}\right)^2 \right).
\end{equation}
In some cases, the constant is being expressed in an alternative way as $K\equiv Q+(L-aE)^2$, however, for our analysis, the form given by \req{carter} remains more appropriate.

Endowed with the four independent integrals of motion which are in involution (i.e., they commute in terms of Poisson brackets) the system becomes fully integrable in the Liouville sense and its Hamilton-Jacobi equation is completely separable \citep[e.g.,][]{goldstein02}. Besides other properties that follow from integrability, we stress that all trajectories in such a system are regular, i.e., no chaotic orbits are present in the whole phase space. 

In the rest of the paper, we switch to specific quantities $E/m\rightarrow E$, $L/m\rightarrow L$, and $Q/m^2\rightarrow Q$ which corresponds to setting the rest mass of the particle $m=1$ in the formulas.

Effective potential expressing the minimal allowed energy of a particle might be expressed as a function of coordinates $r, \theta$ and parameters $a,L$ in the following form \citep{kopacek18b}:

\begin{equation}
\label{eff_pot}
V_{\rm eff}(r,\theta)=\left(-\beta+\sqrt{\beta^2-4\alpha\gamma}\right)/2\alpha,
\end{equation}
where
\begin{equation}
\label{eff_pot_coeff}
\alpha=-g^{tt},\;\;\;\beta=2g^{t\varphi{}}L \;\;\; {\rm and} \;\;\; \gamma=-g^{\varphi\varphi}L^2-1.
\end{equation}
Equipotential curves $V_{\rm eff}=E$ drawn in the $(r,\theta)$ plane connect simultaneous turning points of the particle with energy $E$ in coordinates $r$ and $\theta$.   Analysis of the two-dimensional effective potential may thus be used to localize circular orbits not only in the equatorial plane but also off this plane. Although in the pure Kerr spacetime, the circular geodesics are allowed only in the equatorial plane, if we, for instance, consider electrically charged particles near weakly magnetized Kerr black hole, we also find circular orbits above and below equatorial plane \citep{kovar08,kovar10}, and the method of two-dimensional effective potential also allows to discuss their stability \citep{tursunov16}.

\section{Spherical orbits in Kerr spacetime: Their stability and binding}
\label{spherical}
Spherical orbits are a special class of bound geodesics in Kerr spacetime with a constant radial coordinate $r=\rm{const}$ and latitudinal angle $\theta$ varying around the equatorial plane between the turning points that are positioned symmetrically at $\theta_\star$ and $\pi-\theta_\star$. The value of $\theta_\star$ is determined by the value of Carter constant $Q$ and, in particular, for $Q=0$ the spherical orbit reduces to circular (Keplerian) trajectory in the equatorial plane with $\theta_\star=\pi/2$. An illustration of spherical orbits with different values of Carter constant is provided in \rff{spherical_example}. They can represent trajectories of test particles near the event horizon.

\begin{figure}[ht]
\center
\includegraphics[scale=.75]{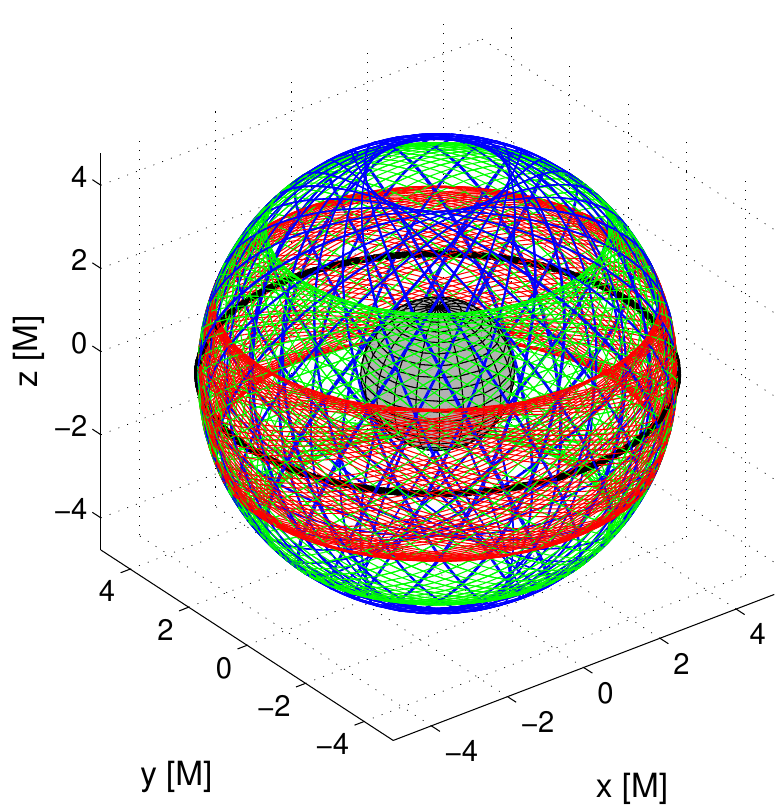}
\caption{Example of a set of spherical orbits launched at the radius $r=5$ from the equatorial plane of a Kerr black hole ($a=0.8$) differing solely by the value of Carter constant $Q$ which specifies the initial value of $p_{\theta}=Q^{1/2}$. In particular, we compare the trajectories with $Q=0$ (black circular orbit), $Q=1$ (red), $Q^{1/2}=2$ (green) and $Q^{1/2}=3$ (blue).}
\label{spherical_example}
\end{figure}

Spherical orbits of massive particles in the Kerr metric were analyzed by \citet{wilkins72} for the special case of extreme Kerr black hole with $a=1$, while \citet{goldstein74} performed the early numerical calculations of such geodesics \citep[see also][]{1986SvPhU..29..215D}. Energy and angular momentum of the spherical orbits may be expressed explicitly as a function of radius, spin, and latitudinal turning point $\theta_\star$ \citep{shakura87}: 
\begin{eqnarray}
 \label{E_shakur}E=\frac{1-\frac{2r}{\Sigma_{\star}}\pm\frac{aq_{\star}}{\Sigma_{\star} \sqrt{r}}\sin{\theta_{\star}}}{\left(1-\frac{3r}{\Sigma_{\star}}\pm \frac{2aq_{\star}}{\Sigma_{\star}\sqrt{r}}\sin{\theta_{\star}} + \frac{a^2}{\Sigma_{\star} r}\cos^2{\theta_{\star}} \right)^{1/2}},\\
 \label{L_shakur}L=\frac{{\pm\frac{q_{\star}(r^2+a^2)}{\Sigma_{\star}\sqrt{r}}\sin\theta_{\star}}-\frac{2ar}{\Sigma_{\star}}\sin^2\theta_{\star}}{\left(1-\frac{3r}{\Sigma_{\star}}\pm \frac{2aq_{\star}}{\Sigma_{\star}\sqrt{r}}\sin{\theta_{\star}} + \frac{a^2}{\Sigma_{\star} r}\cos^2{\theta_{\star}} \right)^{1/2}},
\end{eqnarray}
where $\Sigma_{\star}=r^2+a^2\cos^2\theta_{\star}$ and $q_{\star}=(r^2-a^2\cos^2\theta_{\star})^{1/2}$. Upper signs correspond to co-rotating (direct) orbits while the lower signs are valid for counter-rotating orbits. With the inclination $\theta_{\star}=\pi/2$ the above expressions reduce to familiar formulas for circular (Keplerian) orbits in the equatorial plane \citep{bardeen72}:
\begin{eqnarray}
 \label{kep_E}E_{\rm Kep}&=&\frac{r^2-2r\pm a \sqrt{r}}{r\sqrt{r^2-3r\pm 2a\sqrt{r}}},\\
\label{kep_L}L_{\rm Kep}&=&\frac{\pm(r^2+a^2\mp 2a\sqrt{r})}{\sqrt{r(r^2-3r\pm 2a\sqrt{r})}}.
\end{eqnarray}

\begin{figure}[ht]
\center
\includegraphics[scale=.6]{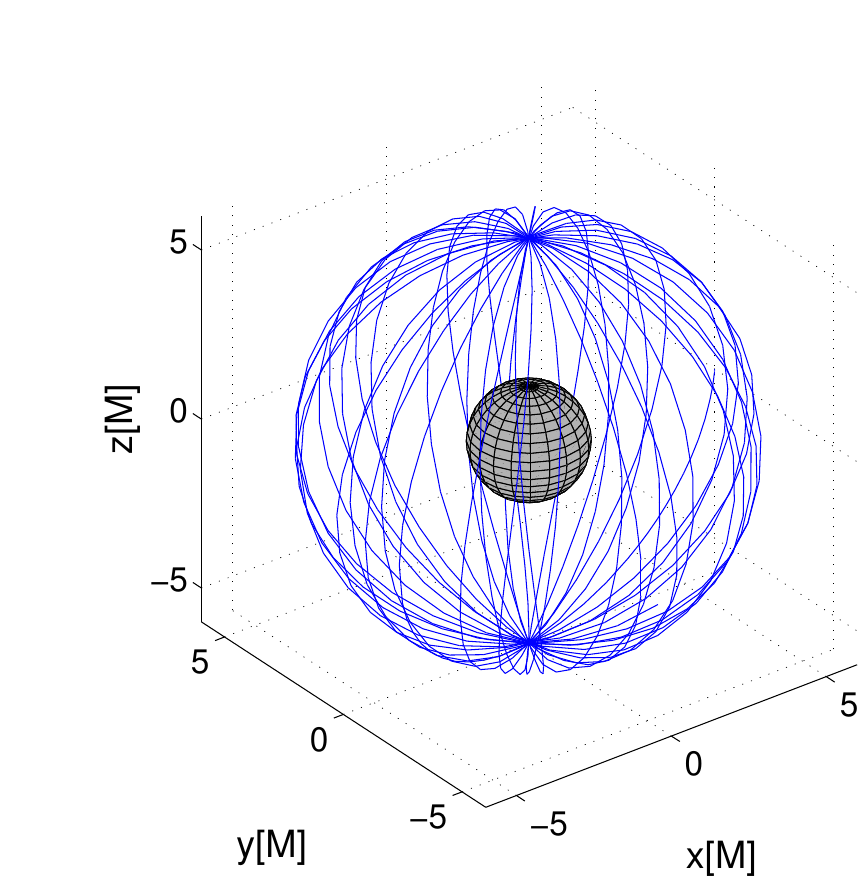}
\includegraphics[scale=.52]{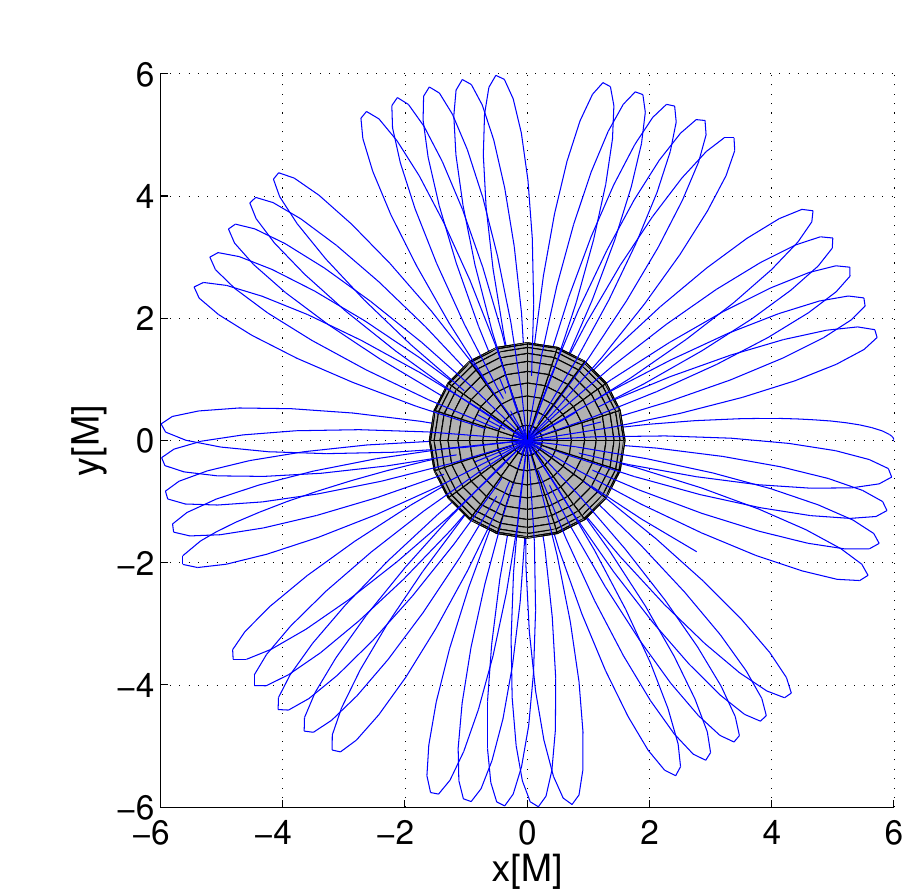}
\caption{Stable polar orbit around Kerr black hole ($a=0.8$) with radius $r=6$.}
\label{polar_example}
\end{figure}

Setting the other limiting value, $\theta_\star=0$, in \reqs{E_shakur} and (\ref{L_shakur}) we obtain polar orbits with $L=0$ and energy given as:
\begin{equation}
E_p=\frac{\Delta \sqrt{r}}{\sqrt{(r^2+a^2)(r^3-3r^2+a^2r+a^2)}}.
\label{polar_energy}
\end{equation}
At a given radius, polar orbits periodically cross both poles while being continuously dragged due to the rotation of the black hole as illustrated in \rff{polar_example}.

\begin{figure}[ht]
\center
\includegraphics[scale=.45]{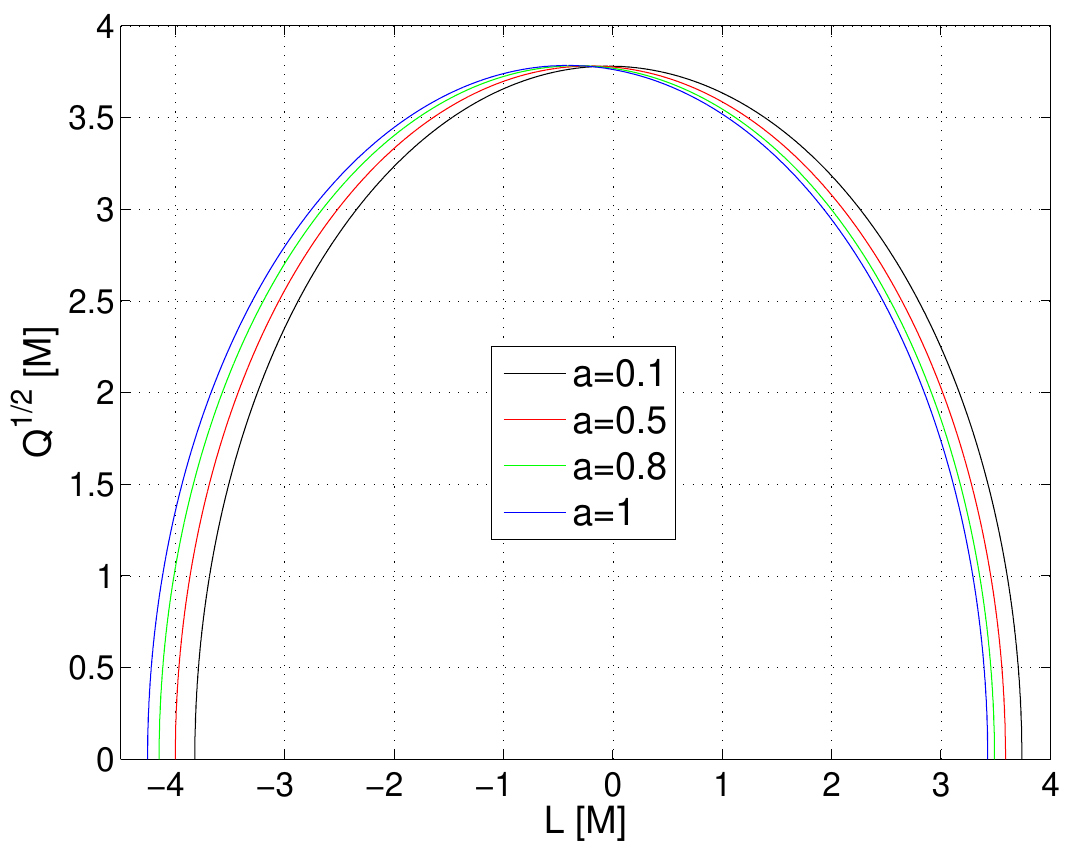}
\includegraphics[scale=.45]{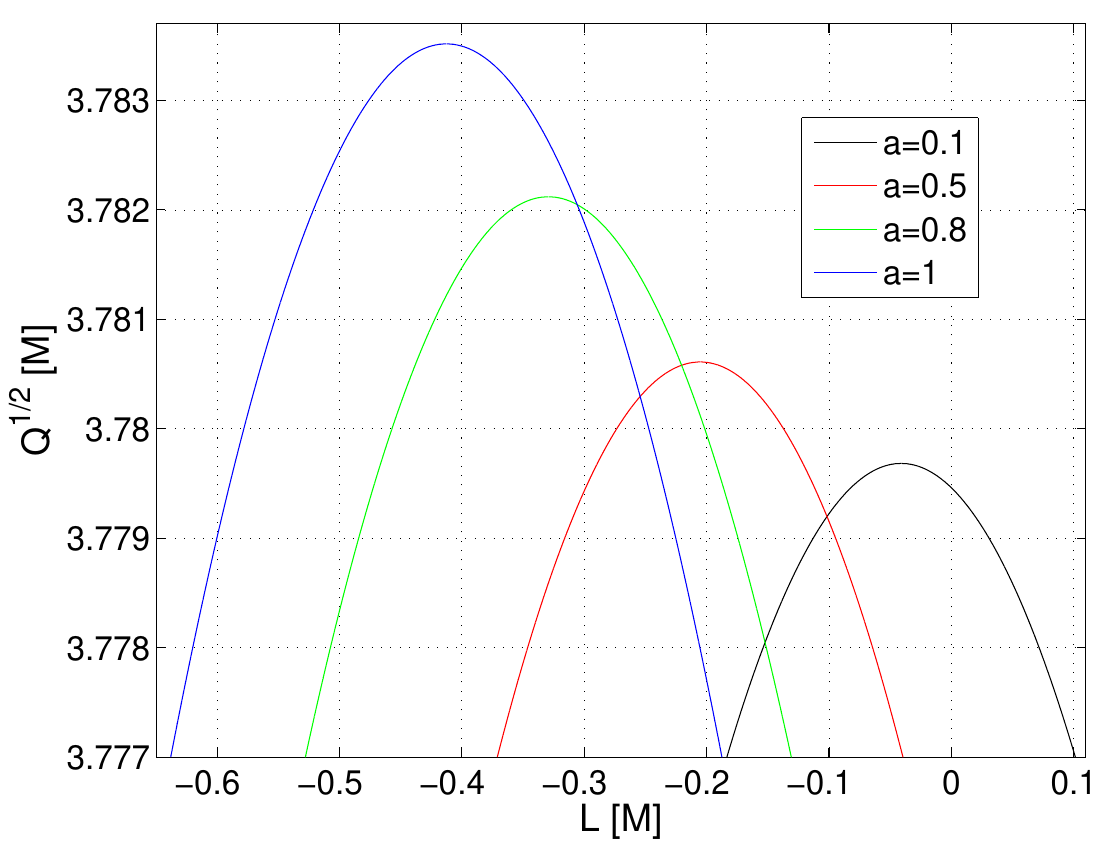}
\caption{Square root of the Carter constant of spherical orbits at $r=10$ as a function of angular momentum $L$ for several values of spin. In the right panel, we zoom the section of the left plot around maximal values.}
\label{diskuze_Lsf}
\end{figure}

Alternatively to the parameterization by the value of the turning point $\theta_\star$, it is also possible to parameterize the integrals of motion $E$ and $L$ by the Carter constant $Q$ \citep{teo21}:

\begin{eqnarray}
\label{spherical_energy}E&=&\frac{r^3\left(r-2\right)-a\left(aQ\mp\sqrt{\Upsilon}\right)}{r^2\sqrt{r^3\left(r-3\right)-2a\left(aQ\mp\sqrt{\Upsilon}\right)}},\\
\label{spherical_angular}L&=&-\frac{2ar^3+\left(r^2+a^2\right)\left(aQ\mp\sqrt{\Upsilon}\right)}{r^2\sqrt{r^3\left(r-3\right)-2a\left(aQ\mp\sqrt{\Upsilon}\right)}},
\end{eqnarray}
where $\Upsilon=r^5-Q(r-3)r^3+a^2Q^2$. Submitting $Q=0$, we may again verify that the above formulas reduce to expressions for circular orbits given by \reqs{kep_E} and (\ref{kep_L}).

In \rff{diskuze_Lsf} we plot the square root of the Carter constant $Q^{1/2}$ (which is equal to the equatorial value of $p_{\theta}$) as a function of the angular momentum given by \req{spherical_angular} for spherical orbits at $r=10$ with several values of spin. In particular, we observe that for a given radius and spin, the maximal value of the Carter constant is attained by the counter-rotating spherical orbit ($L<0$) while the value corresponding to polar orbit with $L=0$ is slightly lower. Each value of $Q^{1/2}$ below its maximum corresponds to the pair of orbits with different $L$, which might be both negative (two counter-rotating orbits) for the Carter constant close to the maximal value, or with mixed signs of $L$ (one co-rotating and one counter-rotating) for lower values. The described behavior of the function $Q^{1/2}(L)$ is not universal and it changes closer to the horizon. As we shall see, neither polar nor spherical orbits may generally exist all the way down to the horizon.

\begin{figure}[ht]
\center
\includegraphics[scale=.48]{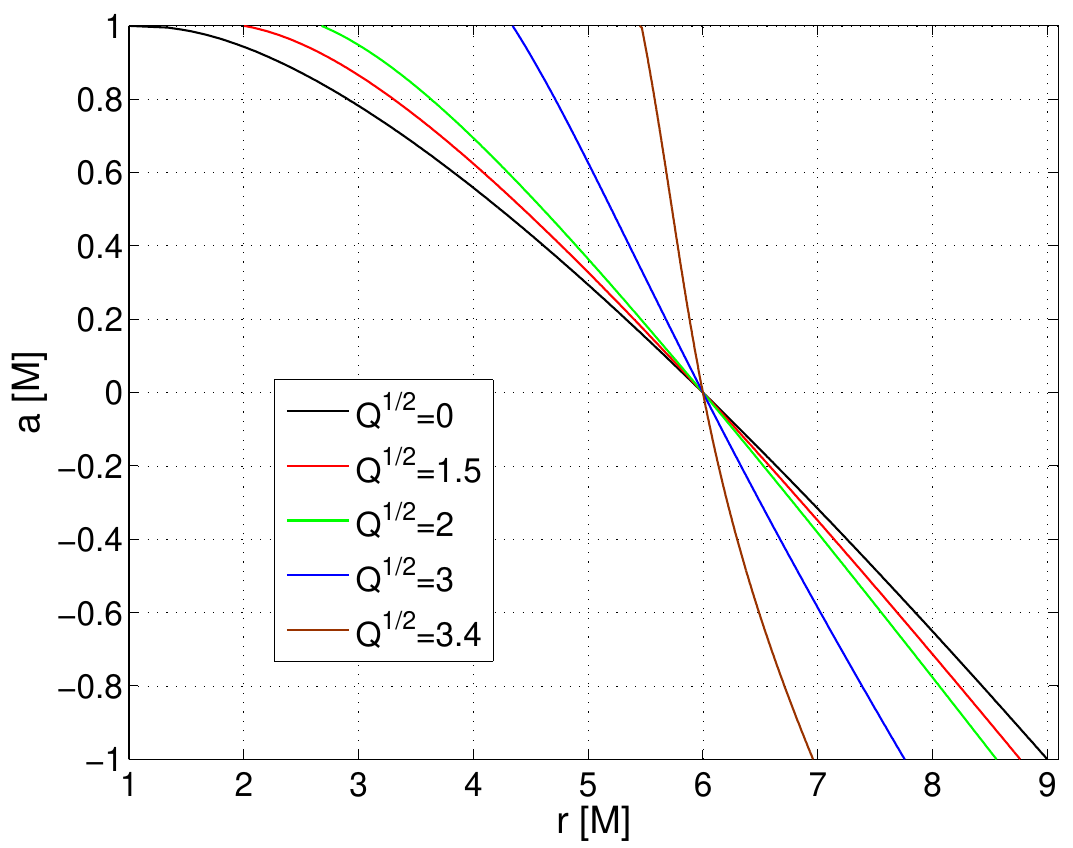}
\includegraphics[scale=.48]{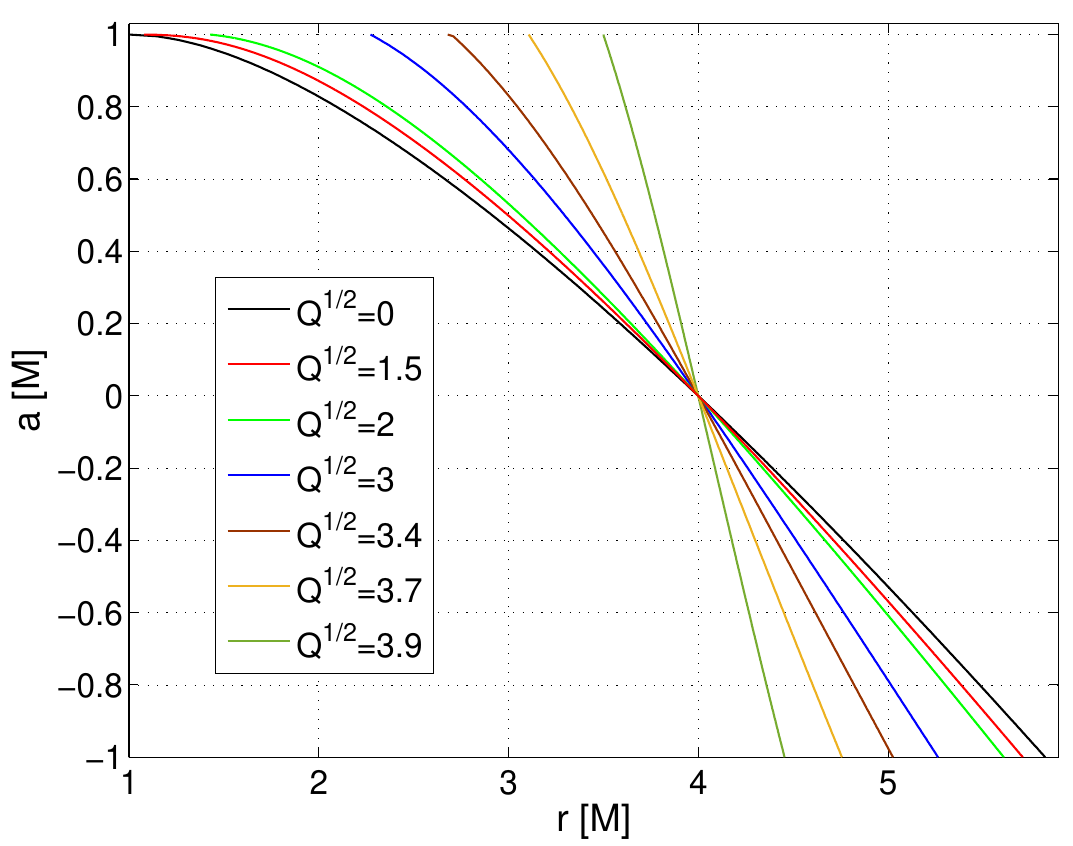}
\caption{Locations of the innermost stable spherical orbits (left panel) and marginally bound spherical orbits (right panel) with different values of the Carter constant and spin. On the vertical axis, parameter $\mbox{a[M]}\equiv\sigma a$ is shown, where $a$ is the Kerr spin parameter, for which $a\geq0$ is assumed throughout the paper, and $\sigma=\pm1$ is a switch that distinguishes prograde vs.\ retrograde motion. Positive values ($\sigma=1$) correspond to orbits co-rotating with the black hole, while negative values ($\sigma=-1$) are for the counter-rotating orbits. These graphs generalize the well-known dependence of the equatorial ISCO radius to the case of orbits inclined with respect to the equatorial plane.}
\label{isso_mbso_carter}
\end{figure}

We note that alternative expressions for the energy and angular momentum of the spherical orbit have also been provided by \citet{rana19} in Eqs.\ (18a), (18b) and (18c) of the referred paper. These equations, however, are not entirely correct. Although they work well for low values of $Q$ for which they reproduce proper parameters of spherical orbits, with higher values of $Q$, they fail to do so and provide erroneous values of the angular momentum  (see the discussion and \rff{L_Rana} in Appendix). 
 
In the equatorial plane, stable circular geodesics are not possible below the radius of Innermost Stable Circular Orbit (ISCO), the position of which is given as \citep{bardeen72}:
\begin{equation}
\label{rms}
r_{\rm{ISCO}}=\left(3+Z_2\mp\sqrt{(3-Z_1)(3+Z_1+2Z_2)}\right),
\end{equation} 
where $Z_1\equiv1+\left(1-a^2\right)^{1/3}\left[\left(1+a\right)^{1/3}+\left(1-a\right)^{1/3}\right]$, $Z_{2} \equiv \sqrt{{3a^2}+Z_1^2}$ and the upper sign corresponds to co-rotating orbit and the lower sign to counter-rotating one. In fact, bound circular orbits exist only above the radius of marginally bound circular orbit (MBCO) given as \citep{bardeen72}:
\begin{equation}
\label{rmb}
r_{\rm MBCO}=2\mp a + 2 \sqrt{1\pm a}.
\end{equation} 
The concept of ISCO has profound importance in our understanding of how standard accretion disks operate in the black hole vicinity. Indeed, spectral characteristics of accreting black holes, both the supermassive ones in cores of Active Galactic Nuclei (AGN) as well as stellar-mass black holes in close binary systems, indicate that the accretion disk often proceeds down to ISCO, although a growing number of counter-examples have been indicated in various sources and related to numerical simulations of truncated accretion disks with strong magnetic fields \citep{2003PASJ...55L..69N,2019MNRAS.487..550L}. Moreover, recently introduced ``puffy'' disks seem to proceed well {\em below} ISCO \citep{2019ApJ...884L..37L,2022MNRAS.514..780W}. Nonetheless, much of the emerging flux is typically released in X-rays that are produced just above the critical ISCO radius. As the black hole spin grows, the ISCO location approaches the event horizon, and the radiation output from the accretion mechanism is also thought to grow in parallel. On the other hand, accretion does not have to be bound to the equatorial plane and, indeed, numerous examples have been reported of objects where the accretion flow extends above the equatorial plane as it proceeds via a non-planar (twisted) accretion disk \citep[e.g.][]{1985MNRAS.213..435K,2007MNRAS.381.1617M}. This brings us to an important question about the analogy of the equatorial ISCO radius outside the equatorial plane. We can introduce the notion of Innermost Stable Spherical Orbit (ISSO) as a natural extension of ISCO. 

\begin{figure}[ht]
\center
\includegraphics[scale=.48]{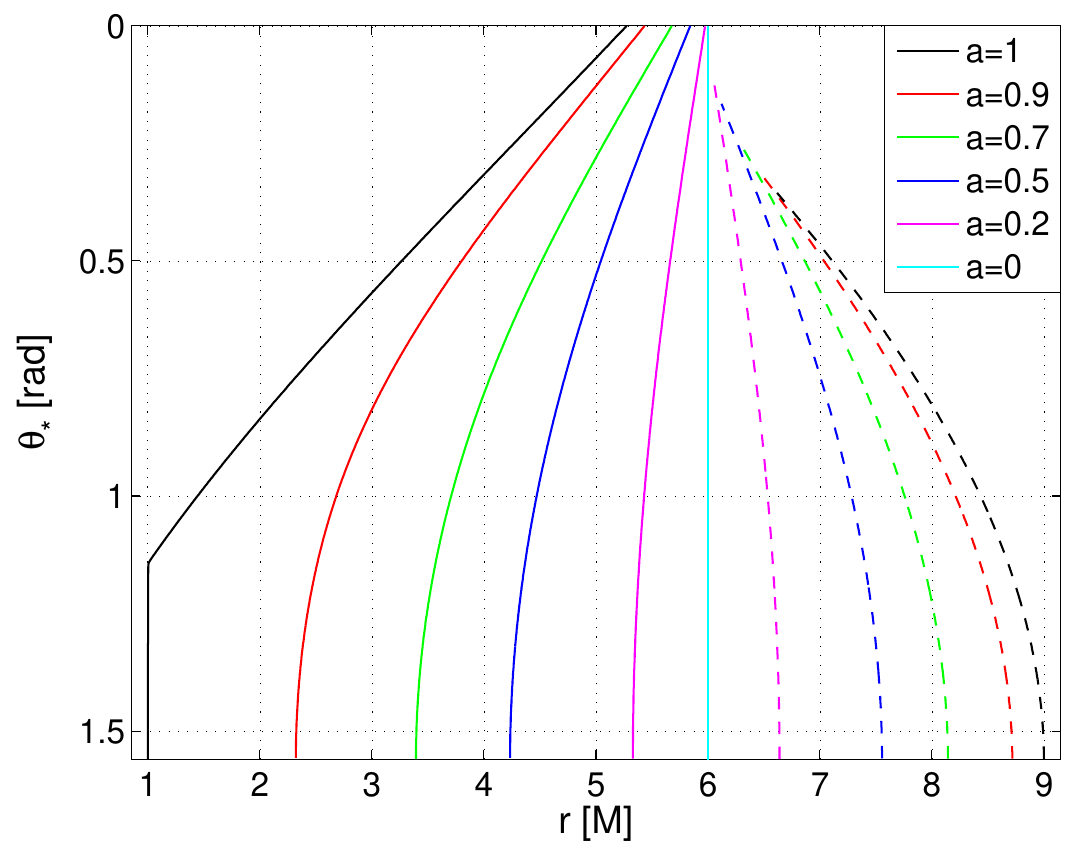}
\includegraphics[scale=.48]{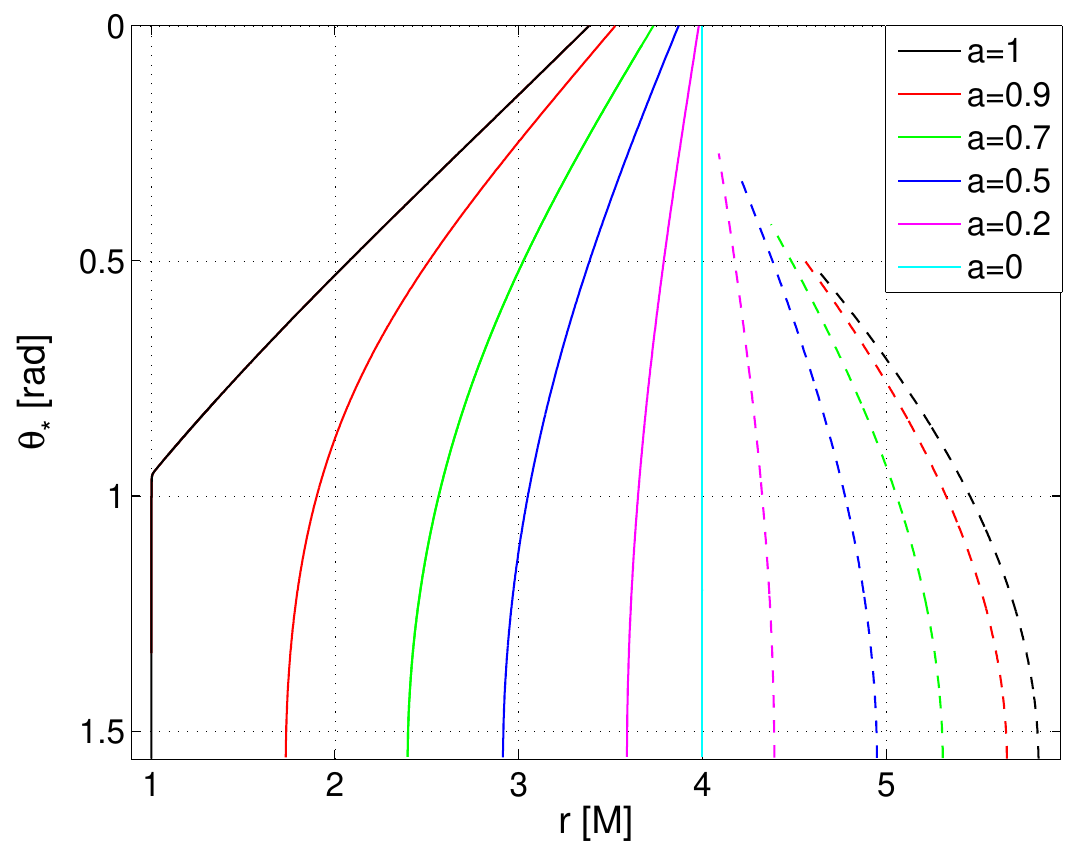}
\caption{Locations of the innermost stable spherical orbits (left panel) and marginally bound spherical orbits (right panel) with respect to the inclination $\theta_{\star}$ for different values of the spin parameter. The meaning of $\theta_{\star}$ is the angular distance of the latitudinal turning point from the zenith or nadir, respectively (see the main text for further details). Dashed lines correspond to the counter-rotating spherical orbits. We notice that the vertical axes of the plots are inverted, i.e., their lower edges ($\theta_{\star}=\pi/2$) correspond to circular orbits in the equatorial plane, while polar orbits with $\theta_{\star}=0$ are found on the upper edges.}
\label{isso_mbso_inclination}
\end{figure}

In a close analogy with the case of circular orbits residing in the equatorial plane, for spherical (generally inclined) orbits the radius $r=r_s$ of ISSO, and the related radius $r=r_b$ of the Marginally Bound Spherical Orbit (MBSO) are given implicitly by the following algebraic relations \citep{rana19}:
\begin{eqnarray}
\nonumber  &&r_{s}^{9}-12 r_{s}^{8}-6 a^{2} r_{s}^{7}+36 r_{s}^{7}+8 a^{2} Q r_{s}^{6}-28 a^{2} r_{s}^{6}-24 a^{2} Q r_{s}^{5}+9 a^{4} r_{s}^{5}-24 a^{4} Q r_{s}^{4}\\
&&+48 a^{2} Q r_{s}^{4}+16 a^{4} Q^{2} r_{s}^{3}-8 a^{4} Q r_{s}^{3}-48 a^{4} Q^{2} r_{s}^{2}+48 a^{4} Q^{2} r_{s}-16 a^{6} Q^{2}=0,
\label{isso}
\end{eqnarray}
and
\begin{eqnarray}
\nonumber  &&r_{b}^{8}-8 r_{b}^{7}-2 a^{2} r_{b}^{6}+16 r_{b}^{6}+2 a^{2} Q r_{b}^{5}-8 a^{2} r_{b}^{5}-6 a^{2} Q r_{b}^{4}+a^{4} r_{b}^{4}-2 a^{4} Q r_{b}^{3}\\
&&+8 a^{2} Q r_{b}^{3}+a^{4} Q^{2} r_{b}^{2}-2 a^{4} Q r_{b}^{2}-2 a^{4} Q^{2} r_{b}+a^{4} Q^{2}=0.
\label{mbso}
\end{eqnarray}

\begin{figure}[ht]
\center
\includegraphics[scale=.48]{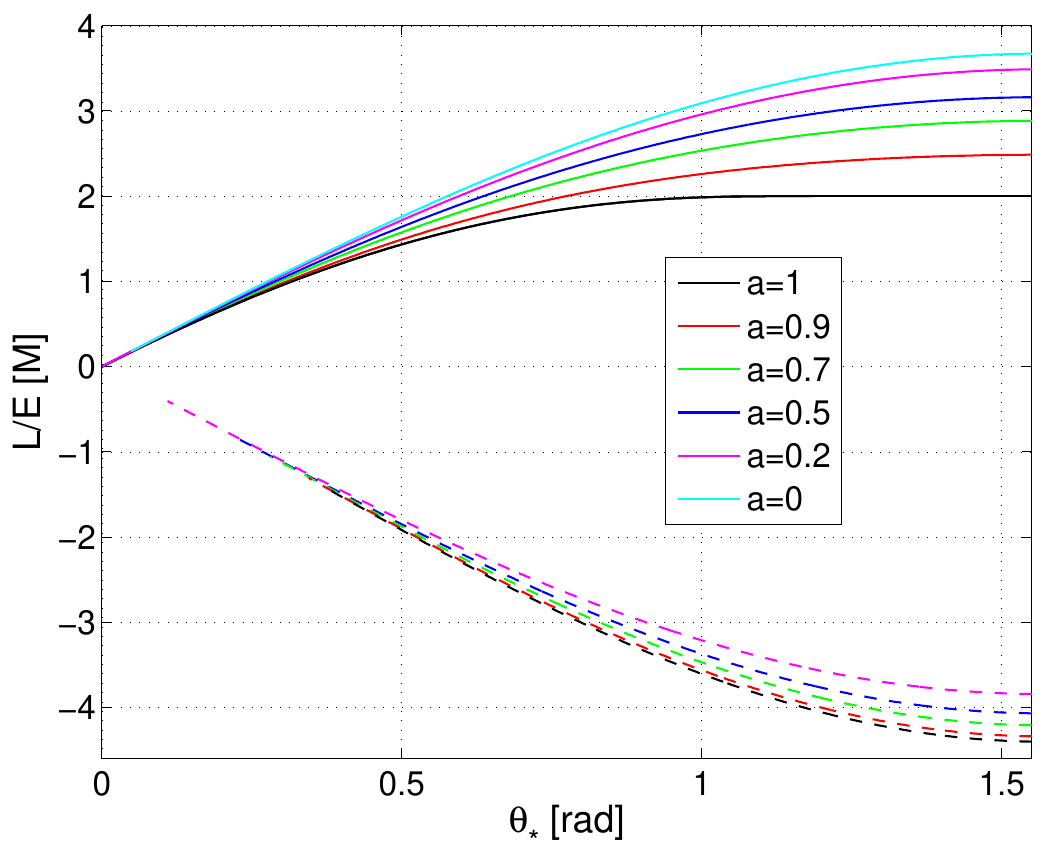}
\includegraphics[scale=.48]{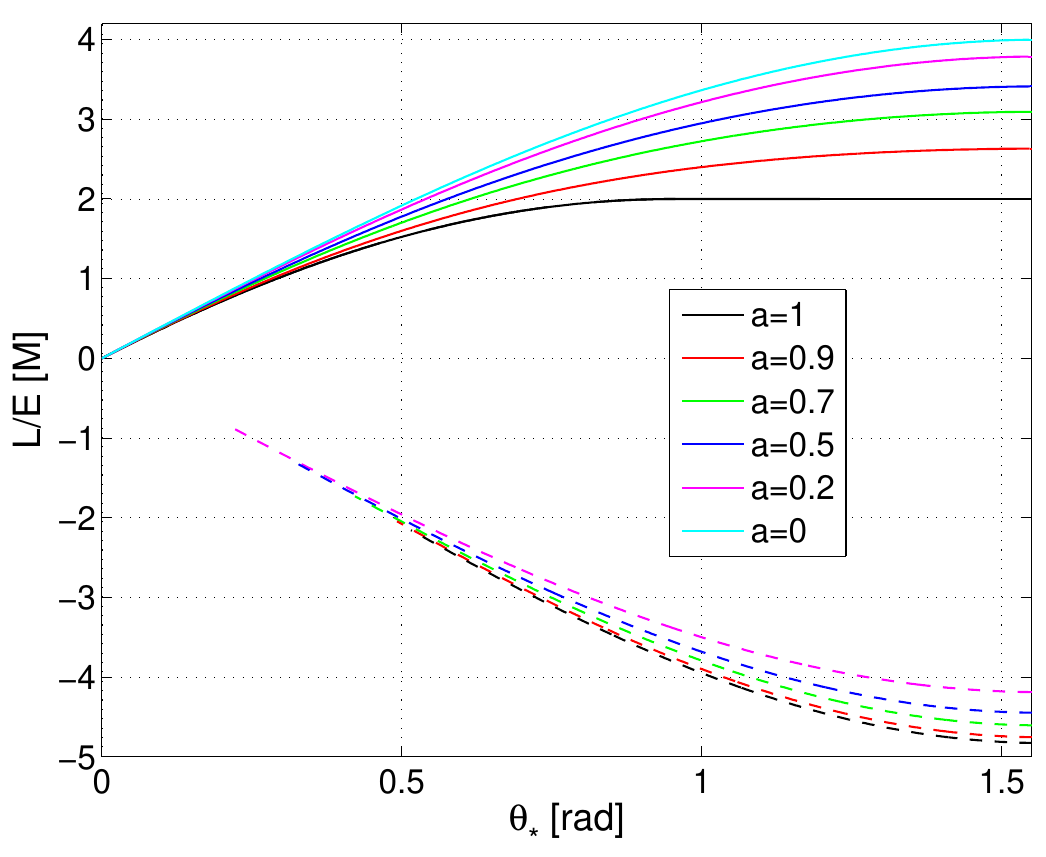}
\caption{Ratio $L/E$ (a.k.a.\ {\em impact parameter}) as a function of the latitudinal turning point $\theta_{\star}$ for spherical orbits at ISSO radius (left panel) and at MBSO radius (right panel) for different values of spin. Solid lines correspond to co-rotating orbits, whereas the dashed lines are for counter-rotating orbits.}
\label{isso_mbso_impact}
\end{figure}

These expressions for ISSO and MBSO radii were obtained using the equation of the separatrix curve in the ($e$, $\rho$) parameter plane of eccentricity $e$ and inverse-latus rectum $\rho$ of the trajectory. In particular, the ISSO radius is derived by inserting $e=0$ and $\rho=1/r_s$, while for the MBSO, the values $e=1$ and $\rho=1/2r_b$ need to be used. The polynomial that defines the separatrix is thus reduced to the degree of nine and eight, respectively. See \citet{rana19} and especially Appendix D therein for further details on the derivation of these relations.

In the relevant range of spin $a\in\left<0,1\right>$, both equations (\ref{isso}) and (\ref{mbso})  have two real roots such that $r>r_+$. The higher value of each pair corresponds to the counter-rotating spherical orbit, while the smaller value locates the co-rotating orbit. In the Schwarzschild limit ($a=0$), the values coincide at $r_{s}=6$ and $r_{b}=4$ corresponding to circular orbit in the static Schwarzschild spacetime. In \rff{isso_mbso_carter}, we plot the radii of ISSO and MBSO as functions of spin for different values of the Carter constant for both directions of orbits. In particular, we note that ISSO and MBSO radii of counter-rotating spherical orbits are located below those of circular orbits while the co-rotating spherical orbits have these radii always higher than circular orbits.

\begin{figure}[ht]
\center
\includegraphics[scale=.6]{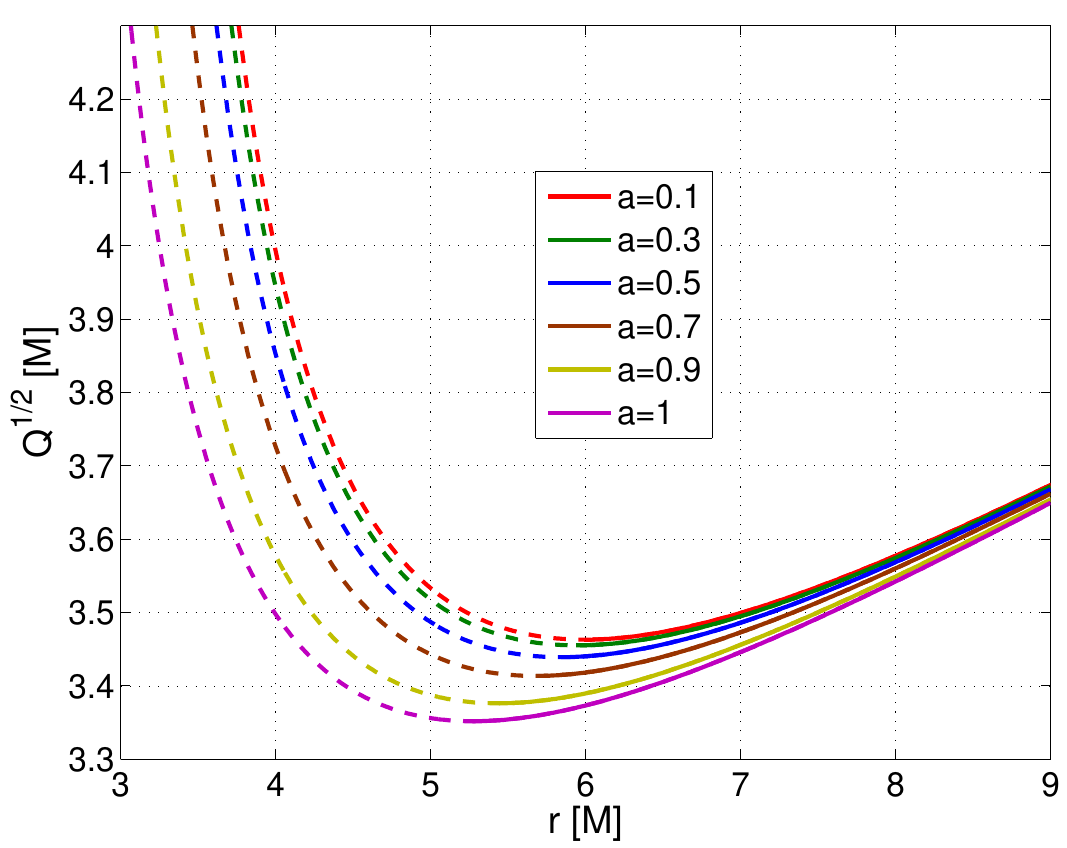}
\caption{Square root of Carter constant $Q_p$ of polar orbits is plotted with respect to their radii for various values of the spin parameter. Solid lines are used for stable orbits above ISPO, while dashed lines correspond to unstable polar orbits below ISPO.}
\label{carter_polar_fig}
\end{figure}

In \rffs{isso_mbso_inclination} and \ref{isso_mbso_impact}, we provide an alternative discussion of ISSO and MBSO radii and corresponding ratio $L/E$ with respect to the inclination angle $\theta_{\star}$ of spherical orbits with different values of spin. In particular, we notice that stable nor bound counter-rotating spherical orbits (dashed curves in \rffs{isso_mbso_inclination} and \ref{isso_mbso_impact}) can never have arbitrarily low inclination $\theta_{\star}$. Indeed, even if we start with the stable counter-rotating orbit (i.e.\ with a radius above the corresponding ISSO), we inevitably lose stability if we gradually decrease the inclination (i.e., increase the Carter constant of the particle) before the orbit becomes polar. On the other hand, in the case of co-rotating orbits, we can, in principle, have a stable orbit of an arbitrary inclination, provided that the radius is sufficiently large to remain above the corresponding ISSO curve. In particular, with the appropriate choice of the Carter constant, $Q=Q_p$, we get the polar orbit with $\theta_{\star}=0$.

The value of $Q_p$ corresponding to a polar orbit may be expressed as:
\begin{equation}
Q_p=\frac{-1-g^{tt}E_p^2}{g^{\theta\theta}},
\label{carter_polar}
\end{equation} 
where $E_p$ is given by \req{polar_energy}, while $g^{tt}$ and $g^{\theta\theta}$ are contravariant components of the Kerr metric (\ref{metric}). Values of $Q_p$ for various spins are plotted in \rff{carter_polar_fig}. Marginal stability and marginal binding of polar orbits may also be investigated using \reqs{isso} and (\ref{mbso}), if the value $Q=Q_p$ is inserted. The resulting radii of the marginally bound polar orbit (MBPO) and innermost stable polar orbit (ISPO) are plotted as a function of spin in \rff{rispo}. In particular, we observe that MBPO changes between $r\approx 3.38$ (with $a=1$) and $r=4$ for $a=0$ (circular orbit around Schwarzschild black hole). Boundary values of ISPO radii are $r\approx 5.28$ (corresponding to $a=1$) and $r=6$ in the Schwarzschild limit.

\begin{figure}[ht]
\center
\includegraphics[scale=.44]{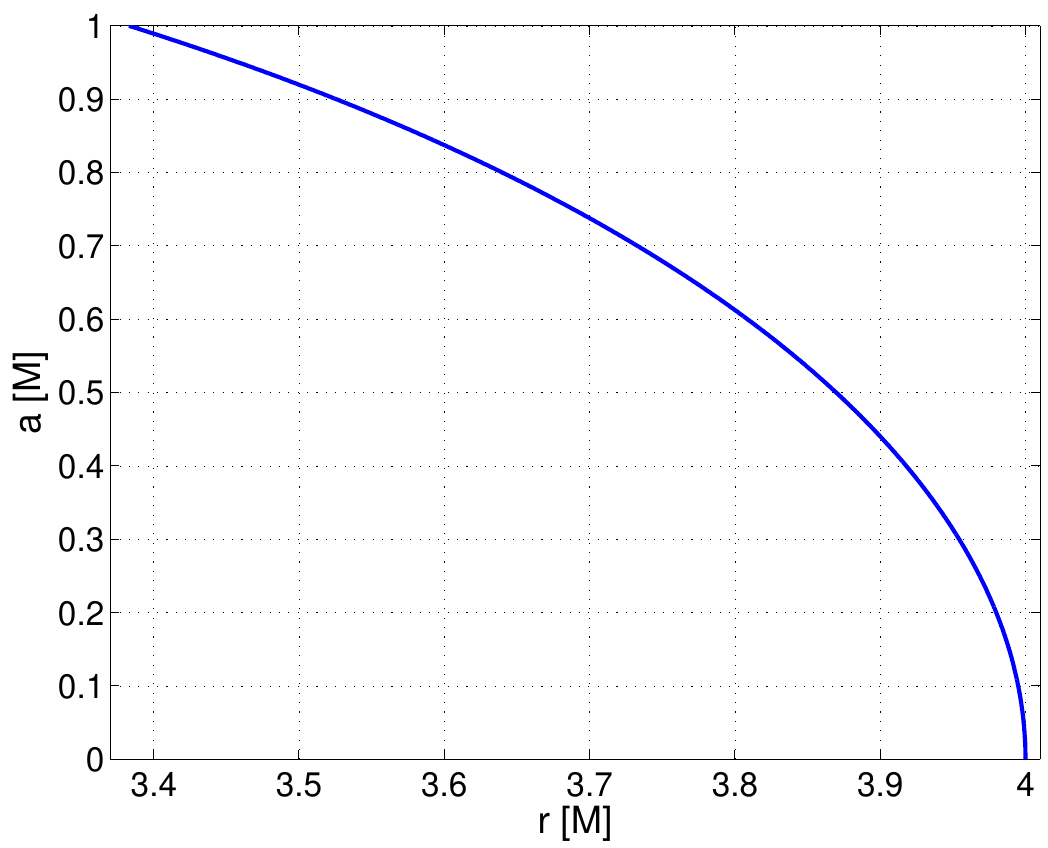}
\includegraphics[scale=.44]{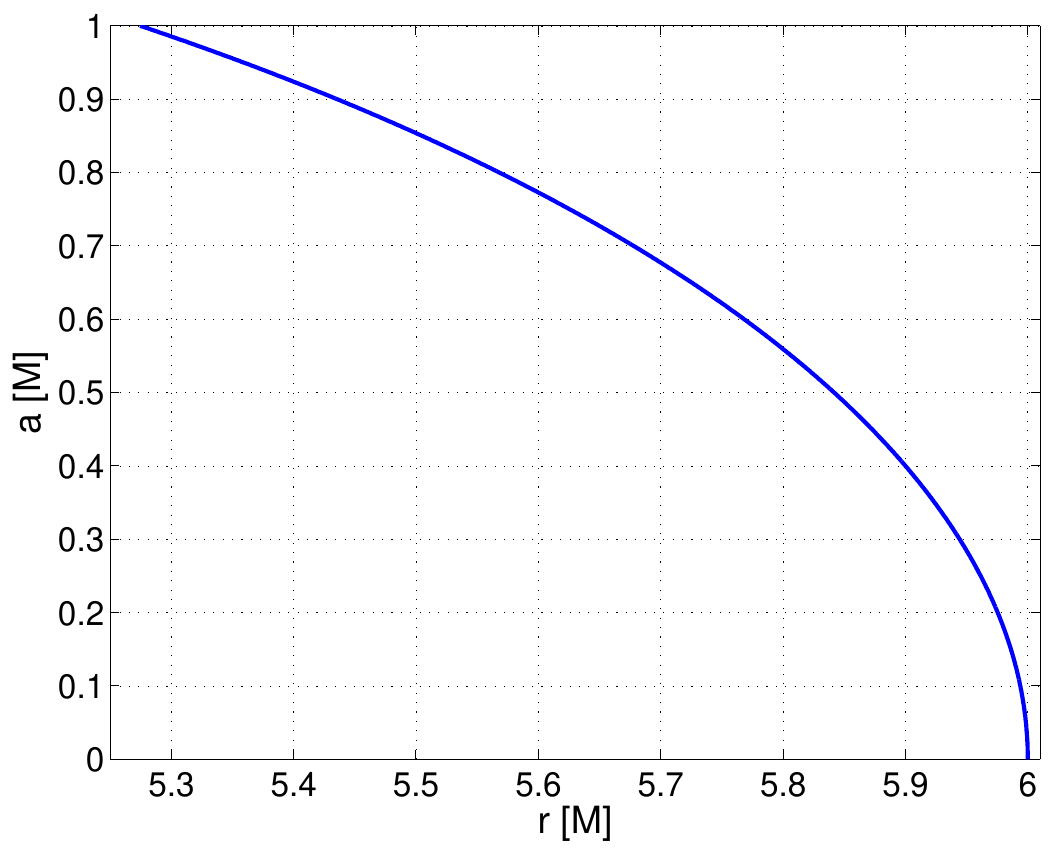}
\caption{Left panel: radii of marginally bound polar orbits (MBPO). Right panel: radii of innermost stable polar orbits (ISPO).}
\label{rispo}
\end{figure}

In the above discussion, we have studied spherical orbits of massive particles using two alternative parameters (with one-to-one correspondence), namely, the Carter constant $Q$ and the inclination angle $\theta_{\star}$. In particular, radii of the innermost stable spherical orbit (ISSO) and marginally bound spherical orbit (MBSO) were discussed for different spin values using both complementary parameterizations. Special attention was paid to the limiting case of polar orbits $\theta_{\star}=0$ for which the corresponding radii of the innermost stable polar orbit (ISPO) and marginally bound polar orbit (MBPO) were also determined. In the following \refsec{numerical}, we shall discuss the behavior of particle orbits below the ISSO (or ISPO), and we will investigate the nature of their instability numerically.

We note that spherical orbits in Kerr spacetime may also be followed by photons \citep{teo03}. It appears, however, that in the region of our interest (i.e., above the outer horizon at $r=r_+$) these are always unstable against perturbations in the radial direction. The discussion of spherical photon orbits is considerably simplified compared to the case of massive particles, nevertheless, both cases still share some interesting common features. Moreover, photon orbits near rotating black holes are also astrophysically relevant as they define the apparent shapes ({\em shadows}) of the black holes for the distant observers \citep{bardeen73,1976PhRvD..14.3281Y}. The concept of black hole shadow has recently been observationally confirmed with spectacular images from Event Horizon Telescope showing the supermassive black hole in M87 galaxy \citep{eht24} and Sagittarius A* in the center of the Milky Way \citep{eht23}.

Unlike massive particles, spherical photon orbits are obtained as a one-parameter family of orbits, i.e., for a given black hole spin $a$ the orbit is fully defined by radius $r$, while in the case of timelike geodesics, we always have one more parameter to select (see \reqs{E_shakur}--(\ref{L_shakur}) and \reqs{spherical_energy}--(\ref{spherical_angular})). Moreover, it appears that for null geodesics described by integrals $E$, $L$ and $Q$, only the two ratios $\phi\equiv L/E$ and $\eta\equiv Q/E^2$ are really independent \citep{bardeen73}. These ratios are directly related to the impact parameters of photons received by a distant observer and are thus relevant for the analysis of black hole shadows and gravitational lensing. In the case of spherical photon orbits their values are given as \citep{teo03}:
\begin{equation}
\phi=-\frac{r^3-3r^2+a^2r+a^2}{a\left(r-1\right)},\;\;\;\;\;\;\;\eta=-\frac{r^3\left(r^3-6r^2+9r-4a^2\right)}{a^2\left(r-1\right)^2}.
\label{impact_parameters}
\end{equation}
Spherical orbits are only allowed in the range of radii $r_1\leq r\leq r_2$, where $r_{1,2}=2\left\{1+\cos\left[\frac{2}{3}\arccos{ \left(\mp a\right) }\right]\right\}$ are the radii of unstable circular photon orbits in the equatorial plane with $\eta=0$. Radius $r_1$ corresponds to a co-rotating (prograde) orbit while at $r_2$ we get a counter-rotating (retrograde) circular orbit. The radial range between $r_1$ and $r_2$ is divided by an intermediate radius $r_3=1+2\sqrt{1-\frac{a^2}{3}}\cos{\left(\frac{1}{3}\arccos{\frac{(1-a^2)}{(1-\frac{a^2}{3})^{\frac{3}{2}}}}\right)}$ corresponding to the polar orbit with $\phi=0$. For $r_1\leq r< r_3$ the orbits are co-rotating ($\phi$ is positive) while for $r_3< r\leq r_2$ we get counter-rotating orbits with negative $\phi$. The value of $\eta$ is maximized within the range of radii of counter-rotating orbits which is analogical to the case of massive particles where also the maximal value of the Carter constant corresponds to the counter-rotating orbit (see \rff{diskuze_Lsf}). In the limit of maximally rotating black hole, $a=1$, the boundary radii of spherical photon orbits become $r_1=1$, $r_2=4$ and $r_3=1+\sqrt{2}$. For  $a=0$, they reduce to $r_1=r_2=r_3=3$, which defines the photon sphere of the Schwarzschild black hole. 

For more details on spherical photon orbits outside the Kerr black holes, we refer to \citet{teo03}. General analysis of photon orbits in Kerr and Kerr-Newman spacetimes is given by \citet{galtsov19}. A detailed review on black hole shadows is provided by \citet{perlick22}. In the rest of the article, we further analyze the spherical orbits of massive particles only.

\section{Numerical analysis of stable and unstable spherical orbits}
\label{numerical}
From the analysis performed in \refsec{spherical} we learn two basic facts regarding the stability of spherical orbits with given values of $Q$ and $a$: i) the radii $r_s$ of innermost stable spherical orbits (ISSO) below which the spherical orbits become unstable, and, ii) the radii $r_b$ of marginally bound spherical orbits (MBSO) below which they become unbound (their energy exceeds the rest energy, i.e., $E>1$ in dimensionless units) and may escape to infinity. Generally, $r_b\leq r_s$, while the both radii only coincide in the case of co-rotating circular orbits ($Q=0$) around maximally spinning black hole for which $r_b=r_s=r_{\pm}=a=1$.  

\begin{figure}[ht]
\center
\includegraphics[scale=.45]{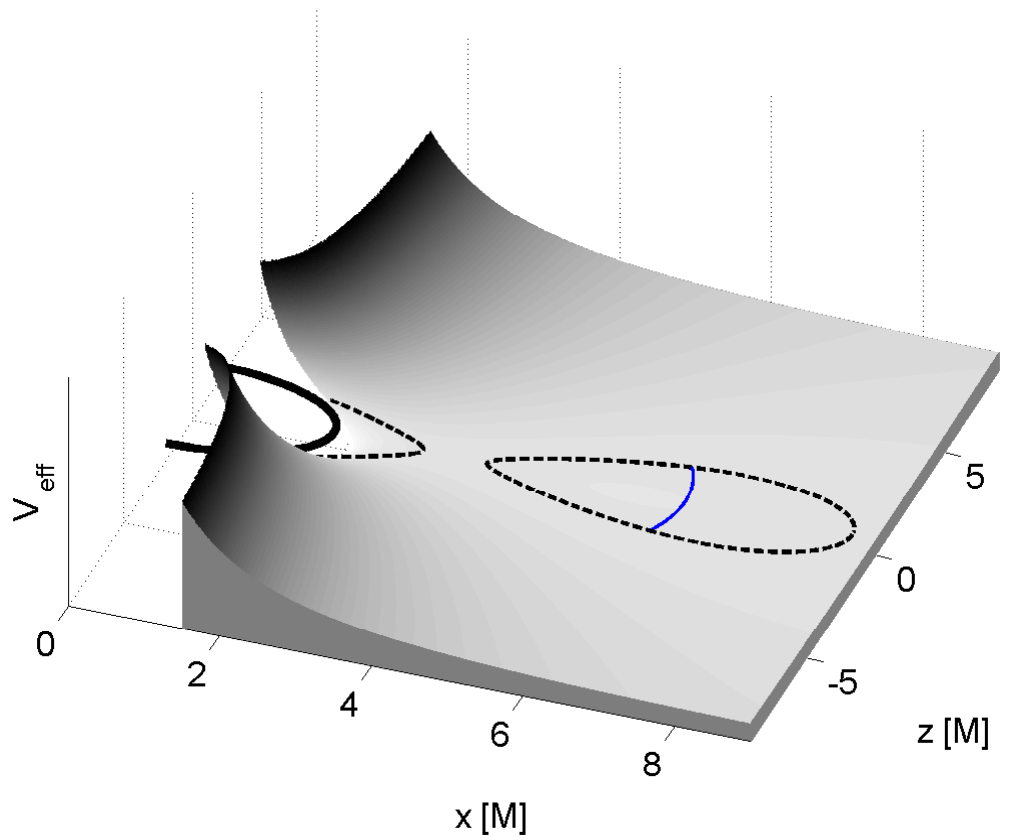}
\includegraphics[scale=.45]{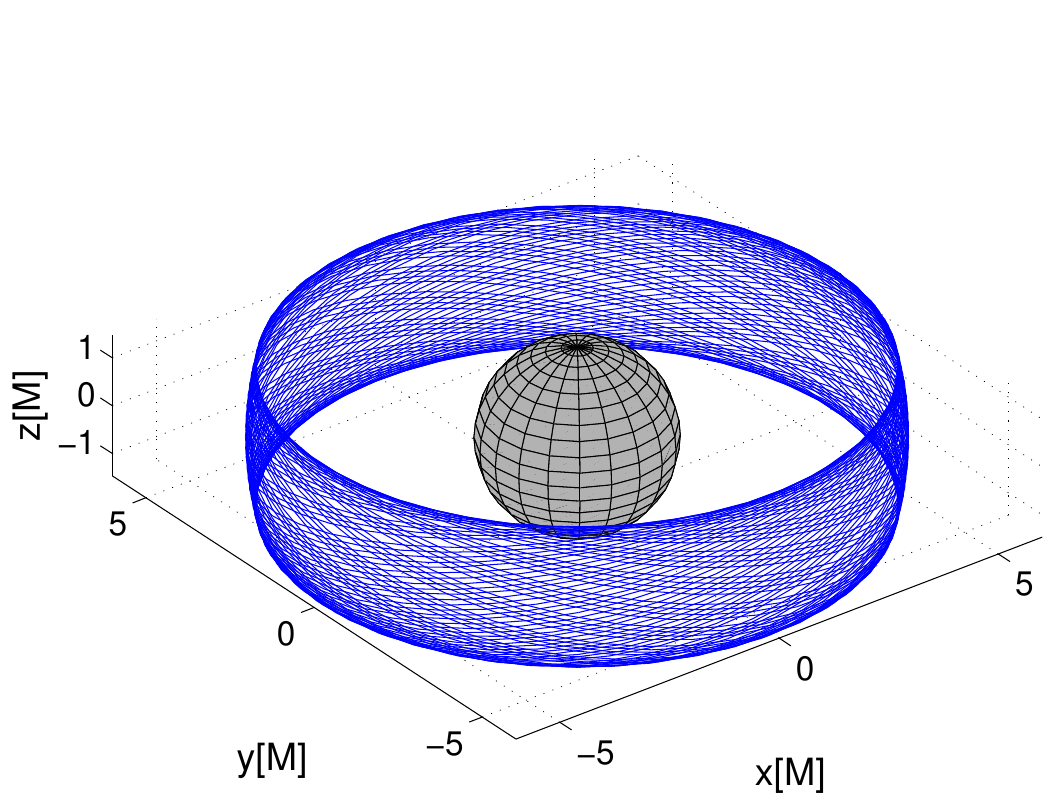}
\caption{Example of a stable spherical orbit above ISSO. The left panel shows the effective potential \req{eff_pot} of the particle with angular momentum given by \req{spherical_angular} while the equipotential curve of corresponding energy given by \req{spherical_energy} is marked by the dashed black line. The solid black line marks the horizon of the black hole. The blue curve shows the projected trajectory. In the right panel, the same trajectory is shown in three dimensions with the horizon marked by the grey sphere. Following parameters were employed: $r_0=6.5$, $Q^{1/2}=0.75$ and $a=0.5$.}
\label{example_stable_above_isso}
\end{figure}

An unstable spherical orbit maintains its initial radius and keeps evolving as a spherical orbit only if it remains completely unperturbed. However, in a real system (physical or numerically simulated), the fluctuations that perturb the dynamics are always present, and unstable orbits are therefore eradicated. As a result, a perturbed spherical orbit with initial radius $r_b<r<r_s$ may, in principle, plunge into the horizon or quasiperiodically oscillate in both radial and latitudinal directions. On the other hand, particles launched with $r_+<r\leq r_b$ may also escape to infinity as they have sufficient energy to surpass the attraction of the center. 

\begin{figure}[ht]
\center
\includegraphics[scale=.45]{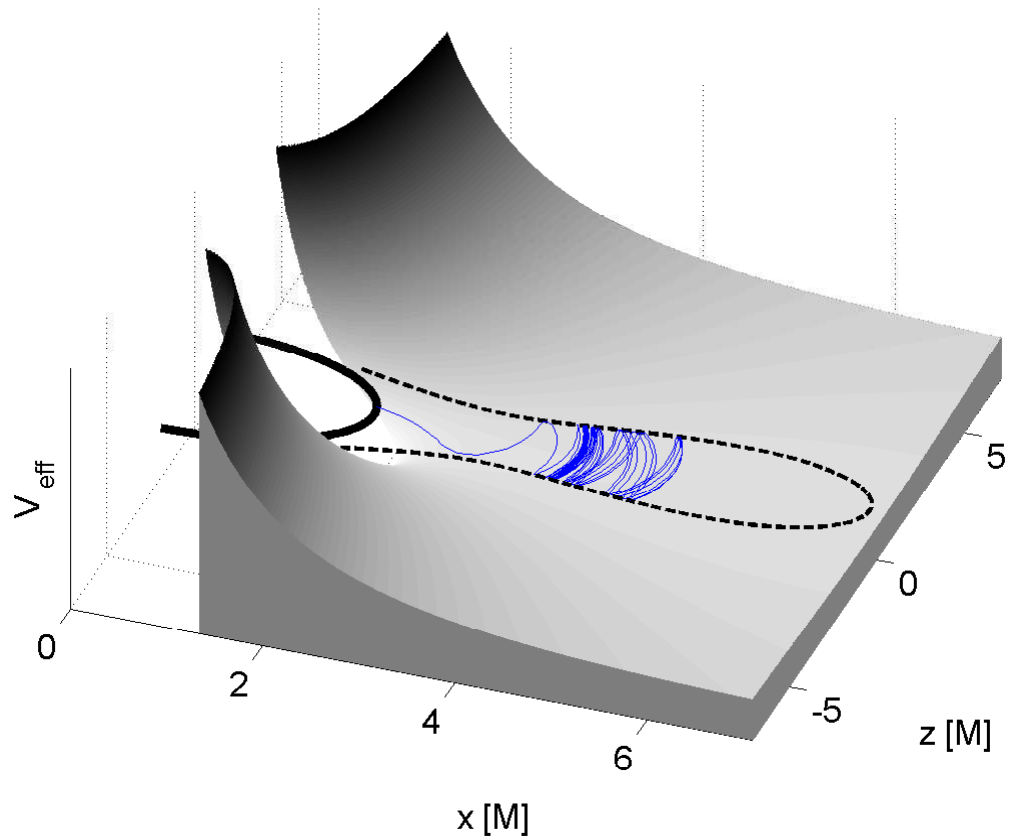}
\includegraphics[scale=.45]{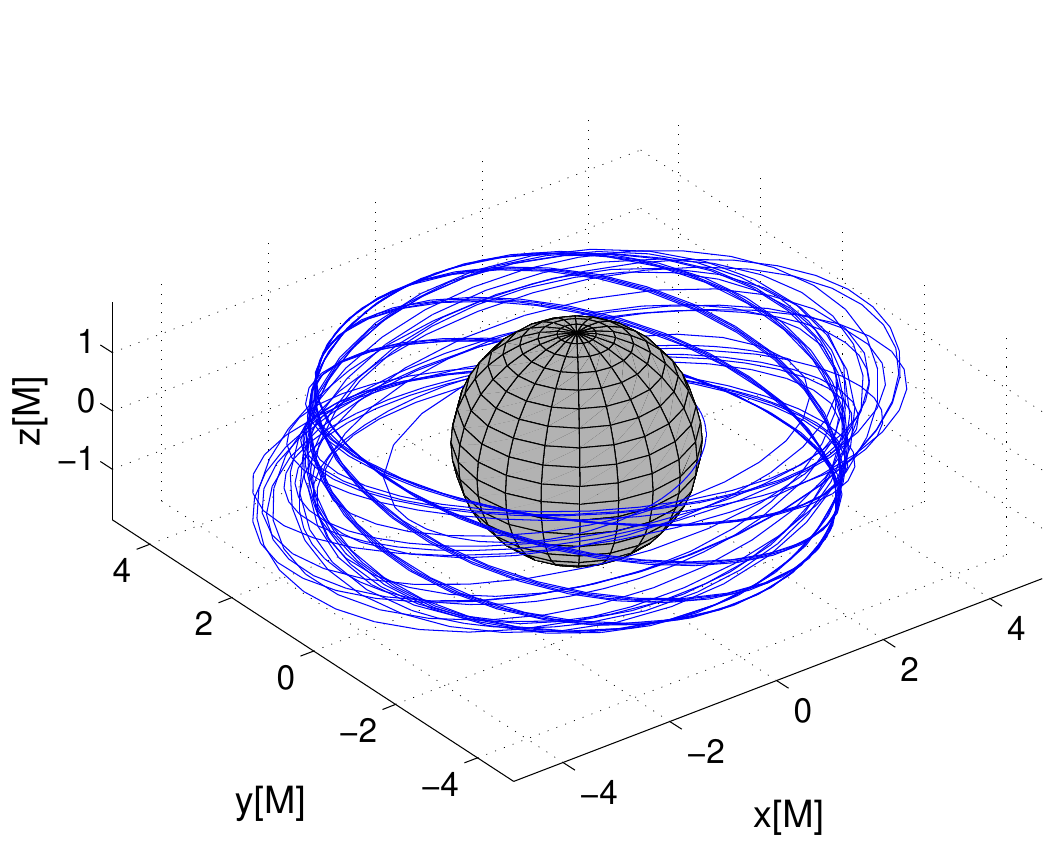}
\caption{Example of an unstable plunging spherical orbit launched below ISSO. Following parameters were employed: $r_0=4$, $Q^{1/2}=0.75$ and $a=0.5$.}
\label{example_plunge_below_isso}
\end{figure}

In \rffs{example_stable_above_isso}-\ref{example_plunge_and_escape_below_mbso} we present typical examples of the above-mentioned trajectories. Each trajectory is visualized in two ways. In the left panels, we show the projections of the trajectories to the poloidal plane $(x,z)$ being bounded by the relevant zero velocity curves (given by the equipotential curves of the effective potential \req{eff_pot}) as well as the shape of the potential surface in its neighborhood, while the right panels of the respective figures show the full 3D trajectory above the black hole horizon. Examples in \rffs{example_stable_above_isso}-\ref{example_plunge_and_escape_below_mbso} share the same value of the Carter constant and spin parameter ($Q^{1/2}=0.75$ and $a=0.5$), while they differ in the initial radii (and thus the values of $E$ and $L$ given by \req{spherical_energy} and \req{spherical_angular}, respectively). 

\begin{figure}[ht]
\center
\includegraphics[scale=.45]{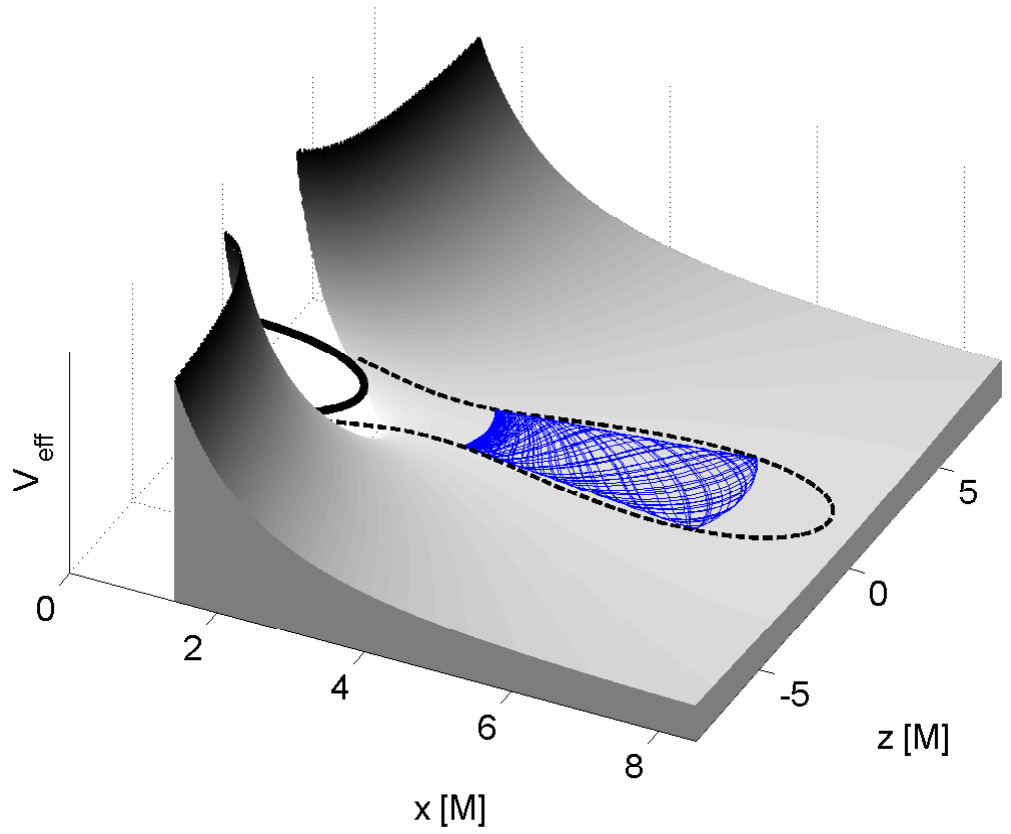}
\includegraphics[scale=.45]{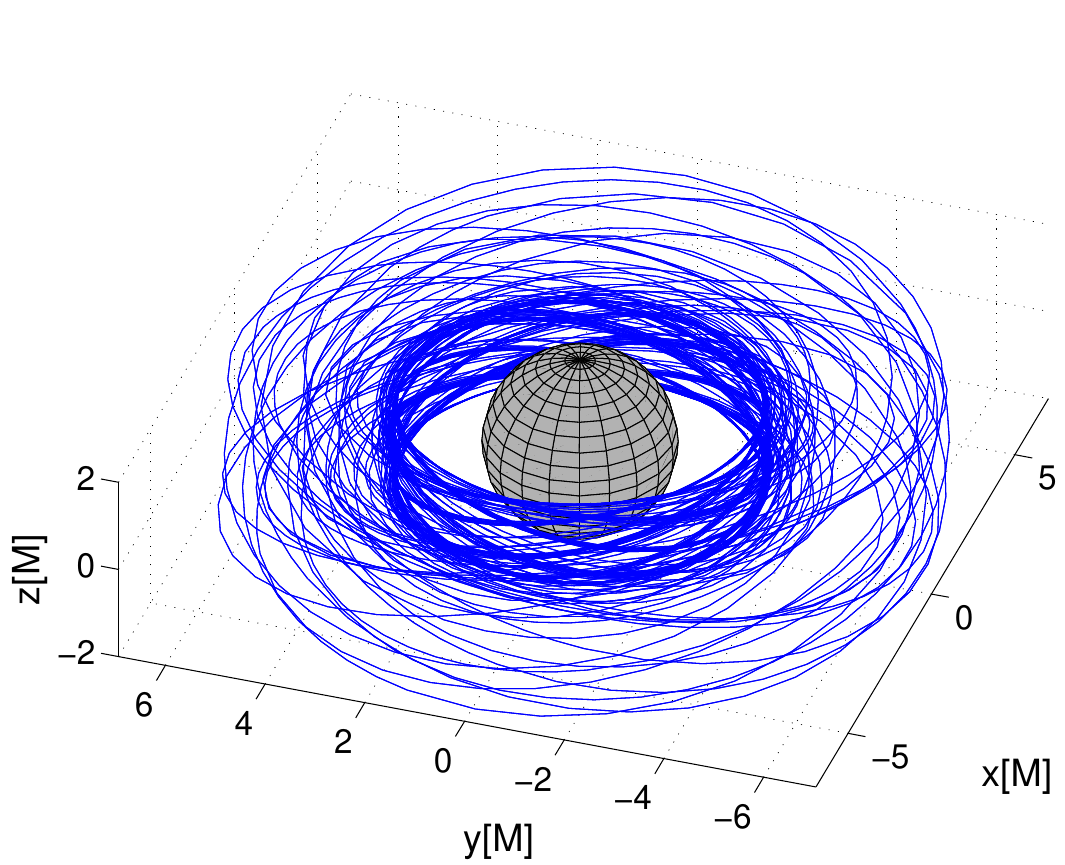}
\caption{Example of an unstable spherical orbit launched below ISSO which evolves into quasiperiodic radially bounded orbit due to numerical perturbation resulting from integration errors. Following parameters were employed: $r_0=3.6$, $Q^{1/2}=0.75$ and $a=0.5$.}
\label{example_stable_below_isso}
\end{figure}

\begin{figure}[ht]
\center
\includegraphics[scale=.45]{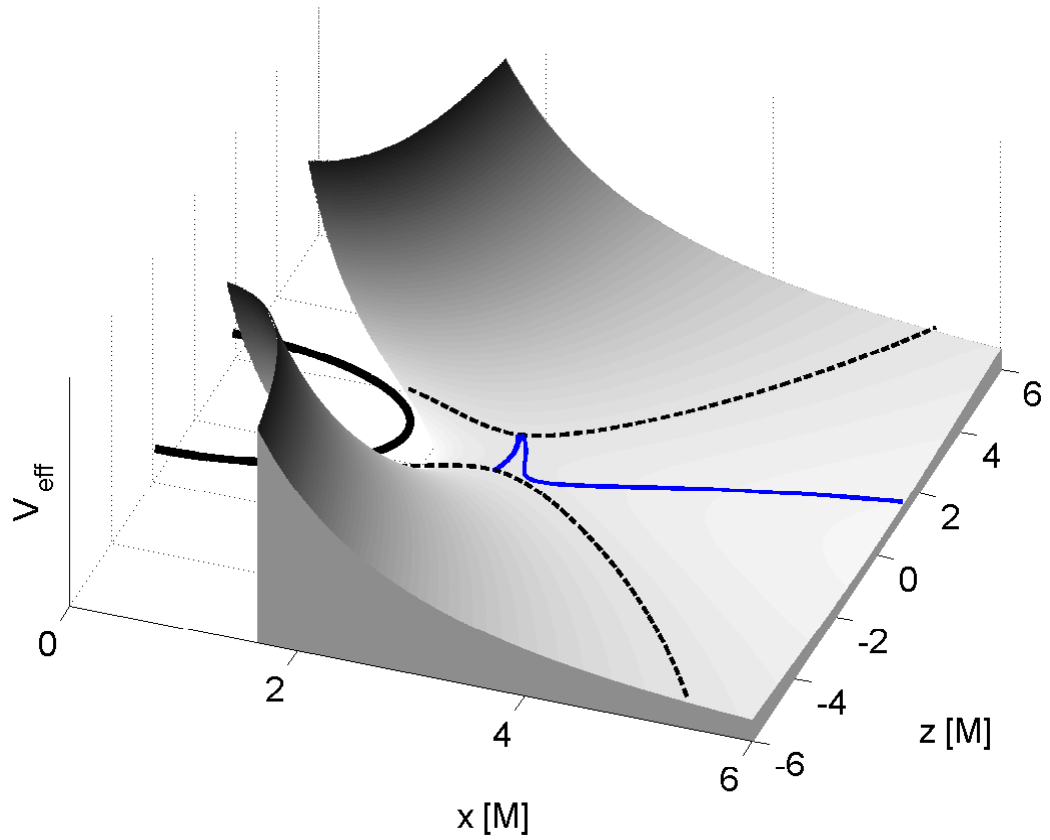}
\includegraphics[scale=.45]{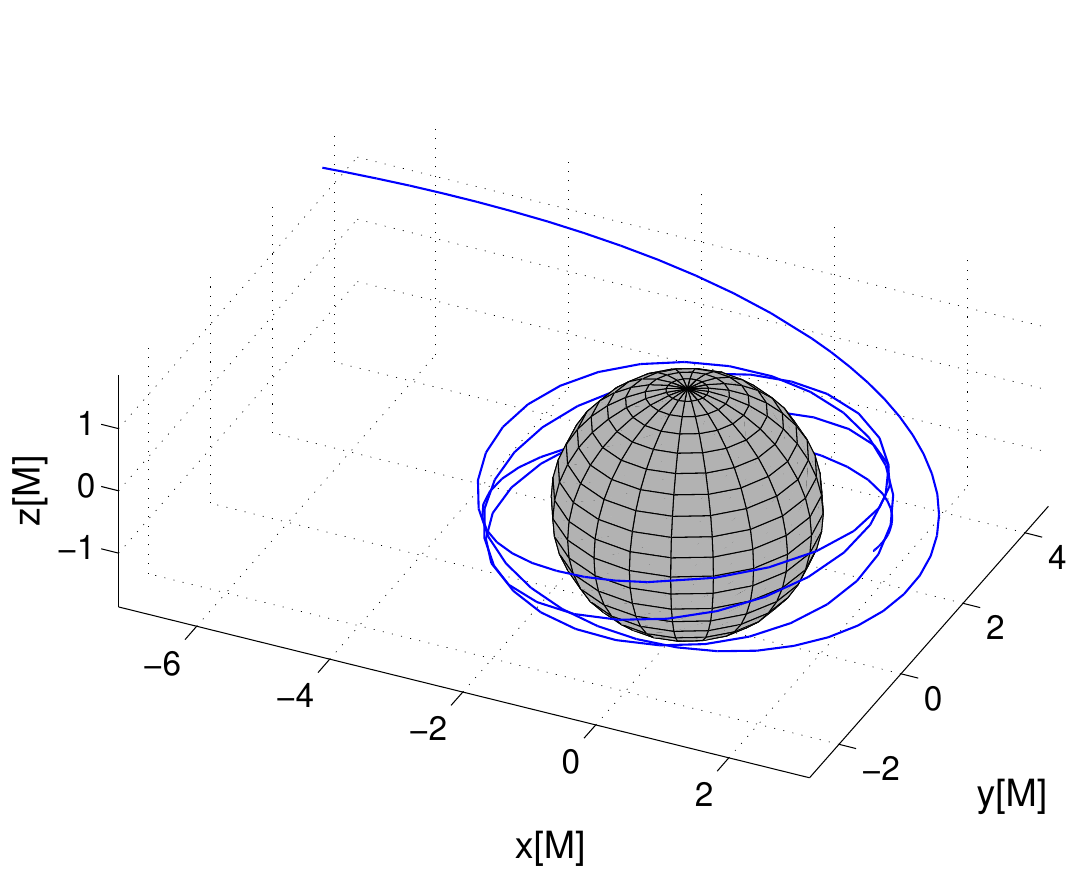}
\includegraphics[scale=.45]{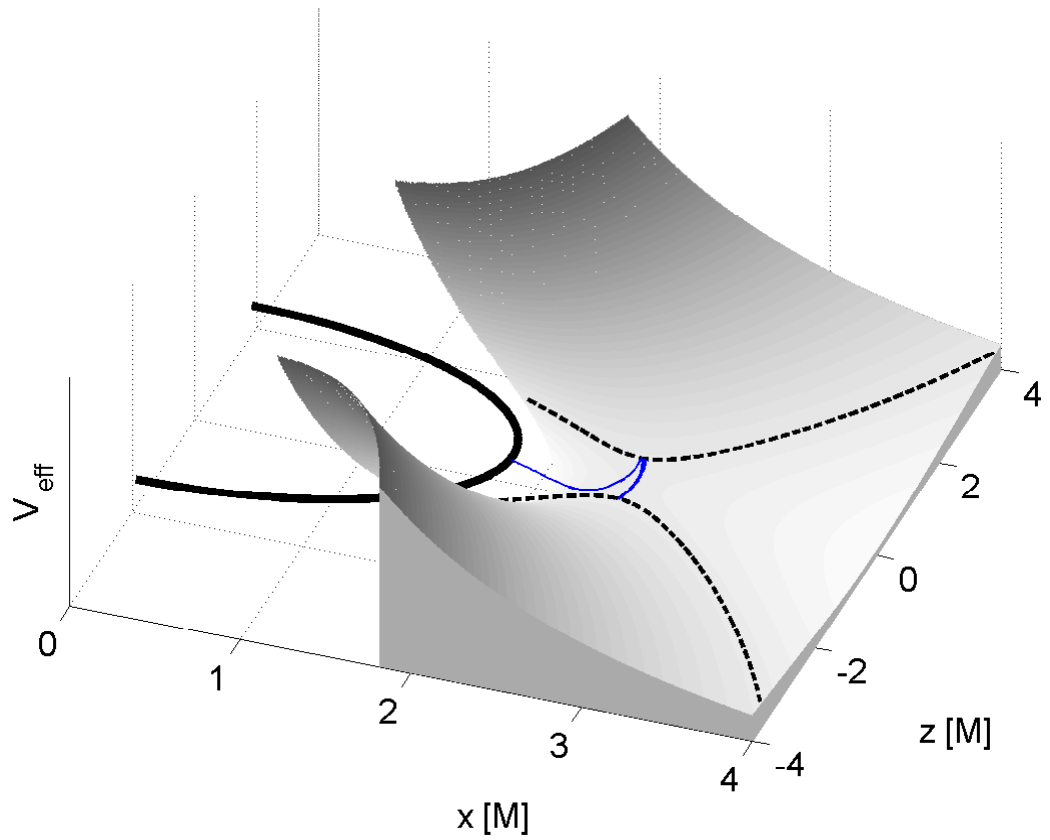}
\includegraphics[scale=.45]{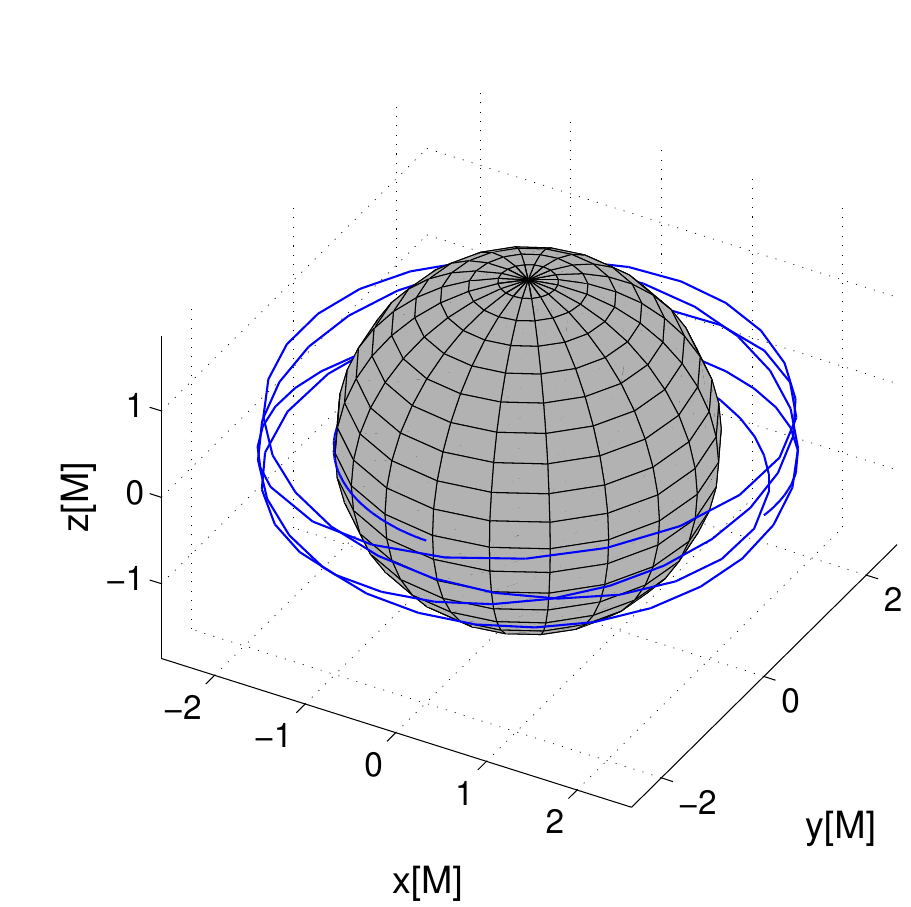}
\caption{Examples of unstable spherical trajectories below MBSO. Upper row: escaping particle launched from $r_0=2.8$. Bottom row: plunging particle launched from $r_0=2.6$. Same parameters as in \rffs{example_plunge_below_isso} and \ref{example_stable_below_isso} are used.}
\label{example_plunge_and_escape_below_mbso}
\end{figure}

A stable spherical orbit above ISSO is shown in \rff{example_stable_above_isso}. For an unstable orbit launched between radii of ISSO and MBSO, there are two options, i.e., plunge into the horizon (\rff{example_plunge_below_isso}) or quasiperiodic oscillations in radial and latitudinal directions (\rff{example_stable_below_isso}) while it largely depends on the initial perturbation of an unstable equilibrium which one will be realized. The particle launched below MBSO also has two options depending on the perturbation of the unstable orbit; it may plunge into the horizon or escape to infinity (both cases are shown in \rff{example_plunge_and_escape_below_mbso}).

The behavior of particles set on unstable geodesics below ISSO  and their possible stabilization on quasiperiodic orbit due to small perturbations represent astrophysically interesting problems that may be tackled by various approaches. In general, the relevant phase space consists of regions corresponding to different modes of motion which are separated by boundary denoted as {\em separatrix}. In particular, here we identify three different modes of behavior, i.e., radially bounded motion, plunge onto the horizon, and escape to infinity. Especially the separatrix between plunging and stable regions of Kerr geodesics is particularly relevant for the discussion of Extreme Mass Ratio Inspirals (EMRIs) which represent one of the key observational targets for the future LISA mission. In this context, it has been recently systematically studied using the parameterization of orbits by eccentricity and semilatus rectum \citep{stein20}. 

\begin{figure}[ht]
\center
\includegraphics[scale=.48]{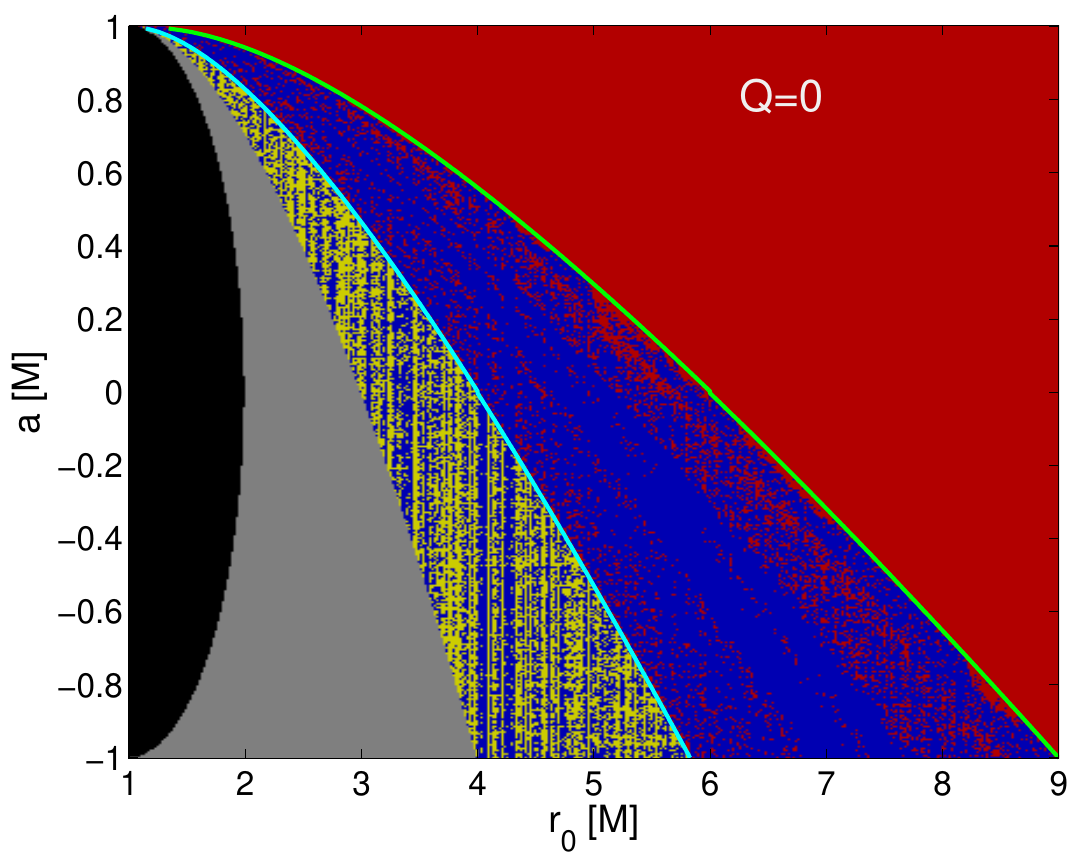}
\includegraphics[scale=.48]{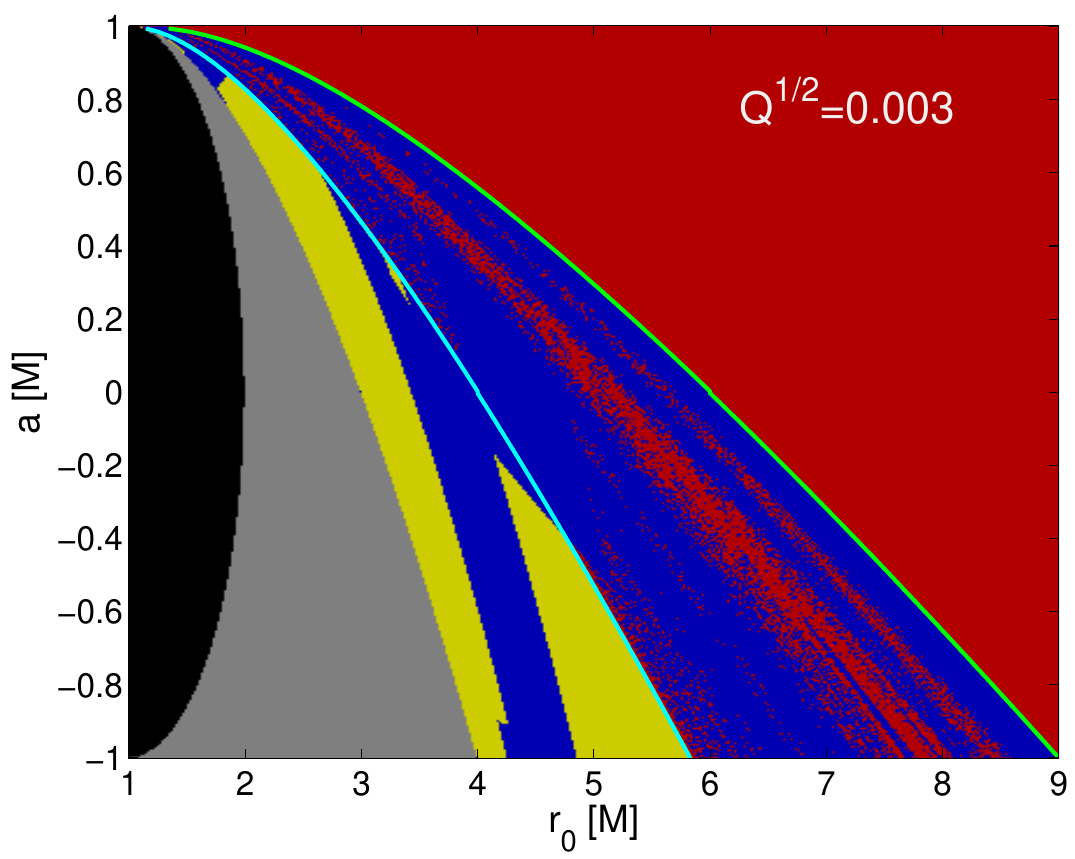}
\includegraphics[scale=.48]{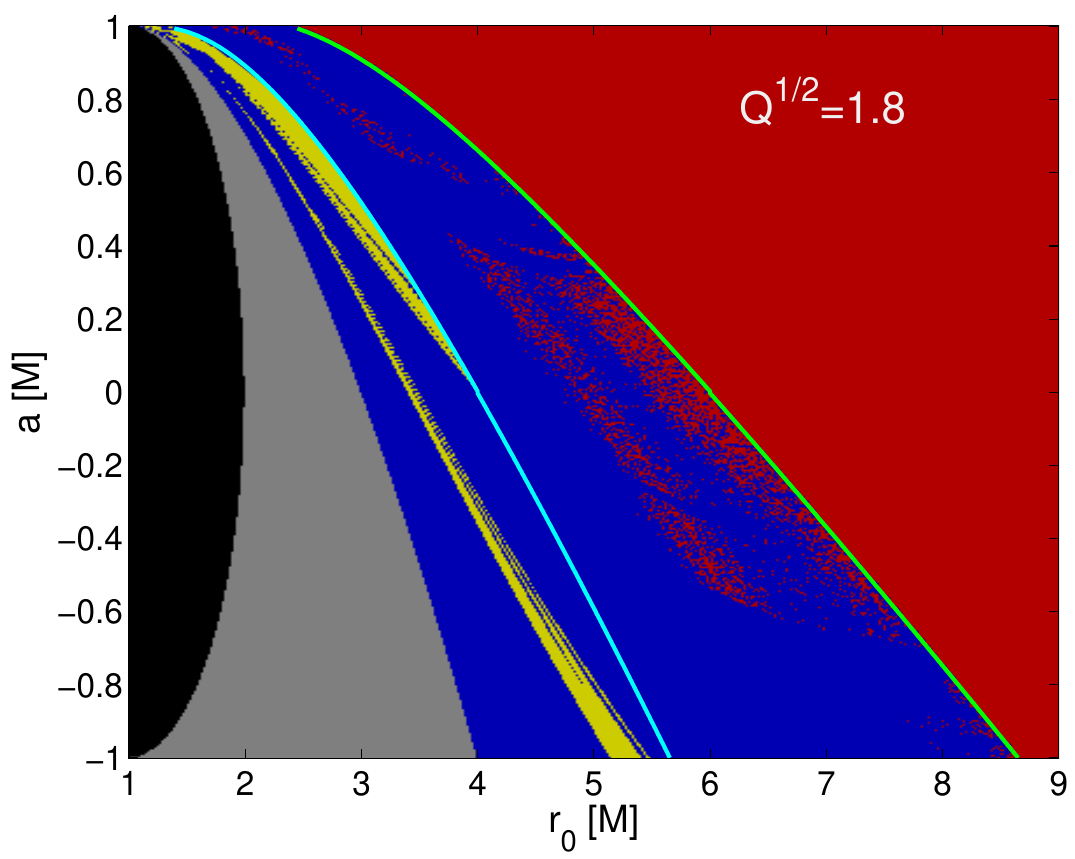}
\includegraphics[scale=.48]{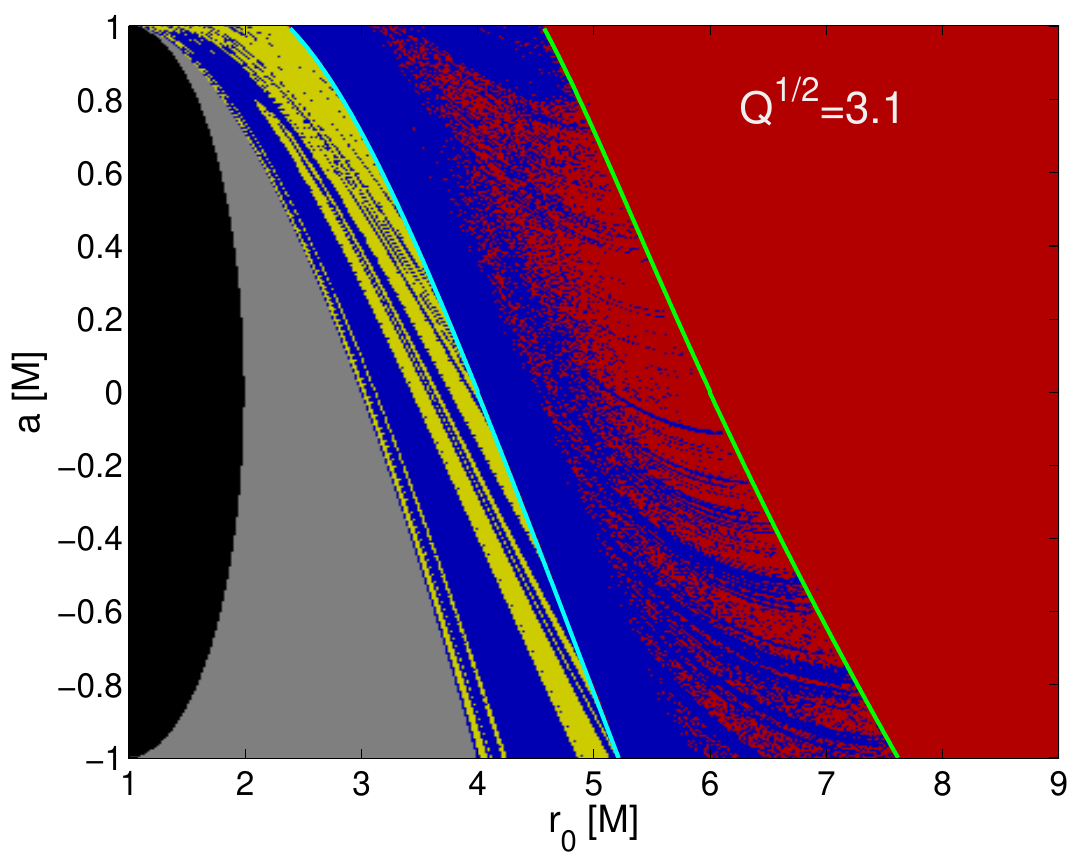}
\caption{Escape-boundary plots with coordinates ($r_0$, a) and fixed values of the Carter constant. Spherical trajectories presented in these plots are color-coded as follows: yellow for escaping, red for bound, and blue for plunging orbits. Black color indicates the black hole interior and grey shows the regions above the horizon where spherical orbits cannot exist. Locations of MBSO are shown by a cyan line and ISSO by a green line. Negative spin values correspond to counter-rotating orbits.}
\label{fixed_carter}
\end{figure}

The class of escaping orbits of particles forming an outflow of matter from the vicinity of the spinning black hole is also of particular astrophysical interest, especially in the context of observed high-energy cosmic rays. While the acceleration mechanism beyond cosmic rays is still a matter of debate and particular mechanisms are being discussed, active galactic nuclei powered by spinning supermassive black holes represent the main suspects for accelerating observed extra-galactic cosmic rays \citep[e.g., ][]{rodrigues21,tursunov20,kolos21}. While the cosmic rays are composed dominantly of charged particles (protons and electrons, in particular), the electrically neutral component is also present, although probably not reaching ultra-high energies \citep{navarro13}. Particles escaping from an unstable equilibrium of spherical orbit below MBSO (illustrated in \rff{example_plunge_and_escape_below_mbso}) may produce an outflow of electrically neutral particles if they are perturbed outward in the radial direction ($\delta(p_r)>0$). 

\begin{figure}[ht]
\center
\includegraphics[scale=.48]{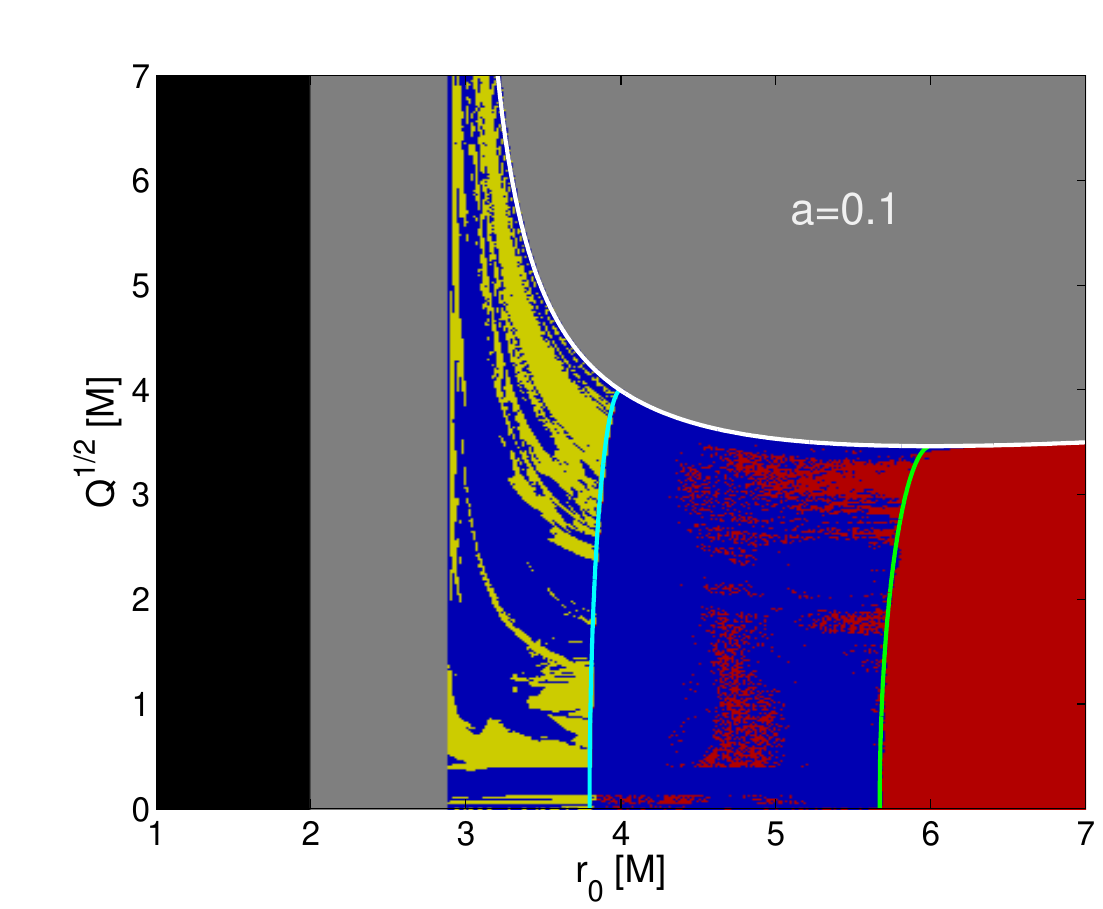}
\includegraphics[scale=.48]{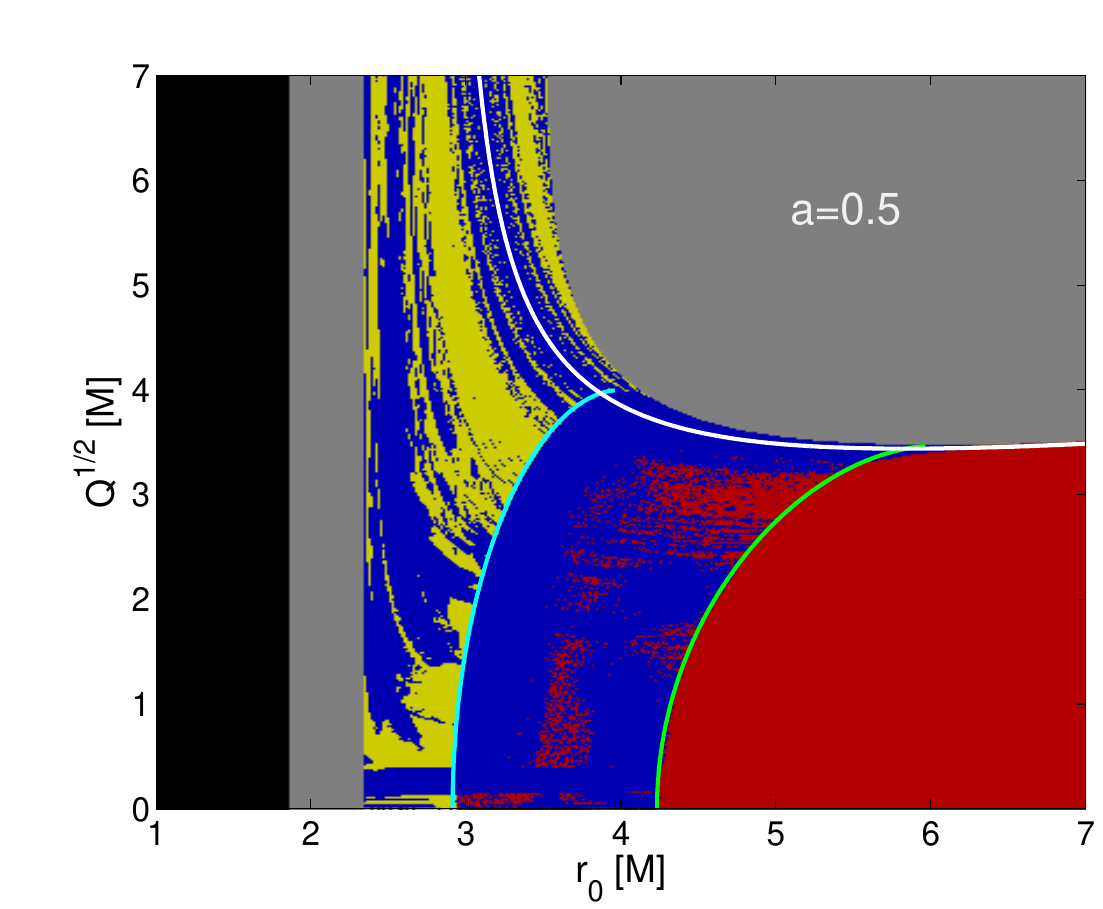}
\includegraphics[scale=.48]{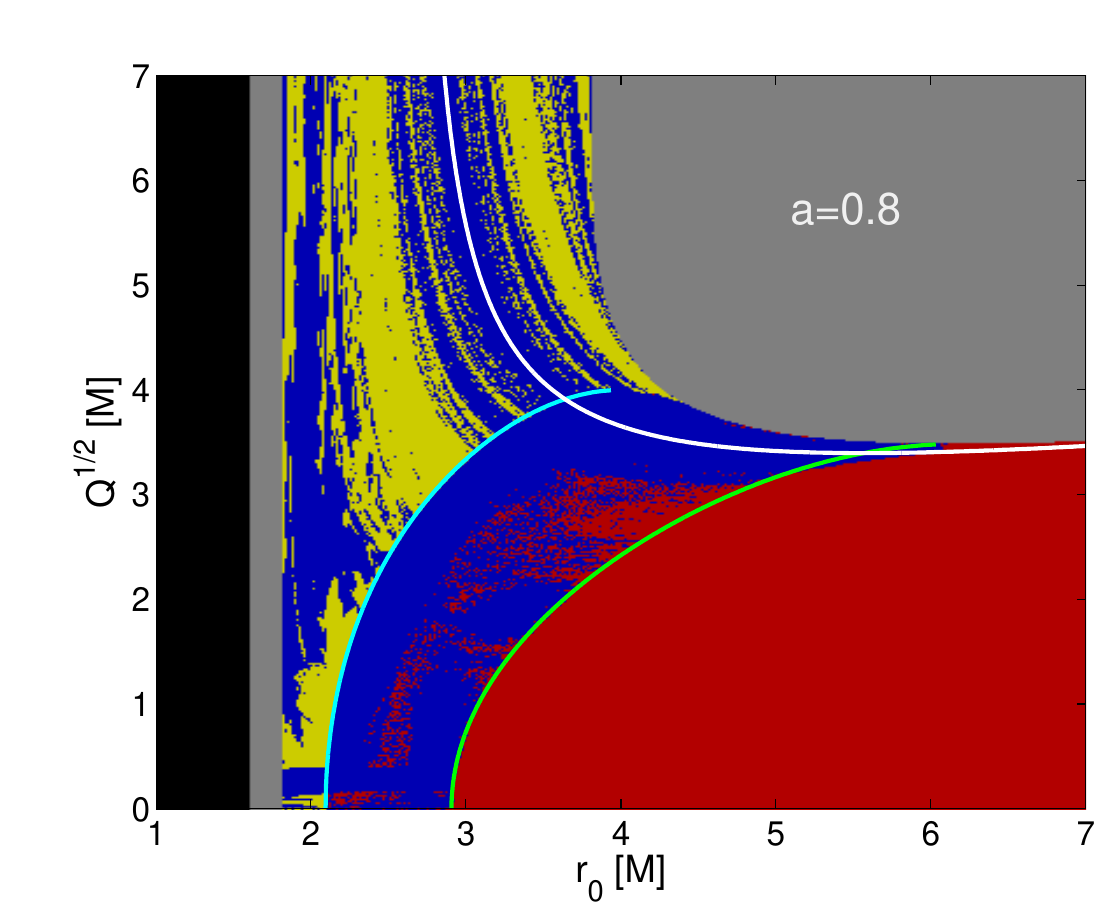}
\includegraphics[scale=.48]{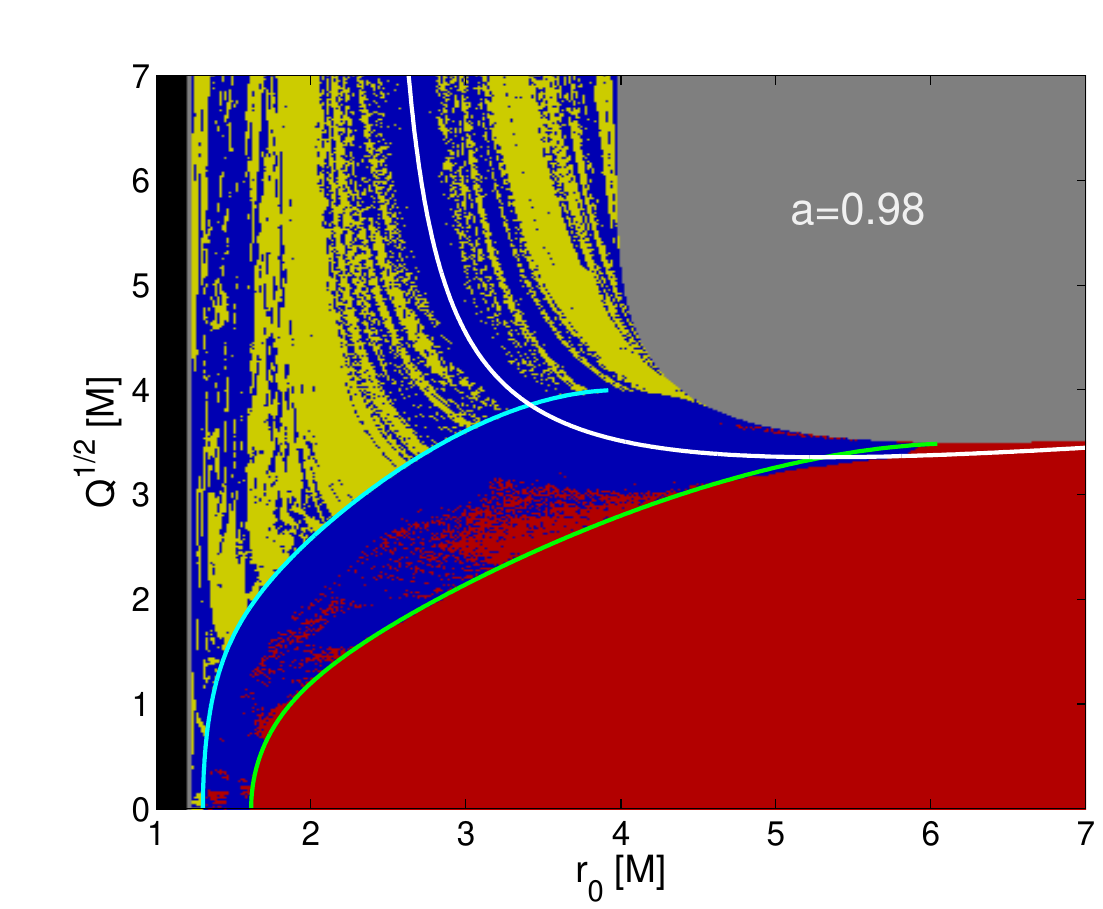}
\caption{Escape-boundary plots with coordinates ($r_0$, $Q^{1/2}$) and fixed values of spin parameter. Spherical trajectories presented in these plots are color-coded as in \rff{fixed_carter}. Besides the MBSO radii (cyan line) and ISSO radii (green line), we also show the locations of polar orbits by a white line.}
\label{fixed_spin}
\end{figure}

\begin{figure}[ht]
\center
\includegraphics[scale=.5]{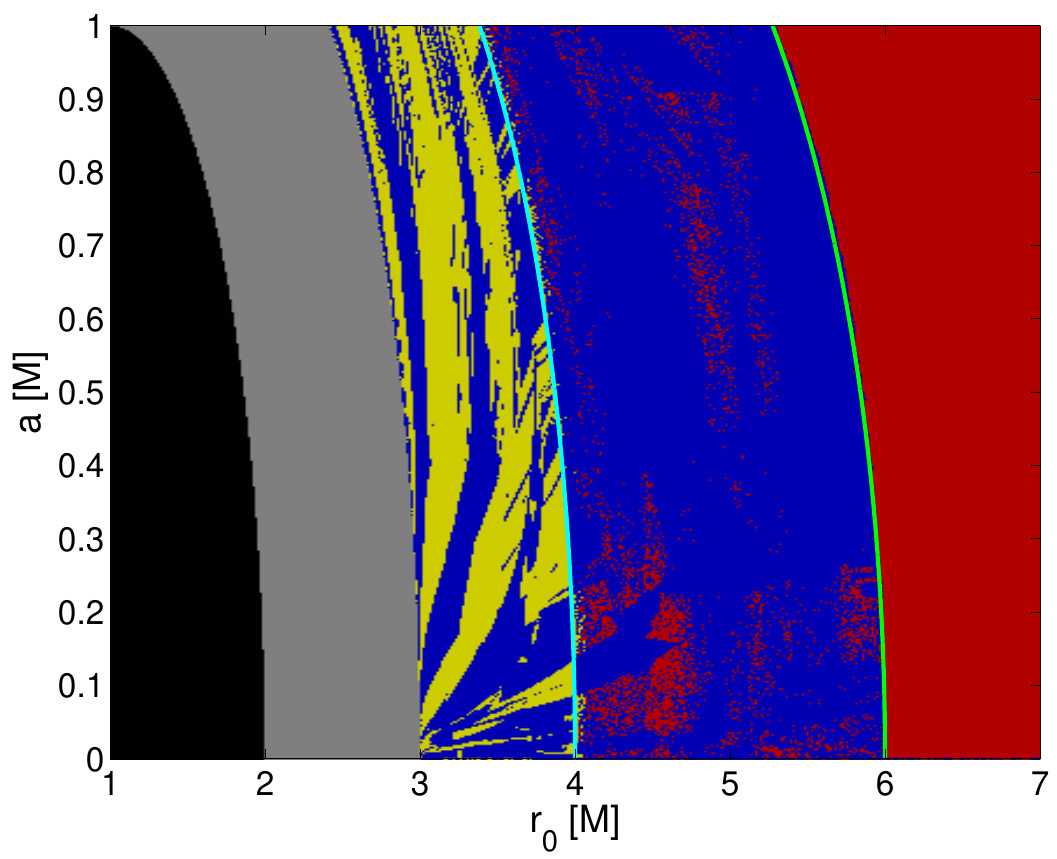}
\caption{Escape-boundary plot of polar orbits is shown in ($r_0$, a) plane. Color coding as in \rffs{fixed_carter} and \ref{fixed_spin}. MBPO radii are indicated by a cyan line and ISPO radii by a green line.}
\label{polar}
\end{figure}

In our previous studies \citep{kopacek20,kopacek18} we investigated a different dynamical system with escapes \citep{karas21,stuchlik16} by a straightforward, yet effective, approach based on escape-boundary plots. Such graphic representation of the dynamics is inspired by standard basin-boundary plots which visualize basins of attraction of attractors in systems with dissipation. Although conservative systems do not have real attractors we may consider escaping orbits (if present) as being attracted by an attractor at infinity \citep{contopoulos10,contopoulos06}. If we pursue this analogy, we may also consider the horizon of the black hole as an attractor for plunging orbits and explore which parts of the phase space belong to which attractor and determine the initial conditions that belong to their basins of attraction. If we visualize a particular set of trajectories (differing in initial conditions or other parameters) in a way that distinguishes them (e.g., by assigning different colors) by their final states (i.e., plunge, escape, or bounded motion, in our case) we obtain an escape-boundary plot as an analogy of basin-boundary plot for the conservative system with escapes. 

The method of escape-boundary plots has been successfully applied in a previously studied non-integrable system of charged particles in magnetized Kerr background \citep{kopacek18,karas13}. Due to the non-integrability of this system \citep[e.g., ][]{mukherjee23}, we have encountered fractal structures of the boundaries with self-similar patterns which were associated with chaotic dynamics \citep{kopacek20} in which case specific methods, e.g., recurrence analysis \citep{marwan07}, become especially useful \citep{kopacek20, glg18}. Nevertheless, in a current study we examine a completely integrable system of Kerr geodesics where the deterministic chaos cannot develop, and the boundaries between domains of stable motion, plunge, and escape should therefore be regular (i.e., with non-fractal dimension) and, in particular, they should respect ISSO and MBSO radii determined in \refsec{spherical}. 

In order to numerically verify the above-given assumptions, we construct the escape-boundary plots of two types: i) with coordinates $r_0$ and $Q^{1/2}$ and fixed values of spin shown in \rff{fixed_carter}, and ii) with coordinates $r_0$ and $a$ and fixed $Q$ presented in \rff{fixed_spin}. The special case of polar orbits is treated separately in \rff{polar}. To obtain these plots we use the grid with resolution $400 \times 400$, where each point represents a spherical trajectory launched at $r_0$ in the equatorial plane with constants $E$ and $L$ given by \reqs{spherical_energy} and (\ref{spherical_angular}). We evolve each trajectory until $\lambda_{\rm{fin}}$ (see Appendix \ref{appa} for details) and assign the color to each point of the grid according to the following convention: yellow for escape ($r_{\rm{fin}}\geq r_e$), red for radially bounded trajectory with $r_+ < r_{\rm{fin}}<r_e$ (i.e., not necessarily with constant $r$), and blue for the plunge. The choice of the escape threshold radius $r_e$ is discussed in Appendix \ref{appa}.  In addition, we use black color for the black hole interior ($r\leq r_+$) and grey for the parts of the parameter space where the spherical orbits are not defined (values given by \reqs {spherical_energy} and (\ref{spherical_angular}) become complex). We also present MBSO (cyan line) and ISSO (green line) radii in these plots. Moreover, for the case of escape boundary plots in ($r_0$, $Q^{1/2}$) plane shown in \rff{fixed_spin}, we highlight the parameters' values corresponding to polar orbits by a white line. The case of polar orbits ($L=0$; $\theta_{\star}=0$) is presented in \rff{polar}, where the energy of the test particles is set according to \req{polar_energy} and the value of the Carter constant is given by \req{carter_polar}.

\begin{figure}[ht]
\center
\includegraphics[scale=.99,trim={1cm -1cm 0cm 0cm}]{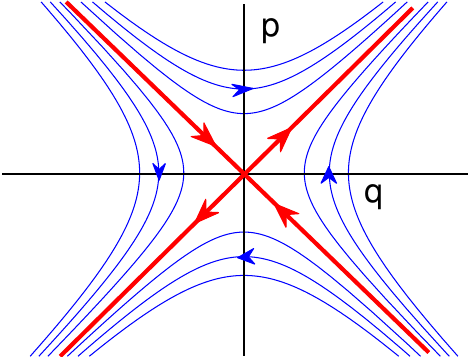}
\includegraphics[scale=.5]{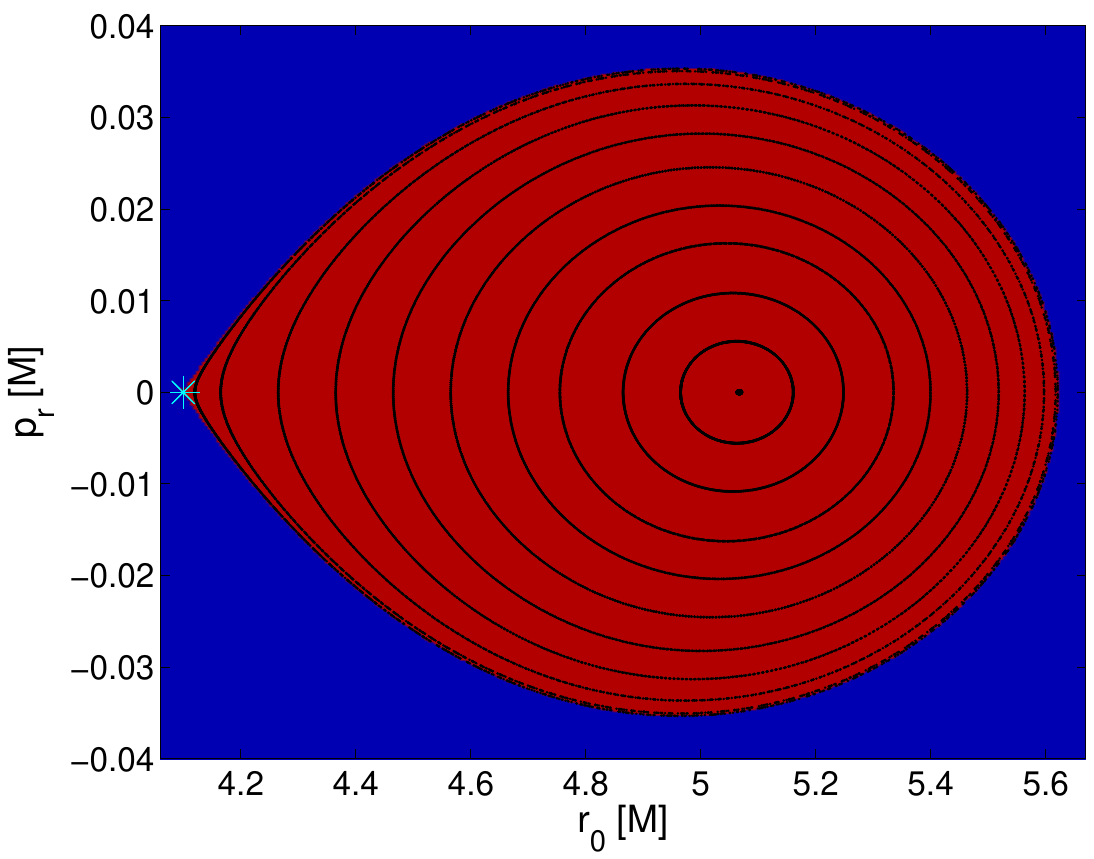}
\caption{Left panel: a sketch of the dynamics in the neighborhood of a saddle-type unstable hyperbolic fixed point in a two-dimensional dynamical system with phase-space coordinates $(q,p)=(0,0)$. Separatrix (red) and flow lines (blue) in different phase-space regions are shown. Right panel: example of actual dynamics near an unstable spherical orbit between MBSO and ISSO radii with parameters $r_0 =4.1$, $Q^{1/2}=3$ and $a=0.9$ (light-blue asterisk). The blue region corresponds to plunging orbits, red region consists of radially bounded quasiperiodic orbits. Poincar\'{e} surface of section (taken at $\theta=\pi/2$ for $\dot{\theta}>0$) with canonical coordinates ($r$, $p_r$) of several quasiperiodic trajectories within the bound region differing in initial radii is shown with black color.}
\label{hyperbolic_point_sketch}
\end{figure}

Inspecting the escape-boundary plots presented in Figs.~\ref{fixed_carter}--\ref{polar} we find that they confirm the previous analysis of ISSO and MBSO locations presented in \refsec{spherical}. Indeed, we observe that: i) the regions above ISSO lines are entirely red since spherical orbits launched here are stable against small numerical perturbations caused by integration errors, and ii) yellow zones of escaping particles only appear below MBSO lines where the energy of particles becomes sufficient for the escape. However, in regions between MBSO and ISSO lines and below the MBSO line, we observe an irregular mixture of different colors; blue and red dots in the former case, and blue and yellow dots in the latter case. 

For the case of circular orbits (upper left panel of \rff{fixed_carter}) the distribution of different orbits in these regions appears random while it becomes more organized for spherical orbits ($Q>0$) and large escape zones also develop here. Especially in Figs.~\ref{fixed_spin} and \ref{polar} the structures in these regions may resemble fractal geometry encountered in analogous plots of escape zones in a previously studied non-integrable system \citep{karas21,kopacek20}. Nevertheless, since here we investigate a fully integrable system, these unexpected features associated with deterministic chaos are necessarily of a numerical origin, i.e., caused by numerical errors of the integration. Although the integration errors may be largely decreased by an appropriate choice of the integrator (discussed in Appendix \ref{appa}), they can never be completely avoided. The reason why these small numerical perturbations are so crucial in the region below ISSO is obviously the fact that in this zone we are dealing with the dynamics of an unstable equilibrium. Dynamical systems close to such an instability must be studied with special caution and appropriate tools must be used.  Artificial (numerically induced) fractal structures observed in Figs.~\ref{fixed_carter}--\ref{polar} show that for the regions below ISSO the escape-boundary plots of this type are not an optimal method and may even become misleading. 

Based on the analysis of escape-boundary plots in Figs.~\ref{fixed_carter}--\ref{polar} we may only conclude, that parameters $a$ and $Q$ strongly affect the behavior of numerically perturbed unstable spherical orbits. In particular, regarding the red-blue zone between MBSO and ISSO, these plots suggest that for co-rotating orbits the Carter constant increases the probability that a perturbed orbit becomes radially bounded, while for counter-rotating orbits it slightly increases the probability of the plunge. The spin parameter appears to have the opposite effect; slightly increasing the probability of plunge for co-rotating orbits while supporting the stabilization of perturbed counter-rotating orbits into quasiperiodic radially bounded orbits. Nevertheless, since the numerical perturbation is not random and non-trivially depends on the integration scheme, such probabilistic interpretation is necessarily vague and needs to be confirmed by other means. Moreover, regarding the yellow-blue zone below MBSO we may draw hardly any conclusion here and irregular fractal-like structures present in the plots make the applicability of this approach in this zone even more questionable. 

\begin{figure}[ht]
\center
\includegraphics[scale=.36]{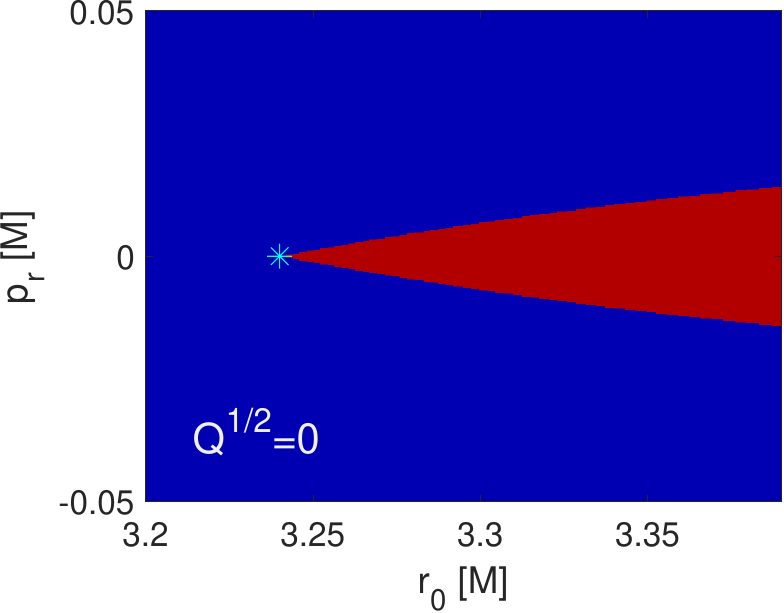}
\includegraphics[scale=.36,trim={0cm 0cm 0cm 0cm}]{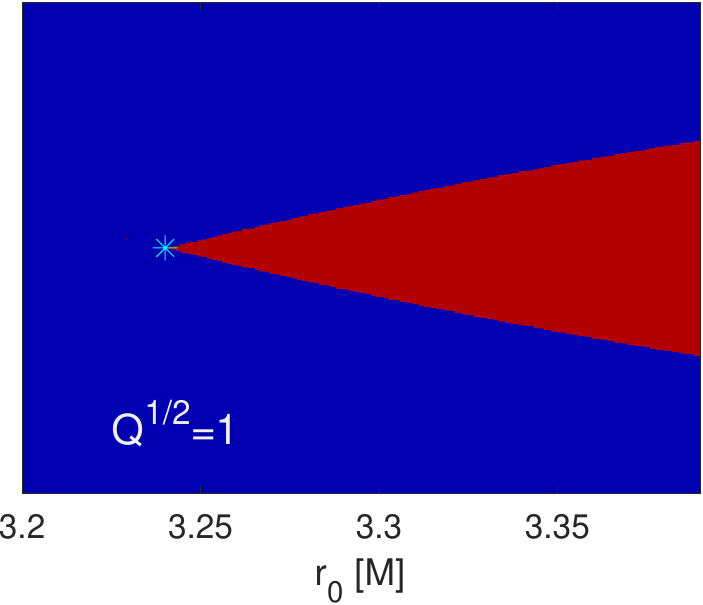}
\includegraphics[scale=.36]{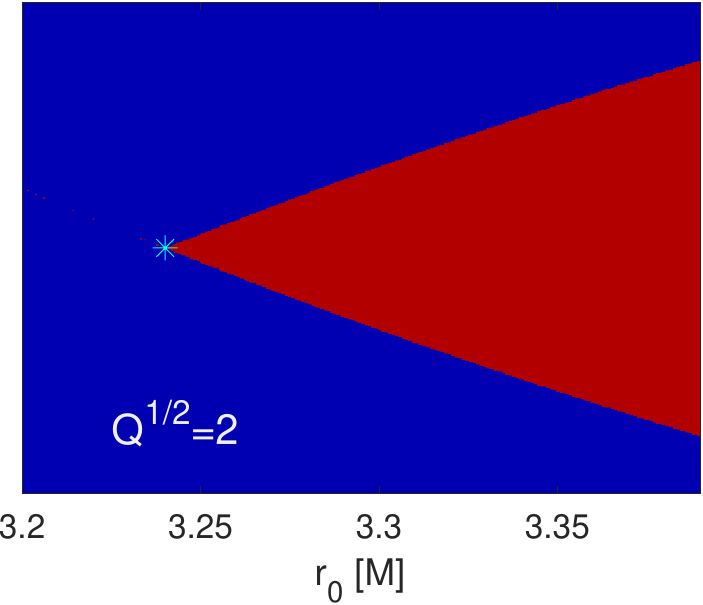}
\includegraphics[scale=.36]{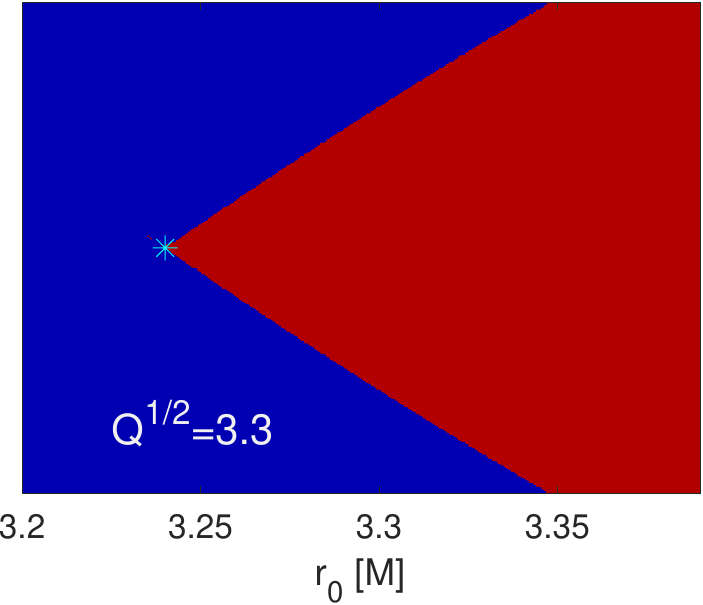}\\
\includegraphics[scale=.352]{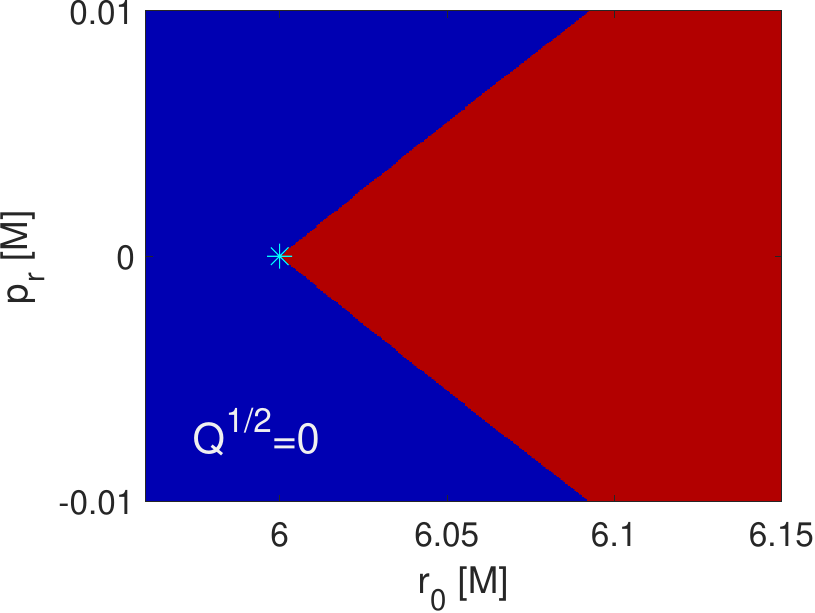}
\includegraphics[scale=.352]{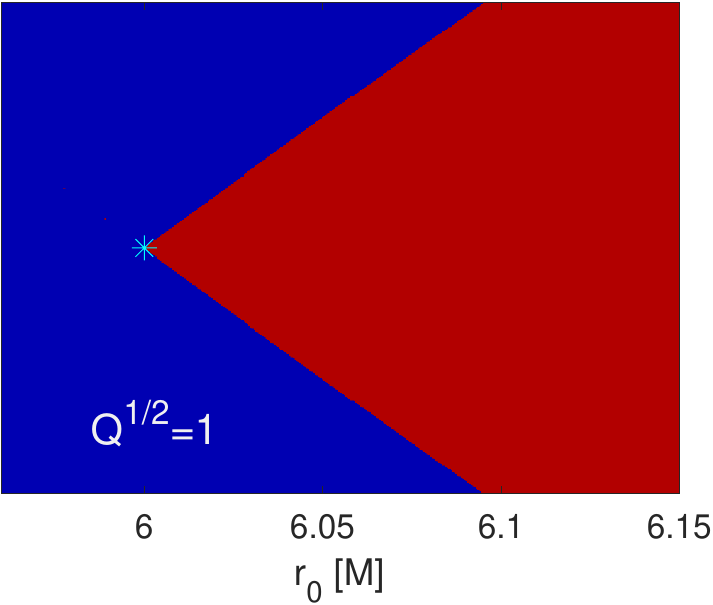}
\includegraphics[scale=.352]{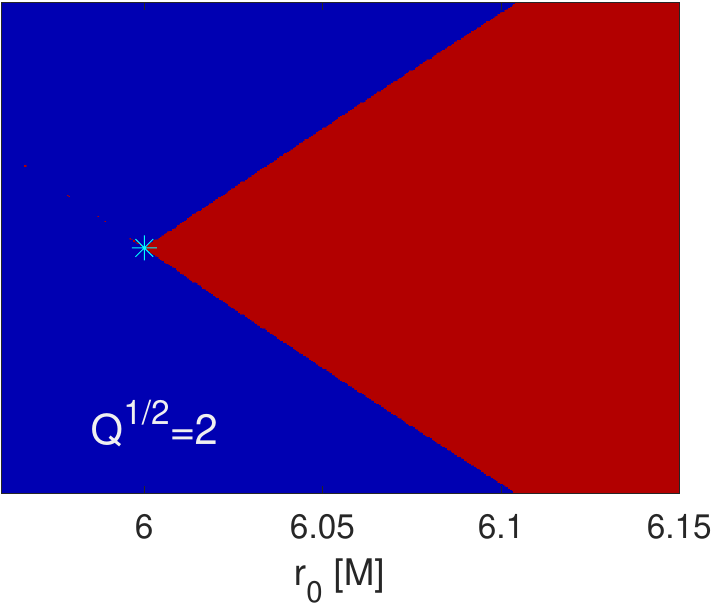}
\includegraphics[scale=.352]{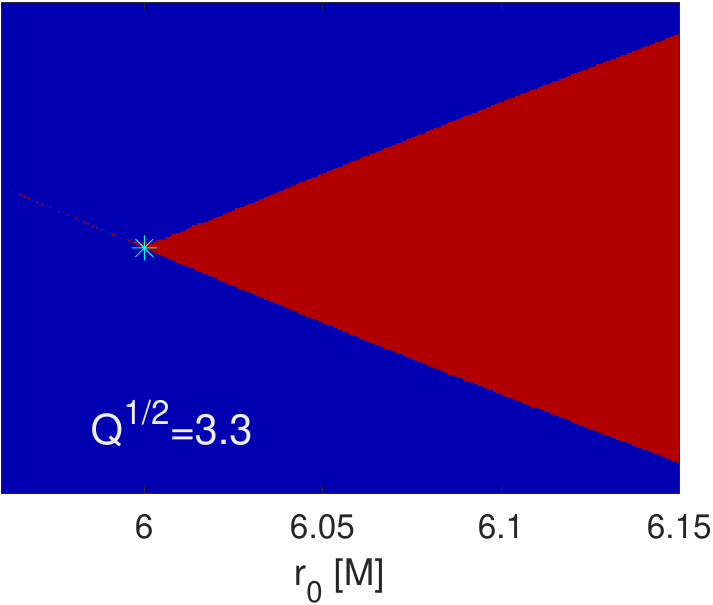}
\caption{Effect of the Carter constant on the dynamics near unstable spherical orbits (marked by light-blue asterisks in the plots) located between MBSO and ISSO radii. The upper row shows the co-rotating orbits near the unstable orbit at $r_0=3.24$ with spin $a=0.7$, while the bottom row reveals the counter-rotating orbits around unstable orbit $r_0=6$ with the spin $a=0.8$. The range of $p_r$ values (vertical axis) is different in each row in order to make the effect of $Q^{1/2}$ more apparent.}
\label{hyperbolic_point_carter}
\end{figure}

In order to assess the stability of a periodic orbit (or a corresponding fixed point of the  Poincar\'{e} map) and the dynamics in its neighborhood, the usual procedure is to linearize the system (considering only the first-order terms of the corresponding Taylor expansion) and solve the eigenvalue problem. \citep[for details see, e.g., ][]{strogatz19,tabor89}. Based on the eigenvalues of the relevant Jacobian matrix (also denoted as stability matrix in this context), which in our case consists of the second derivatives of the Hamiltonian (\ref{hamiltonian}) with respect to the phase-space variables $x^{\mu}$ and $p_{\mu}$, we may introduce a following classification and terminology of the fixed points \citep{wiggins06}. Namely, if all the eigenvalues have nonzero real parts, the fixed point is denoted as hyperbolic. If some, but not all, of the eigenvalues have real parts greater than zero (resp., moduli greater than one) and the rest of the eigenvalues have real parts less than zero (resp., moduli less than one), then the hyperbolic point is called a saddle. If all the eigenvalues have negative real parts (resp., moduli less than one), then the hyperbolic fixed point is called a stable node or sink. If all of the eigenvalues have positive real parts (resp., moduli greater than one), then the hyperbolic fixed point is called an unstable node or source. For each type of hyperbolic fixed point, the structure of the (linearized) flow in its vicinity significantly differs. Moreover, for the case of a purely imaginary set of eigenvalues of the stability matrix, we obtain a fixed point of different class, namely, an elliptic fixed point. For further details and precise mathematical treatment of the problem of linearized stability, we refer to \citet{wiggins06}. For our purpose, the basic classification of the fixed points associated with the periodic spherical orbits remains sufficient.

\begin{figure}[ht]
\center
\includegraphics[scale=.35]{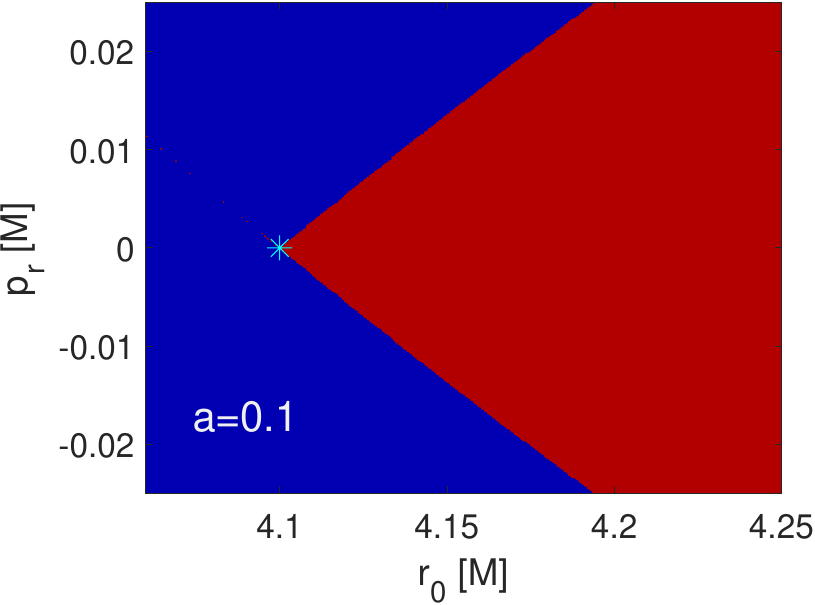}
\includegraphics[scale=.35,trim={0cm 0cm 0cm 0cm}]{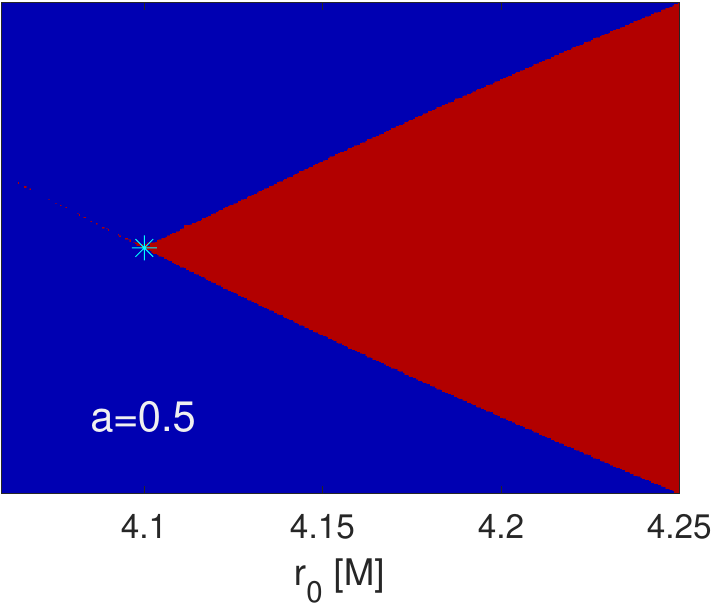}
\includegraphics[scale=.35]{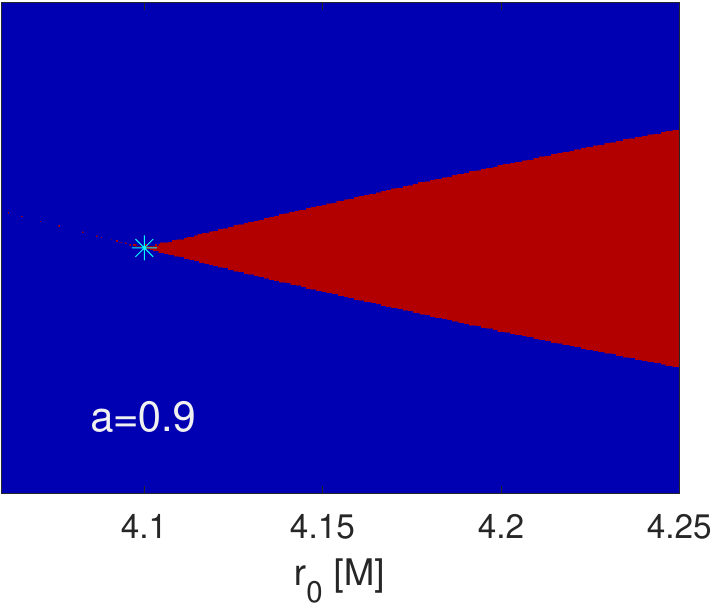}
\includegraphics[scale=.35]{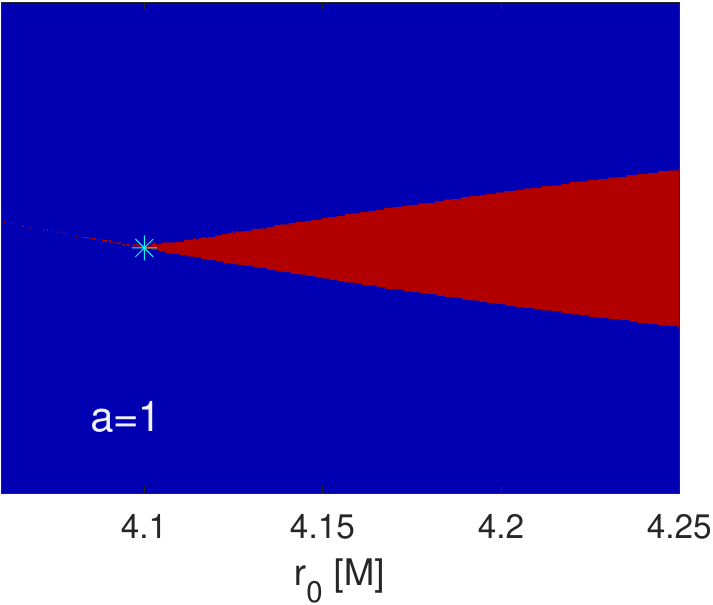}\\
\includegraphics[scale=.35]{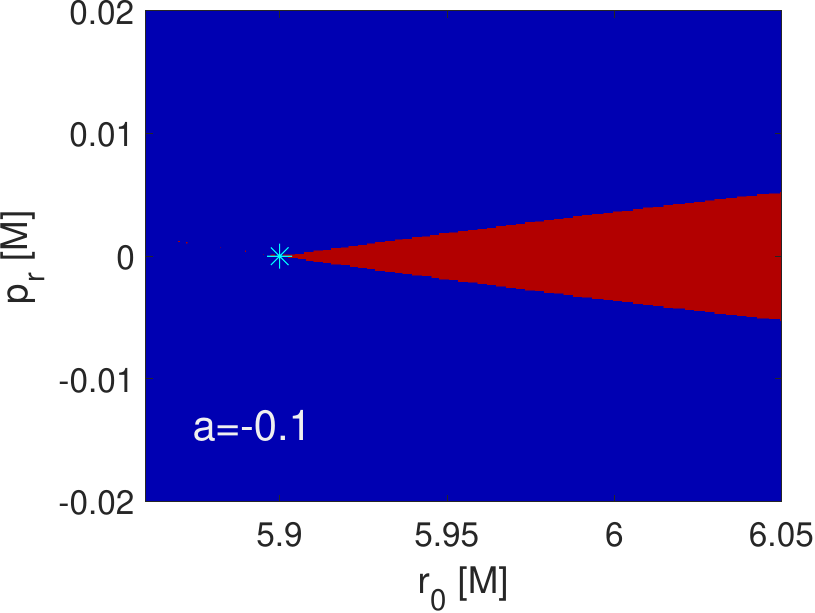}
\includegraphics[scale=.35,trim={0cm 0cm 0cm 0cm}]{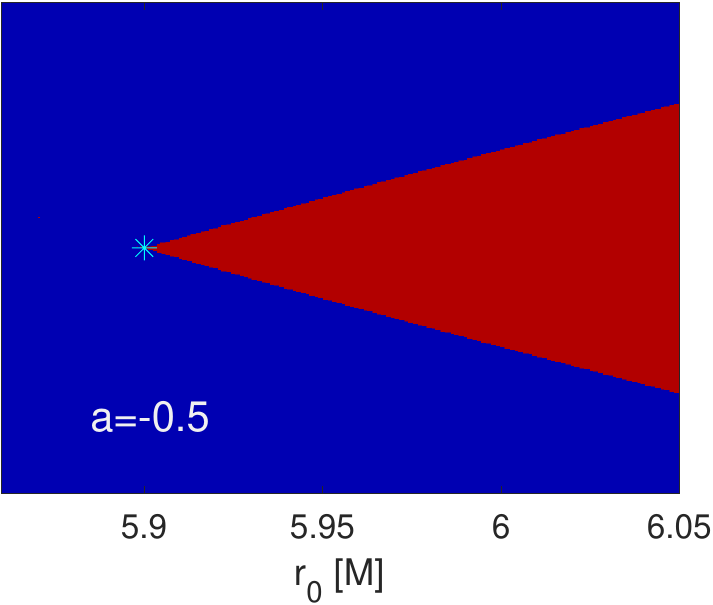}
\includegraphics[scale=.35]{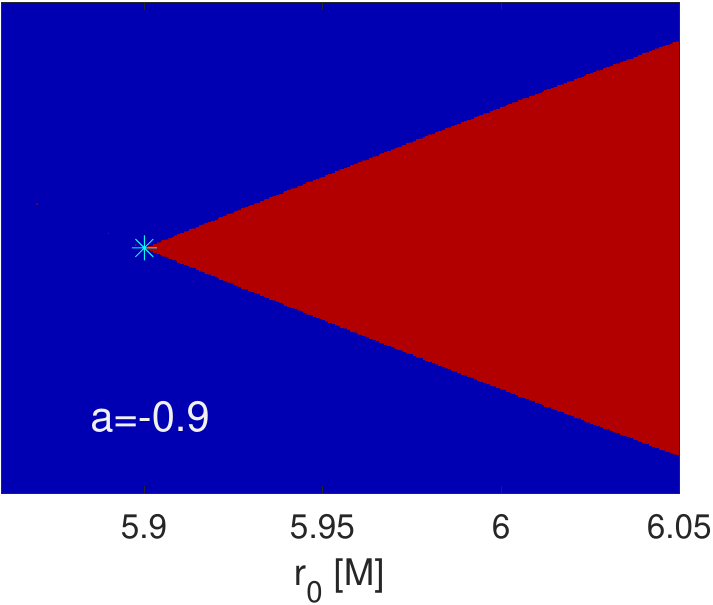}
\includegraphics[scale=.35]{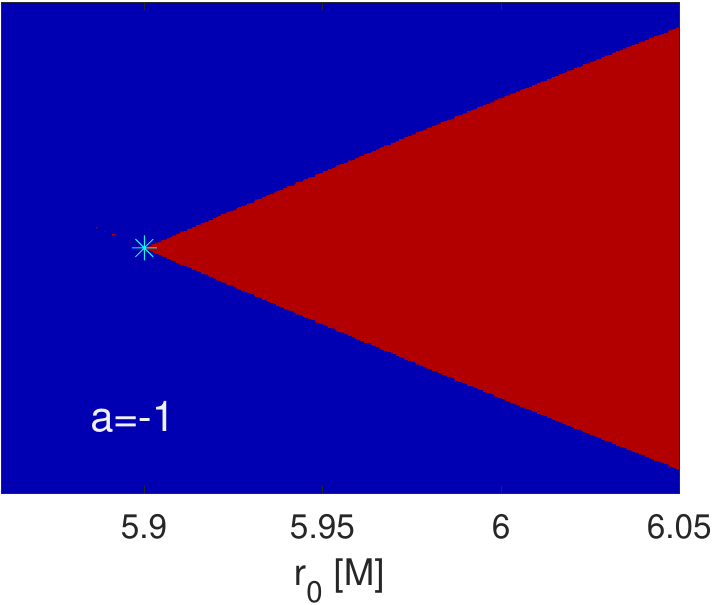}
\caption{Effect of the spin parameter on the dynamics near unstable spherical orbits (marked by blue asterisks in the plots) located between MBSO and ISSO radii. The upper row shows the co-rotating orbits near the unstable orbit at $r_0=4.1$, with $Q^{1/2}=3$, while the bottom row reveals the counter-rotating orbits around the unstable orbit at $r_0=5.9$ with $Q^{1/2}=1.8$.}
\label{hyperbolic_point_spin}
\end{figure}

In particular, regarding the zone between MBSO and ISSO the above analysis of linearized dynamics indicates that initial conditions of a spherical orbit given by \reqs{spherical_energy} and (\ref{spherical_angular}) correspond to an unstable saddle-type hyperbolic fixed point of the relevant Poincar\'{e} map. A sketch of the dynamics near such a point in a phase space of a two-dimensional dynamical system is presented in the left panel of \rff{hyperbolic_point_sketch}. In that case, the trajectories near the fixed point are exactly hyperbolic (blue curves in the sketch) while in the fixed point itself (located at the origin of the plot), the stable ({\em attracting}) and unstable ({\em repelling}) manifolds (indicated by red lines) intersect. Stable manifolds of the fixed point consist of orbits that reach it asymptotically (as time $t\to \infty$) while those of unstable manifolds do so as $t\to -\infty$. Together they form a boundary that separates different types of trajectories (different modes of behavior of a dynamical system) and is thus usually denoted as {\em separatrix}. For precise definitions of the above terms and more detailed treatment of the topic, we refer to standard textbooks on nonlinear dynamical systems \citep[e.g., ][]{ott93,lichtenberg92}.

In the case of saddle-type unstable fixed points, we may observe a special class of orbits, namely {\em homoclinic} orbits, which are found at the intersections of the stable and unstable manifolds and connect the fixed point to itself (while the {\em heteroclinic} orbit connects different fixed points). For the investigated system, however, only the former is relevant. In particular, for the case of circular geodesics in Kerr spacetime, it has been previously shown, that there is a one-to-one correspondence between energetically-bound unstable circular orbits (i.e., circular orbits with radii between MBCO and ISCO) and homoclinic orbits \citep{levin09,perez-giz09}. Moreover, it has been demonstrated that homoclinic orbits represent a limiting case of zoom-whirl behavior of orbits with extreme perihelion precession occurring in a strong-field regime. 

The zoom-whirl behavior of geodesics in Kerr background is obviously not restricted to unstable circular orbits and may also be observed in connection with unstable spherical orbits (see, e.g., \rff{example_stable_below_isso}). To our knowledge, the properties of homoclinic orbits associated with energetically bound unstable spherical orbits (i.e., with radii between MBSO and ISSO) have not been studied in detail. Nevertheless, there are further qualitative indications based on numerical analysis which allow us to establish such correspondence. In particular, in the right panel of \rff{hyperbolic_point_sketch},  we combine the escape-boundary technique to distinguish plunging initial conditions (blue) from those leading to radially bounded motion (red), and thus locate the separatrix of the fixed point (light-blue asterisk), with the method of Poincar\'{e} surface of section recording the one-way intersections of trajectories with the equatorial plane $\theta=\pi/2$ and $\dot{\theta}>0$ (black). These trajectories, differing in $r_0$, are launched from the equatorial plane with $u^r(0)=p_r(0)=0$ and they share the same values of energy and angular momentum given by \reqs{spherical_energy} and (\ref{spherical_angular}) at the unstable point at $r_0=4.1$. For sufficiently long integration, the intersection points draw closed curves typical for quasiperiodic orbits which are characterized by incommensurable frequencies. Only at $r_0\approx5.1$ we get the stable periodic orbit of constant $r$, i.e., stable spherical orbit above ISSO which corresponds to a fixed point of a Poincar\'{e} map.

\begin{figure}[ht]
\center
\includegraphics[scale=.36]{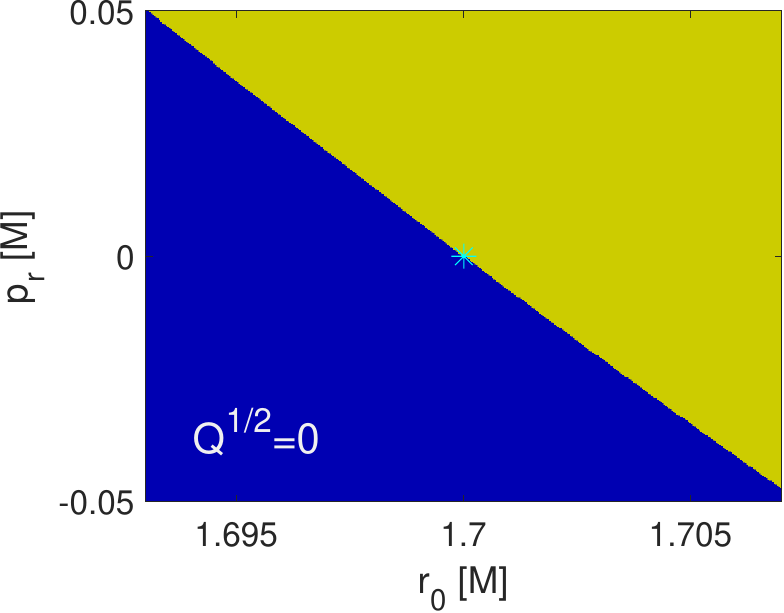}
\includegraphics[scale=.36,trim={0cm 0cm 0cm 0cm}]{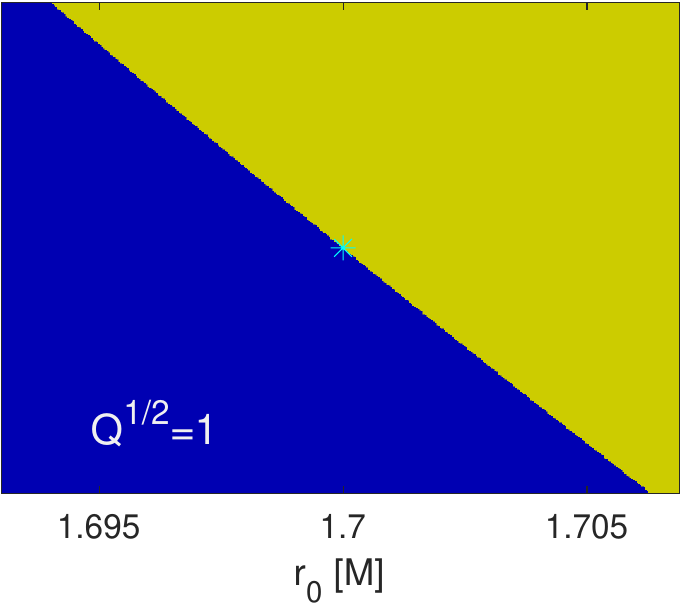}
\includegraphics[scale=.36]{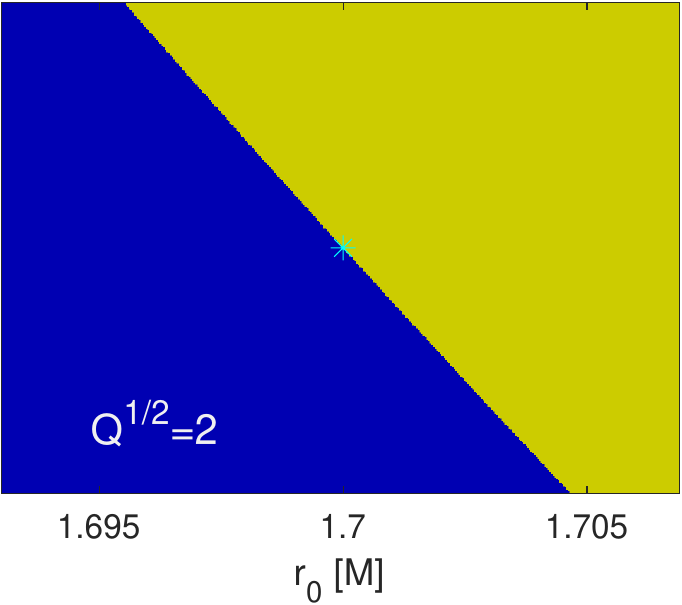}
\includegraphics[scale=.36]{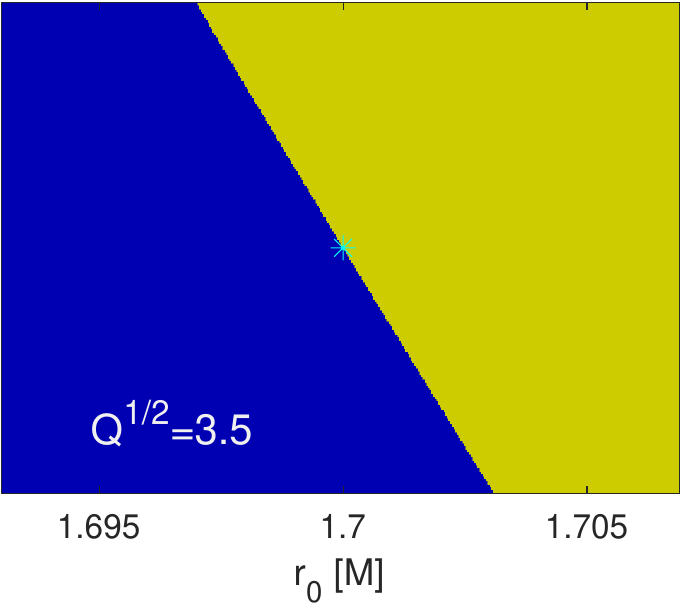}\\
\includegraphics[scale=.36]{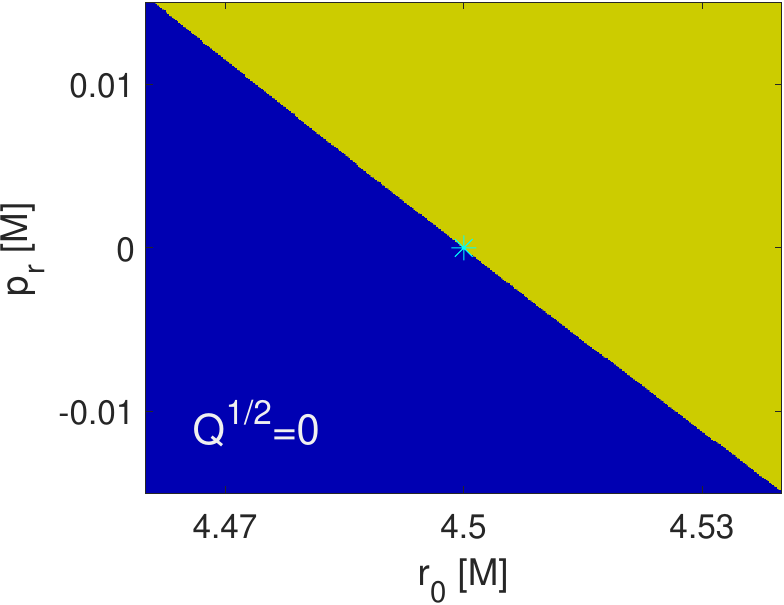}
\includegraphics[scale=.36]{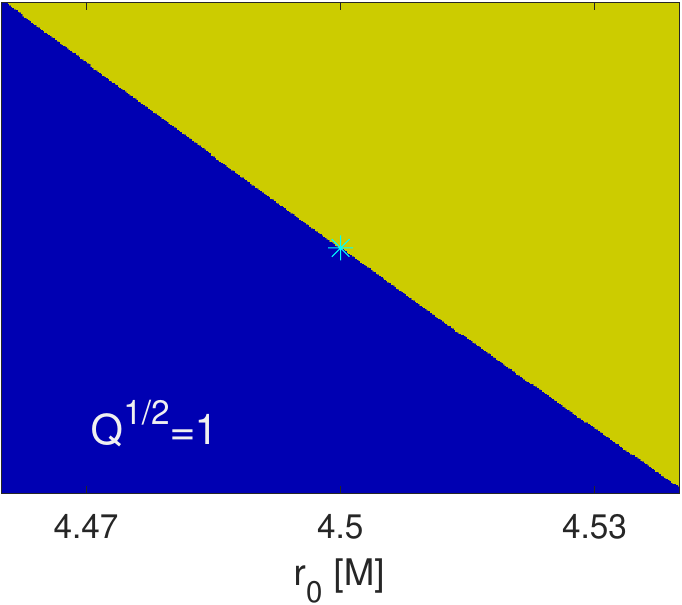}
\includegraphics[scale=.36]{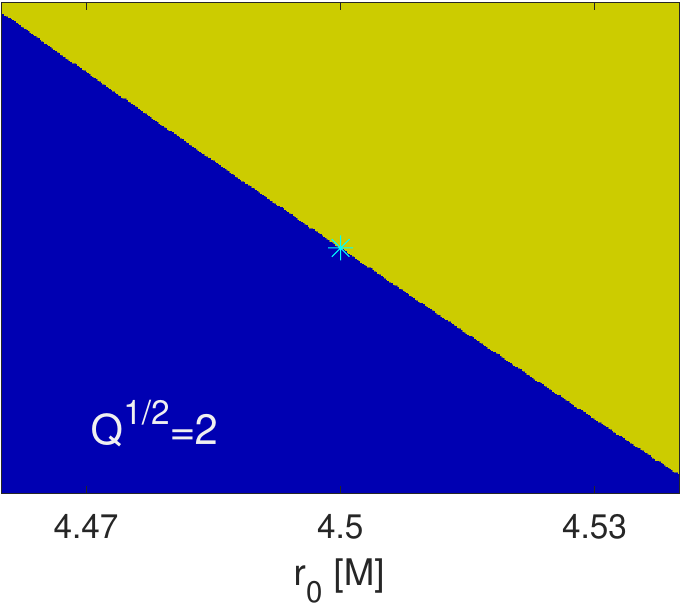}
\includegraphics[scale=.36]{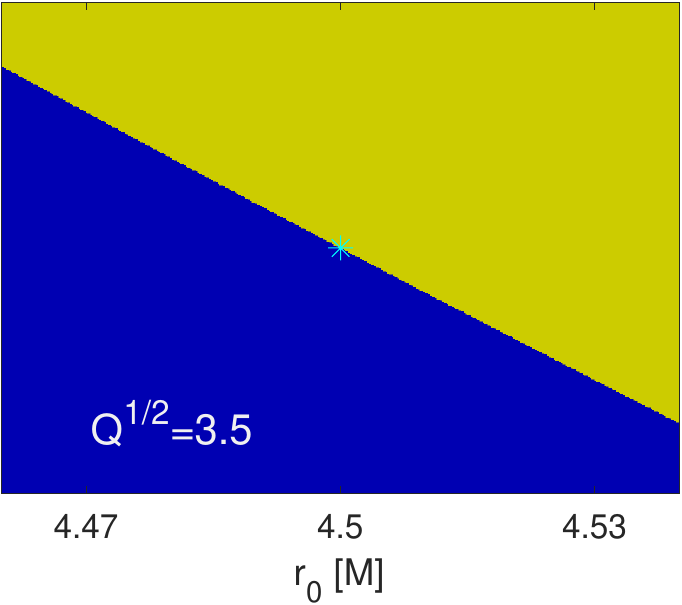}
\caption{The effect of the Carter constant on the dynamics near unstable spherical orbits (marked by blue asterisks in the plots) that are located below MBSO radius. The upper row shows co-rotating orbits near the unstable orbit at $r_0=1.7$ for the spin $a=0.9$. The bottom row reveals the counter-rotating orbits around the unstable orbit $r_0=4.5$ with the spin $a=0.6$. Note that the range of $p_r$ values (vertical axis) is different in each row.}
\label{hyperbolic_point_mbso}
\end{figure}

In order to determine the role of the Carter constant and spin in the evolution of unstable spherical orbit under perturbation, we employ the technique of escape-boundary plots once again. However, unlike the application in \rffs{fixed_carter}--\ref{polar}, now we shall study the vicinity of the particular unstable orbit in the plane $(r_0,p_r)$. In \rff{hyperbolic_point_carter} we observe how the slope of separatrix, which divides the blue plunging zone from the red wedge of bounded motion, changes with different values of the Carter constant (while all other parameters of the orbits are kept constant). In the first row of \rff{hyperbolic_point_carter} we show an unstable co-rotating orbit ($r_0=3.24$, $a=0.7$) with different values of the Carter constant. The slope of the separatrix rises and the opening angle of the red stable zone gradually and significantly increases as the Carter constant increases. If we consider the direction of the perturbation of the unstable equilibrium to be random in the $(r_0,p_r)$ plane, then the probability of the stabilization of the perturbed co-rotating orbit clearly increases with the Carter constant. On the other hand, in the case of counter-rotating orbits (with parameters $r_0=6$ and $a=0.8$), the probability of stabilization is decreased by the Carter constant, although the effect is not as prominent as in the co-rotating case (see the second row of \rff{hyperbolic_point_carter}).

In \rff{hyperbolic_point_spin} we perform an analogous discussion of the role of the spin parameter. Therefore for a particular unstable orbit, we fix the Carter constant (here we choose a co-rotating orbit with $r_0=4.1$, $Q^{1/2}=3$, and counter-rotating orbit with $r_0=5.9$, $Q^{1/2}=1.8$) and vary the spin. As previously, for each case, we evolve a grid of $400 \times 400$ initial conditions in $(r_0,p_r)$ plane around the unstable point. Unlike the effect of the Carter constant, the spin is observed to decrease the probability of the stabilization of co-rotating orbits (first row of \rff{hyperbolic_point_spin}) and increase it for counter-rotating orbits (second row). As for the Carter constant, the effect is stronger for the co-rotating orbits, although the difference is not that significant. 

We may conclude that both parameters (the Carter constant and the black hole spin) play an important role in the evolution of unstable spherical orbits with radii between MBSO and ISSO. In the case of co-rotating orbits, the Carter constant increases the probability of stabilization, while the spin suppresses it. For the counter-rotating orbits, the role of the parameters is reversed -- the Carter constant supports the tendency to plunge while the spin tends to stabilize the perturbed orbits. Of course, the probabilistic interpretation of \rffs{hyperbolic_point_carter} and \rff{hyperbolic_point_spin} is based on the assumption that the direction of the perturbation in the $(r_0,p_r)$ plane is random, and thus the probability of stabilization in the form of a quasiperiodic eccentric orbit is proportional to the opening angle of the red wedge of stable motion. These observations directly and more clearly confirm the assumptions previously inferred from Figs.~\ref{fixed_carter}--\ref{polar}.

The above results apply to the region between MBSO and ISSO. Below MBSO the orbits become energetically unbound allowing the particles to escape to infinity and the nature of instability changes from the saddle type to an unstable node. Nevertheless, we may use the analogous approach as previously to study this part of the phase space. In \rff{hyperbolic_point_mbso} we study the effect of the Carter constant on a particular spherical orbit below MBSO. Namely, in the upper row of \rff{hyperbolic_point_mbso} we pick co-rotating orbits with $r_0=1.7$ and $a=0.9$, while the bottom row reveals the counter-rotating orbits with $r_0=4.5$ and $a=0.6$. It appears that in this case, the Carter constant has a rather negligible effect on the destiny of particles. Indeed, we observe that it only slightly rotates the separatrix between the yellow escape zone and the blue plunging region. With increasing Carter constant the rotation of the separatrix is clockwise for co-rotating orbits and counterclockwise for counter-rotating orbits. Nevertheless, the separatrix remains straight and the probability of the escape or plunge for the random perturbation therefore does not change. The fractal-like structures observed below MBSO in Figs.~\ref{fixed_carter}--\ref{polar} were therefore a product of numerical noise lacking physical significance.

\section{Conclusions}
\label{conclusions}
Spherical orbits in Kerr spacetime are a special class of geodesics with constant radii. They are defined with respect to a suitably selected radial coordinate and generalize the equatorial circular orbits by allowing non-zero values of the Carter constant. These orbits can be interpreted as inclined (by the angle $\theta_\star$ with respect to the rotation axis) circular geodesics; the orbital plane is continuously rotated by the frame-dragging effect. The latter effect due to the spinning black hole has no classical analog and it represents a purely relativistic phenomenon. The spherical orbits were first studied in the early 1970s \citep{wilkins72} and have recently attracted a renewed interest \citep{teo21,stein20,rana19} due to their relevance for EMRIs \citep[e.g., ][]{amaro18}, which represent a key target for the future gravitational observatory LISA \citep{amaro23}. In this paper, we have revisited the topic, focusing on the dynamics of both stable and unstable spherical orbits.

In \refsec{spherical} we first discussed the radii of the Innermost Stable Spherical Orbit (ISSO) and the Marginally Bound Spherical Orbit (MBSO) with respect to the values of the spin parameter and the Carter constant. We introduced the parameterization of orbits by the turning (or inclination) angle $\theta_\star$ as an alternative to the Carter constant. The parameterization by $\theta_\star$ provides an intuitive picture of the geometric shape of the orbit, and the two limiting special cases, namely, polar orbits ($\theta_\star=0$) and circular equatorial orbits ($\theta_\star=\pi/2$) are easily identified. For a given orientation of the orbit (co-rotating or counter-rotating), there is a one-to-one correspondence between $\theta_\star$ and the Carter constant. The analysis shows that for co-rotating orbits, both ISSO and MBSO radii are gradually shifted to higher radii compared to ISCO and MBCO of the corresponding circular orbit as the Carter constant increases. On the other hand, for counter-rotating spherical orbits, ISSO and MBSO are always smaller than those of circular orbits, and they gradually shrink as the Carter constant is increased.

Spherical orbits below ISSO become unstable, and the fate of a particle, i.e., whether it plunges to the black hole, stabilizes on a quasiperiodic orbit, or escapes (from below MBSO), depends primarily on the phase-space direction of the perturbation of the unstable equilibrium. In \refsec{numerical} we numerically study the nature of the instability using the technique of escape-boundary plots \citep{kopacek23,kopacek20} to reveal how the parameters, namely the spin and the Carter constant, affect the probability of each possible outcome. It appears that a straightforward application of this method may become questionable since the system near instability is prone to numerical noise. Therefore, we adapt the method and plot the neighborhood of the unstable spherical orbit in the plane of the canonical variables ($r$, $p_r$), which allows us to combine the plot with the Poincar\'{e} sections of particular trajectories and, more importantly, we can thus directly trace the separatrix between the regions of the plunge and escape (occurring below MBSO) or between the plunge and radially bounded motion (above MBSO). Comparing the results for different values of the parameters (and assuming that the perturbation has a random direction in the ($r$, $p_r$) plane), allows us to conclude that the spin parameter decreases the probability of the stabilization for co-rotating unstable spherical orbits while it increases this probability for counter-rotating orbits. The Carter constant has an opposite role: it contributes to the stabilization of co-rotating orbits while it makes counter-rotating orbits more prone to plunge. Nevertheless, below the MBSO radii, the nature of the instability changes. Namely, it becomes an unstable node for which the probability of plunge or escape does not change with these parameters since the separatrix remains a straight line that only slightly rotates as the spin or Carter constant increases. 

Our results have been obtained with a general numerical approach of escape-boundary plots, whose applicability is not limited to a fully integrable problem of Kerr geodesics and can also be used to study a chaotic system under non-integrable perturbations \citep[e.g.][]{kopacek20,kopacek18}. On the other hand, the integrable case allows to proceed analytically. In particular, with an appropriate parameterization, it is possible to find an implicit formula for the separatrix between the plunging and quasiperiodic regions of the phase space \citep{stein20}. More recently, an analytical solution for plunging geodesics in Kerr has also been discussed \citep{dyson23}. In this context, the numerical method used in our analysis provides a complementary and intuitive approach that allows to verify the analytical calculations, while effectively tackling an astrophysically relevant problem of the stability of spherical orbits in Kerr spacetime.

\acknowledgements
The authors acknowledge continued support from the Research Infrastructure CTA-CZ (LM2023047) of the Czech Ministry of Education, Youth and Sports. OK is supported by the Lumina Quaeruntur Fellowship No.\ LQ100032102 of the Czech Academy of Sciences. VK also thanks the Czech Science Foundation grant (GA\v{C}R 21-06825X). 

\appendix
\section*{Integrating the equations of motion: precision versus numerical noise}
\label{appa}
Integration of non-linear first-order differential equations of motion (\ref{hameq}) is performed with a multistep Adams-Bashforth-Moulton integrator based on the pre\-dic\-tor-corrector (PECE) method which is implemented in \texttt{Matlab} as a function \texttt{ode113}. The stepsize is adaptive and local truncation error is controlled by relative tolerance specified by the parameter \texttt{RelTol} which we set to the highest allowed precision ($\texttt{RelTol}\approx10^{-14}$). Using this setting we obtain trajectories with final values of relative error of energy $E_{rr}$ not exceeding the level of $\approx10^{-8}$. 

To verify the results obtained with \texttt{ode113} we also employ a high-order explicit Runge-Kutta scheme, namely a \texttt{Matlab} built-in function \texttt{ode89} which is an implementation of Verner's ``most robust'' Runge-Kutta 9(8) pair with an 8th-order continuous extension \citep{verner10}. The solution is advanced with the 9th-order result. While the results of both methods are in general qualitatively comparable for our application (i.e., delivering equivalent escape-boundary plots in similar computational times), we also notice some differences. In particular, it appears that \texttt{ode89} is superior to \texttt{ode113} in terms of precision when integrating periodic orbits (namely, stable spherical orbits). On the other hand, \texttt{ode113} is more precise when integrating quasiperiodic trajectories near unstable spherical orbits. This allows us to trace the separatrix in the phase space with a considerably better resolution compared to \texttt{ode89} and makes the \texttt{ode113} a method of choice for this task.     

Even higher precision in terms of conservation of integrals of motion of particular trajectories may be achieved with Dormand-Prince 8th - 7th order explicit scheme \texttt{ode87}\footnote{Vasiliy Govorukhin: ode87 Integrator, available at \href{https://www.mathworks.com/matlabcentral/fileexchange/3616-ode87-integrator}{MATLAB Central File Exchange (File ID: \#3616)}} \citep{dormand80} which is a high precision integrator of embedded Runge-Kutta family with the local error of order $\mathcal{O}(h^9)$. Nevertheless, the computation time steeply rises compared to \texttt{ode113} while the increase of accuracy is usually not crucial here. We have only employed \texttt{ode87} to obtain Poincar\'{e} sections presented in the right panel of \rff{hyperbolic_point_sketch} which require very long integration with high precision.

Besides routines described above, we have previously also tested an implicit $s$-stage Gauss-Legendre symplectic solver \texttt{gls} which proved superior to \texttt{ode87} in terms of relative error on the long time-scale but also computationally very expensive \citep{kopacek14b}. We conclude that for a current application (not-so-long integration of numerous trajectories) the accuracy of fast multistep integrator \texttt{ode113} remains sufficient in most cases. 

\begin{figure}
\center
\includegraphics[scale=.25,trim={1.2cm 0 .2cm 0},clip]{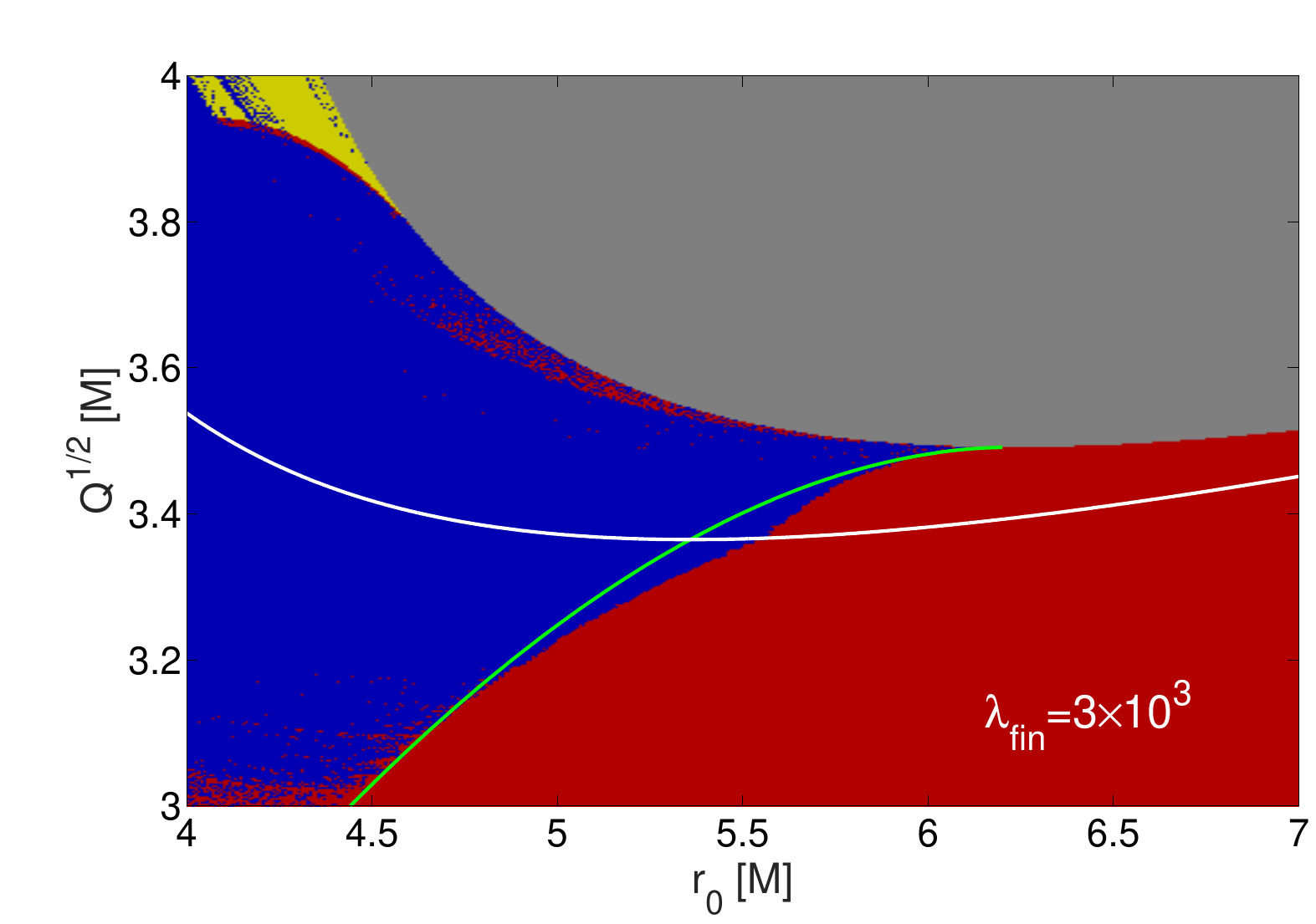}
\includegraphics[scale=.25,trim={3.6cm 0 0.2cm 0},clip]{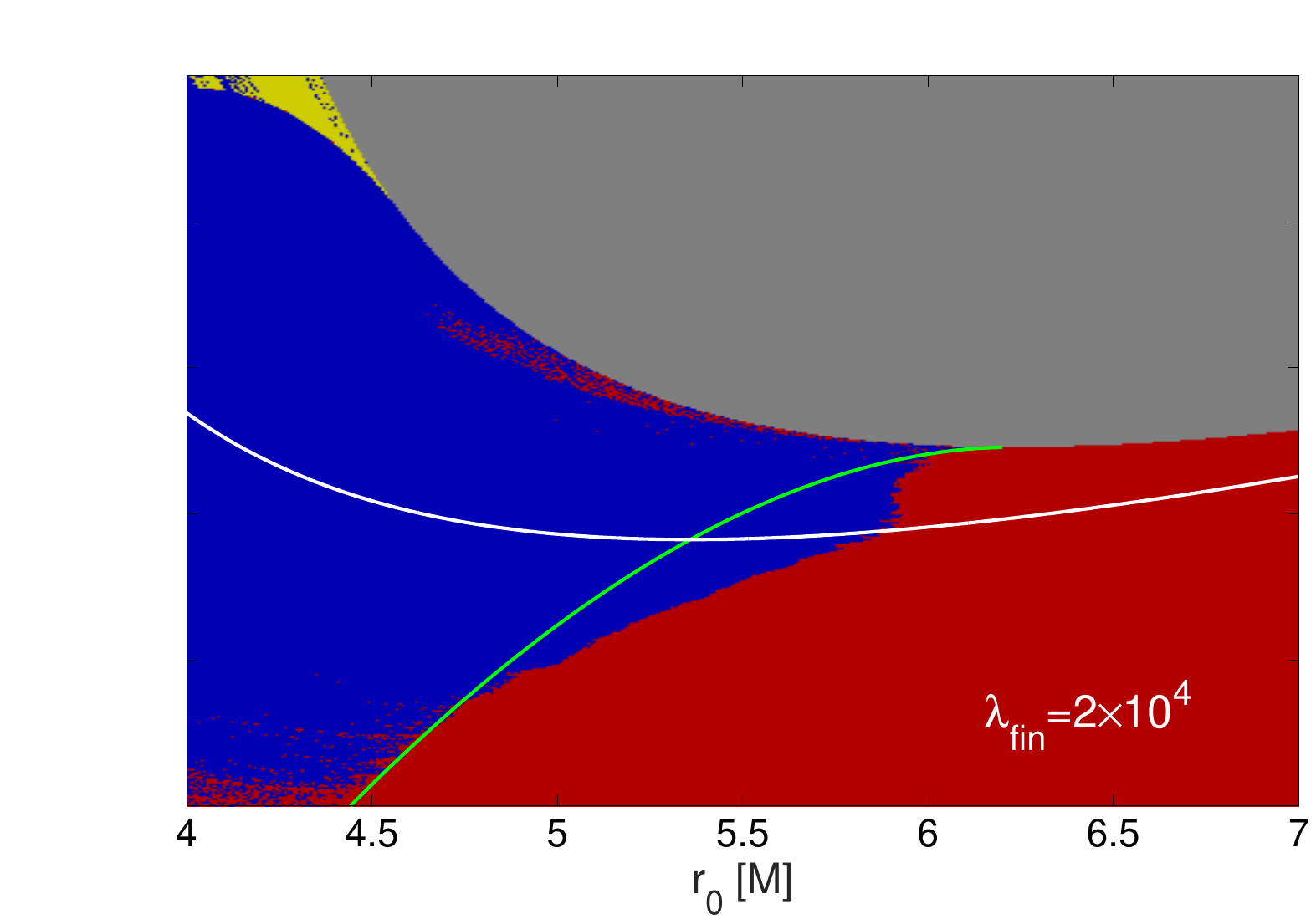}
\includegraphics[scale=.25,trim={0cm 0 0.2cm 0},clip]{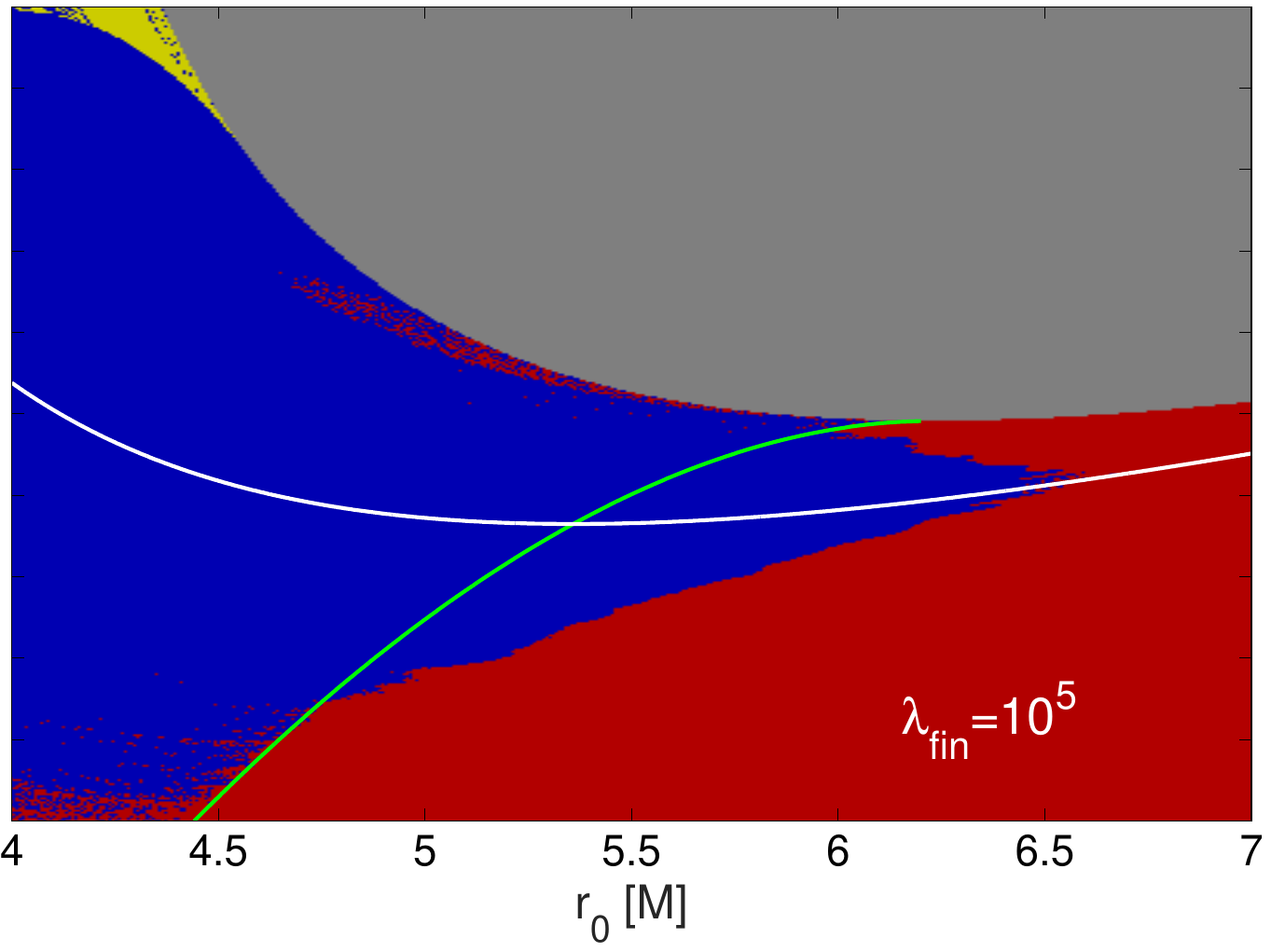}
\caption{The apparent region of near-polar plunging spherical orbits occurring just above ISSO is in fact a numerical artifact caused by integration errors which generally grow with the integration time $\lambda_{\rm fin}$. In this example, spin has been set to the value of $a=0.95$. The color scheme is as in Fig.\ \ref{fixed_carter} (plunging trajectories are indicated in blue).}
\label{artifacts}
\end{figure}

Nevertheless, we still need to be aware of numerical errors when interpreting the results and constructing the escape-boundary plots. In particular, the parameter region corresponding to polar and near-polar orbits appears to be especially sensitive to errors. In \rff{artifacts}, we observe that secular numerical error may introduce here an artificial plunging zone above ISSO as an obvious numerical artifact which grows with the integration time $\lambda_{\rm fin}$. Nevertheless, the integration time must be sufficient for the escaping orbits to reach the escape threshold $r_{\rm e}$ which is another crucial parameter of escape-boundary plots which needs to be set with caution. In \rff{artifacts2} we observe, how the improper choice of $r_{\rm e}$ may lead to incorrect identification of orbits. In the first two panels the underestimated value of $r_e$ leads to the false identification of highly eccentric quasiperiodic orbits near MBPO boundary as escaping. On the other hand, an overestimated escape threshold may cause misinterpretation of escaping orbits below MBPO as quasiperiodic (shown in the third panel of \rff{artifacts2}). 

\begin{figure}
\center
\includegraphics[scale=.335,trim={.2cm 0 .2cm 0},clip]{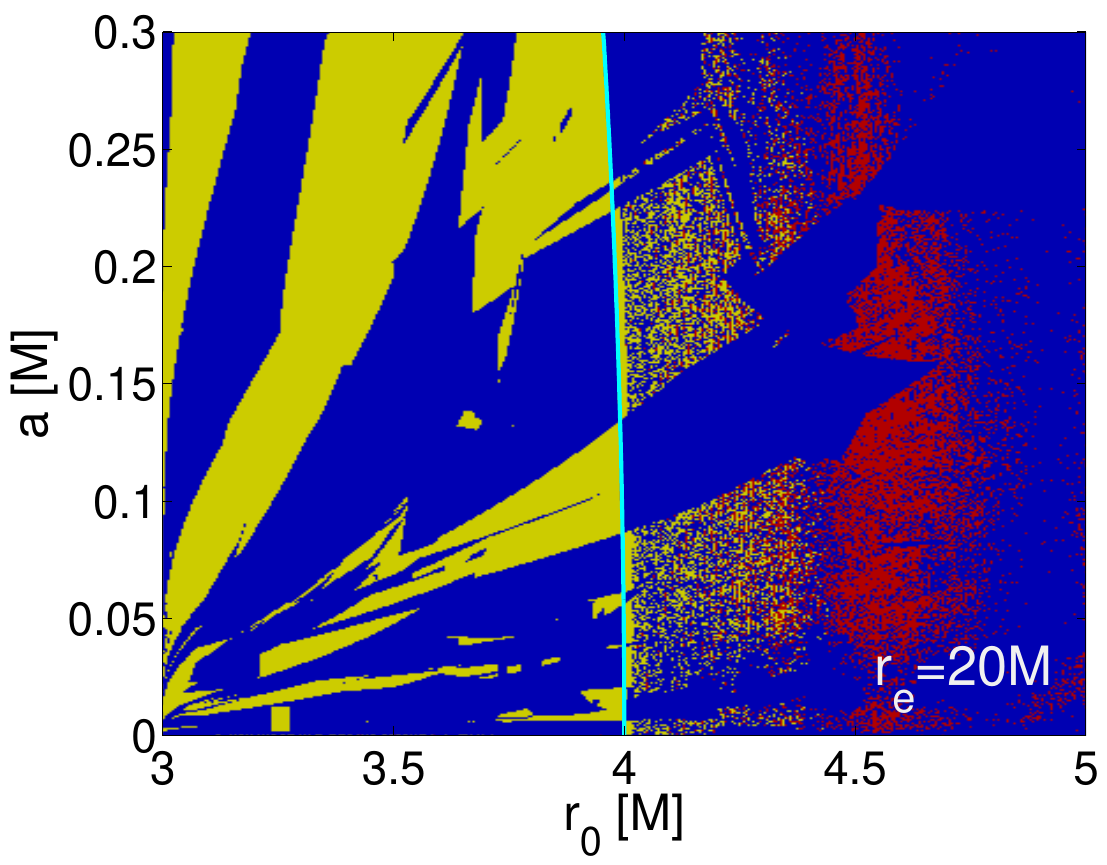}
\includegraphics[scale=.335,trim={2.2cm 0 0.2cm 0},clip]{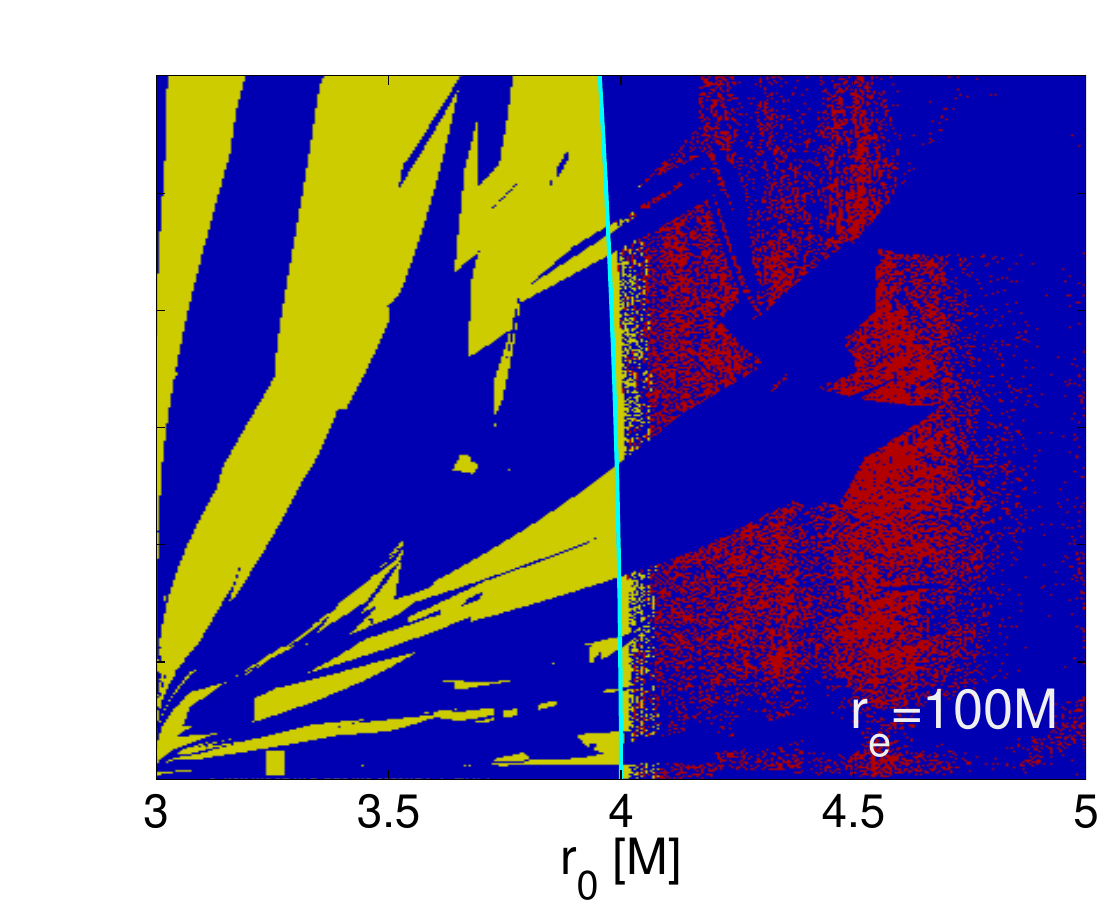}
\includegraphics[scale=.335,trim={0cm 0 0.2cm 0},clip]{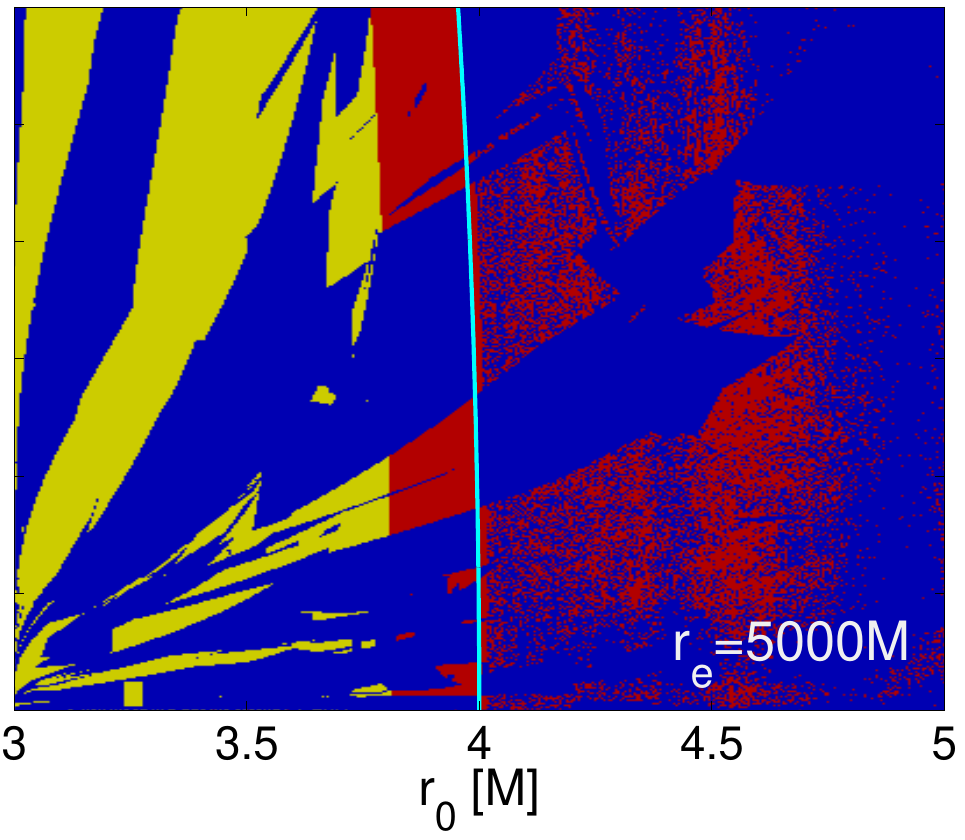}
\caption{Seemingly escaping polar orbits may appear above MBPO if the escape threshold radius $r_e$ is set too low (first two panels). Overestimating its value may, on the other hand, lead to a false interpretation of escaping orbits as radially bounded (third panel). The same magnified section of \rff{polar} is shown in all panels which only differ in $r_{e}$.}
\label{artifacts2}
\end{figure}

\begin{figure}
\center
\includegraphics[scale=.45,trim={.2cm 0 .2cm 0},clip]{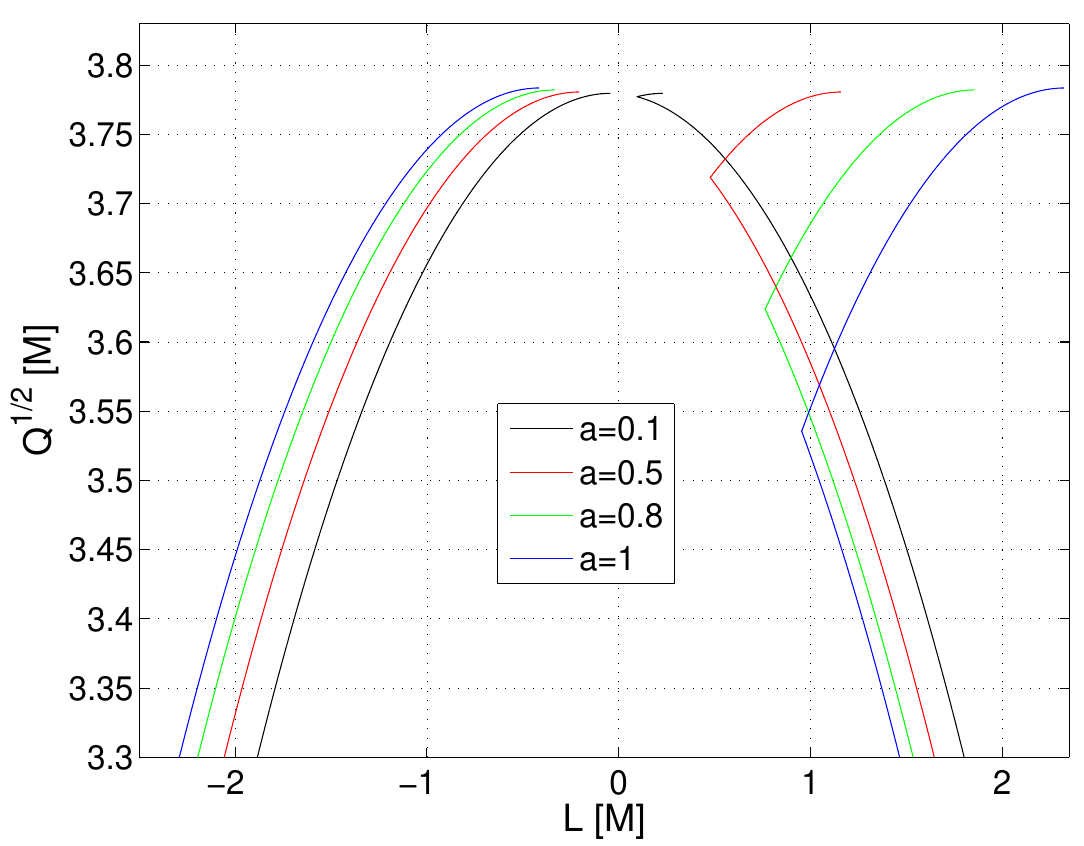}
\caption{The formula relating the angular momentum of spherical orbit $L$ to the Carter constant $Q$ as given in \citet{rana19} fails to provide correct values of $L$ of near-polar co-rotating orbits. This figure is to be compared with the left panel of \rff{diskuze_Lsf} computed with the correct formula \req{spherical_angular}.}
\label{L_Rana}
\end{figure}

After testing different values of the parameters $\lambda_{\rm fin}$ and $r_e$, we found the optimal setting for the escape-boundary plots in \rffs{fixed_carter}--\ref{polar} to be $\lambda_{\rm fin}=2\times10^4$ with $r_e=1000$. For the survey of the vicinity of the unstable fixed point in \rffs{hyperbolic_point_carter}--\ref{hyperbolic_point_mbso}, a shorter integration remains sufficient, for which also the threshold needs to be adjusted. Namely, we set $\lambda_{\rm fin}=2000$ and $r_e=100$ for these calculations.

In addition to the numerical problems mentioned above, we have also come across an erroneous result in the literature. Namely, the expression of the angular momentum of a spherical orbit as a function of the Carter constant published by \citet{rana19} in formulas (18a)--(18c) of the cited paper gives incorrect results for the co-rotating near-polar spherical orbits, although it works well for counter-rotating orbits and for co-rotating orbits with lower values of the Carter constant (see \rff{L_Rana}). Therefore, for our calculations, we employ an alternative expression \req{spherical_angular} by \citet{teo21}, which works correctly regardless of the value of the Carter constant (see \rff{diskuze_Lsf}).

\bibliography{article}

\begin{thebibliography}{}
\expandafter\ifx\csname natexlab\endcsname\relax\def\natexlab#1{#1}\fi
\providecommand{\url}[1]{\href{#1}{#1}}
\providecommand{\dodoi}[1]{doi:~\href{http://doi.org/#1}{\nolinkurl{#1}}}
\providecommand{\doeprint}[1]{\href{http://ascl.net/#1}{\nolinkurl{http://ascl.net/#1}}}
\providecommand{\doarXiv}[1]{\href{https://arxiv.org/abs/#1}{\nolinkurl{https://arxiv.org/abs/#1}}}

\bibitem[{{Amaro-Seoane}(2018)}]{amaro18}
{Amaro-Seoane}, P. 2018, Living Reviews in Relativity, 21, 4,
  \dodoi{10.1007/s41114-018-0013-8}

\bibitem[{{Amaro-Seoane} {et~al.}(2023){Amaro-Seoane}, {Andrews}, {Arca Sedda},
  {Askar}, {Baghi}, {Balasov}, {Bartos}, {Bavera}, {Bellovary}, {Berry},
  {Berti}, {Bianchi}, {Blecha}, {Blondin}, {Bogdanovi{\'c}}, {Boissier},
  {Bonetti}, {Bonoli}, {Bortolas}, {Breivik}, {Capelo}, {Caramete},
  {Cattorini}, {Charisi}, {Chaty}, {Chen}, {Chru{\'s}li{\'n}ska}, {Chua},
  {Church}, {Colpi}, {D'Orazio}, {Danielski}, {Davies}, {Dayal}, {De Rosa},
  {Derdzinski}, {Destounis}, {Dotti}, {Dutan}, {Dvorkin}, {Fabj}, {Foglizzo},
  {Ford}, {Fouvry}, {Franchini}, {Fragos}, {Fryer}, {Gaspari}, {Gerosa},
  {Graziani}, {Groot}, {Habouzit}, {Haggard}, {Haiman}, {Han}, {Istrate},
  {Johansson}, {Khan}, {Kimpson}, {Kokkotas}, {Kong}, {Korol}, {Kremer},
  {Kupfer}, {Lamberts}, {Larson}, {Lau}, {Liu}, {Lloyd-Ronning}, {Lodato},
  {Lupi}, {Ma}, {Maccarone}, {Mandel}, {Mangiagli}, {Mapelli}, {Mathis},
  {Mayer}, {McGee}, {McKernan}, {Miller}, {Mota}, {Mumpower}, {Nasim},
  {Nelemans}, {Noble}, {Pacucci}, {Panessa}, {Paschalidis}, {Pfister},
  {Porquet}, {Quenby}, {Ricarte}, {R{\"o}pke}, {Regan}, {Rosswog}, {Ruiter},
  {Ruiz}, {Runnoe}, {Schneider}, {Schnittman}, {Secunda}, {Sesana}, {Seto},
  {Shao}, {Shapiro}, {Sopuerta}, {Stone}, {Suvorov}, {Tamanini}, {Tamfal},
  {Tauris}, {Temmink}, {Tomsick}, {Toonen}, {Torres-Orjuela}, {Toscani},
  {Tsokaros}, {Unal}, {V{\'a}zquez-Aceves}, {Valiante}, {van Putten}, {van
  Roestel}, {Vignali}, {Volonteri}, {Wu}, {Younsi}, {Yu}, {Zane}, {Zwick},
  {Antonini}, {Baibhav}, {Barausse}, {Bonilla Rivera}, {Branchesi},
  {Branduardi-Raymont}, {Burdge}, {Chakraborty}, {Cuadra}, {Dage}, {Davis}, {de
  Mink}, {Decarli}, {Doneva}, {Escoffier}, {Gandhi}, {Haardt}, {Lousto},
  {Nissanke}, {Nordhaus}, {O'Shaughnessy}, {Portegies Zwart}, {Pound},
  {Schussler}, {Sergijenko}, {Spallicci}, {Vernieri}, \&
  {Vigna-G{\'o}mez}}]{amaro23}
{Amaro-Seoane}, P., {Andrews}, J., {Arca Sedda}, M., {et~al.} 2023, Living
  Reviews in Relativity, 26, 2, \dodoi{10.1007/s41114-022-00041-y}

\bibitem[{{Bardeen}(1973)}]{bardeen73}
{Bardeen}, J.~M. 1973, in Black Holes (Les Astres Occlus), 215--239

\bibitem[{{Bardeen} {et~al.}(1972){Bardeen}, {Press}, \&
  {Teukolsky}}]{bardeen72}
{Bardeen}, J.~M., {Press}, W.~H., \& {Teukolsky}, S.~A. 1972, \apj, 178, 347,
  \dodoi{10.1086/151796}

\bibitem[{{Carter}(1968)}]{carter68}
{Carter}, B. 1968, Physical Review, 174, 1559, \dodoi{10.1103/PhysRev.174.1559}

\bibitem[{{Chandrasekhar}(1998)}]{1998mtbh.book.....C}
{Chandrasekhar}, S. 1998, {The Mathematical Theory of Black Holes} (Oxford
  University Press, New York)

\bibitem[{Contopoulos(2006)}]{contopoulos06}
Contopoulos, G. 2006, International Journal of Bifurcation and Chaos, 16, 1795,
  \dodoi{10.1142/S0218127406015684}

\bibitem[{Contopoulos \& Harsoula(2010)}]{contopoulos10}
Contopoulos, G., \& Harsoula, M. 2010, International Journal of Bifurcation and
  Chaos, 20, 2005, \dodoi{10.1142/S0218127410026915}

\bibitem[{Dormand \& Prince(1980)}]{dormand80}
Dormand, J., \& Prince, P. 1980, Journal of Computational and Applied
  Mathematics, 6, 19, \dodoi{https://doi.org/10.1016/0771-050X(80)90013-3}

\bibitem[{{Dymnikova}(1986)}]{1986SvPhU..29..215D}
{Dymnikova}, I.~G. 1986, Soviet Physics Uspekhi, 29, 215,
  \dodoi{10.1070/PU1986v029n03ABEH003178}

\bibitem[{{Dyson} \& {van de Meent}(2023)}]{dyson23}
{Dyson}, C., \& {van de Meent}, M. 2023, Classical and Quantum Gravity, 40,
  195026, \dodoi{10.1088/1361-6382/acf552}

\bibitem[{{Event Horizon Telescope Collaboration} {et~al.}(2022){Event Horizon
  Telescope Collaboration}, {Akiyama}, {Alberdi}, {Alef}, {Algaba}, {Anantua},
  {Asada}, {Azulay}, {Bach}, {Baczko}, {Ball}, {Balokovi{\'c}}, {Barrett},
  {Baub{\"o}ck}, {Benson}, {Bintley}, {Blackburn}, {Blundell}, {Bouman},
  {Bower}, {Boyce}, {Bremer}, {Brinkerink}, {Brissenden}, {Britzen},
  {Broderick}, {Broguiere}, {Bronzwaer}, {Bustamante}, {Byun}, {Carlstrom},
  {Ceccobello}, {Chael}, {Chan}, {Chatterjee}, {Chatterjee}, {Chen}, {Chen},
  {Cheng}, {Cho}, {Christian}, {Conroy}, {Conway}, {Cordes}, {Crawford},
  {Crew}, {Cruz-Osorio}, {Cui}, {Davelaar}, {De Laurentis}, {Deane}, {Dempsey},
  {Desvignes}, {Dexter}, {Dhruv}, {Doeleman}, {Dougal}, {Dzib}, {Eatough},
  {Emami}, {Falcke}, {Farah}, {Fish}, {Fomalont}, {Ford}, {Fraga-Encinas},
  {Freeman}, {Friberg}, {Fromm}, {Fuentes}, {Galison}, {Gammie}, {Garc{\'\i}a},
  {Gentaz}, {Georgiev}, {Goddi}, {Gold}, {G{\'o}mez-Ruiz}, {G{\'o}mez}, {Gu},
  {Gurwell}, {Hada}, {Haggard}, {Haworth}, {Hecht}, {Hesper}, {Heumann}, {Ho},
  {Ho}, {Honma}, {Huang}, {Huang}, {Hughes}, {Ikeda}, {Impellizzeri}, {Inoue},
  {Issaoun}, {James}, {Jannuzi}, {Janssen}, {Jeter}, {Jiang},
  {Jim{\'e}nez-Rosales}, {Johnson}, {Jorstad}, {Joshi}, {Jung}, {Karami},
  {Karuppusamy}, {Kawashima}, {Keating}, {Kettenis}, {Kim}, {Kim}, {Kim},
  {Kim}, {Kino}, {Koay}, {Kocherlakota}, {Kofuji}, {Koch}, {Koyama}, {Kramer},
  {Kramer}, {Krichbaum}, {Kuo}, {La Bella}, {Lauer}, {Lee}, {Lee}, {Leung},
  {Levis}, {Li}, {Lico}, {Lindahl}, {Lindqvist}, {Lisakov}, {Liu}, {Liu},
  {Liuzzo}, {Lo}, {Lobanov}, {Loinard}, {Lonsdale}, {Lu}, {Mao}, {Marchili},
  {Markoff}, {Marrone}, {Marscher}, {Mart{\'\i}-Vidal}, {Matsushita},
  {Matthews}, {Medeiros}, {Menten}, {Michalik}, {Mizuno}, {Mizuno}, {Moran},
  {Moriyama}, {Moscibrodzka}, {M{\"u}ller}, {Mus}, {Musoke}, {Myserlis},
  {Nadolski}, {Nagai}, {Nagar}, {Nakamura}, {Narayan}, {Narayanan},
  {Natarajan}, {Nathanail}, {Fuentes}, {Neilsen}, {Neri}, {Ni}, {Noutsos},
  {Nowak}, {Oh}, {Okino}, {Olivares}, {Ortiz-Le{\'o}n}, {Oyama}, {{\"O}zel},
  {Palumbo}, {Paraschos}, {Park}, {Parsons}, {Patel}, {Pen}, {Pesce},
  {Pi{\'e}tu}, {Plambeck}, {PopStefanija}, {Porth}, {P{\"o}tzl}, {Prather},
  {Preciado-L{\'o}pez}, {Psaltis}, {Pu}, {Ramakrishnan}, {Rao}, {Rawlings},
  {Raymond}, {Rezzolla}, {Ricarte}, {Ripperda}, {Roelofs}, {Rogers}, {Ros},
  {Romero-Ca{\~n}izales}, {Roshanineshat}, {Rottmann}, {Roy}, {Ruiz},
  {Ruszczyk}, {Rygl}, {S{\'a}nchez}, {S{\'a}nchez-Arg{\"u}elles},
  {S{\'a}nchez-Portal}, {Sasada}, {Satapathy}, {Savolainen}, {Schloerb},
  {Schonfeld}, {Schuster}, {Shao}, {Shen}, {Small}, {Sohn}, {SooHoo},
  {Souccar}, {Sun}, {Tazaki}, {Tetarenko}, {Tiede}, {Tilanus}, {Titus},
  {Torne}, {Traianou}, {Trent}, {Trippe}, {Turk}, {van Bemmel}, {van
  Langevelde}, {van Rossum}, {Vos}, {Wagner}, {Ward-Thompson}, {Wardle},
  {Weintroub}, {Wex}, {Wharton}, {Wielgus}, {Wiik}, {Witzel}, {Wondrak},
  {Wong}, {Wu}, {Yamaguchi}, {Yoon}, {Young}, {Young}, {Younsi}, {Yuan},
  {Yuan}, {Zensus}, {Zhang}, {Zhao}, {Zhao}, {Agurto}, {Allardi}, {Amestica},
  {Araneda}, {Arriagada}, {Berghuis}, {Bertarini}, {Berthold}, {Blanchard},
  {Brown}, {C{\'a}rdenas}, {Cantzler}, {Caro}, {Castillo-Dom{\'\i}nguez},
  {Chan}, {Chang}, {Chang}, {Chang}, {Chang}, {Chen}, {Chilson}, {Chuter},
  {Ciechanowicz}, {Colin-Beltran}, {Coulson}, {Crowley}, {Degenaar},
  {Dornbusch}, {Dur{\'a}n}, {Everett}, {Faber}, {Forster}, {Fuchs}, {Gale},
  {Geertsema}, {Gonz{\'a}lez}, {Graham}, {Gueth}, {Halverson}, {Han}, {Han},
  {Hasegawa}, {Hern{\'a}ndez-Rebollar}, {Herrera}, {Herrero-Illana},
  {Heyminck}, {Hirota}, {Hoge}, {Hostler Schimpf}, {Howie}, {Huang}, {Jiang},
  {Jinchi}, {John}, {Kimura}, {Klein}, {Kubo}, {Kuroda}, {Kwon}, {Lacasse},
  {Laing}, {Leitch}, {Li}, {Liu}, {Liu}, {Lin}, {Lu}, {Mac-Auliffe},
  {Martin-Cocher}, {Matulonis}, {Maute}, {Messias}, {Meyer-Zhao},
  {Monta{\~n}a}, {Montenegro-Montes}, {Montgomerie}, {Moreno Nolasco},
  {Muders}, {Nishioka}, {Norton}, {Nystrom}, {Ogawa}, {Olivares}, {Oshiro},
  {P{\'e}rez-Beaupuits}, {Parra}, {Phillips}, {Poirier}, {Pradel}, {Qiu},
  {Raffin}, {Rahlin}, {Ram{\'\i}rez}, {Ressler}, {Reynolds},
  {Rodr{\'\i}guez-Montoya}, {Saez-Madain}, {Santana}, {Shaw}, {Shirkey},
  {Silva}, {Snow}, {Sousa}, {Sridharan}, {Stahm}, {Stark}, {Test},
  {Torstensson}, {Venegas}, {Walther}, {Wei}, {White}, {Wieching}, {Wijnands},
  {Wouterloot}, {Yu}, {Yu (于威)}, \& {Zeballos}}]{eht23}
{Event Horizon Telescope Collaboration}, {Akiyama}, K., {Alberdi}, A., {et~al.}
  2022, \apjl, 930, L12, \dodoi{10.3847/2041-8213/ac6674}

\bibitem[{{Event Horizon Telescope Collaboration} {et~al.}(2024){Event Horizon
  Telescope Collaboration}, {Akiyama}, {Alberdi}, {Alef}, {Algaba}, {Anantua},
  {Asada}, {Azulay}, {Bach}, {Baczko}, {Ball}, {Balokovi{\'c}},
  {Bandyopadhyay}, {Barrett}, {Baub{\"o}ck}, {Benson}, {Bintley}, {Blackburn},
  {Blundell}, {Bouman}, {Bower}, {Boyce}, {Bremer}, {Brissenden}, {Britzen},
  {Broderick}, {Broguiere}, {Bronzwaer}, {Bustamante}, {Carlstrom}, {Chael},
  {Chan}, {Chang}, {Chatterjee}, {Chatterjee}, {Chen}, {Chen}, {Cheng}, {Cho},
  {Christian}, {Conroy}, {Conway}, {Crawford}, {Crew}, {Cruz-Osorio}, {Cui},
  {Dahale}, {Davelaar}, {De Laurentis}, {Deane}, {Dempsey}, {Desvignes},
  {Dexter}, {Dhruv}, {Dihingia}, {Doeleman}, {Dzib}, {Eatough}, {Emami},
  {Falcke}, {Farah}, {Fish}, {Fomalont}, {Ford}, {Foschi}, {Fraga-Encinas},
  {Freeman}, {Friberg}, {Fromm}, {Fuentes}, {Galison}, {Gammie}, {Garc{\'\i}a},
  {Gentaz}, {Georgiev}, {Goddi}, {Gold}, {G{\'o}mez-Ruiz}, {G{\'o}mez}, {Gu},
  {Gurwell}, {Hada}, {Haggard}, {Hesper}, {Heumann}, {Ho}, {Ho}, {Honma},
  {Huang}, {Huang}, {Hughes}, {Ikeda}, {Violette Impellizzeri}, {Inoue},
  {Issaoun}, {James}, {Jannuzi}, {Janssen}, {Jeter}, {Jiang},
  {Jim{\'e}nez-Rosales}, {Johnson}, {Jorstad}, {Jones}, {Joshi}, {Jung},
  {Karuppusamy}, {Kawashima}, {Keating}, {Kettenis}, {Kim}, {Kim}, {Kim},
  {Kim}, {Kino}, {Koay}, {Kocherlakota}, {Kofuji}, {Koch}, {Koyama}, {Kramer},
  {Kramer}, {Kramer}, {Krichbaum}, {Kuo}, {La Bella}, {Lee}, {Levis}, {Li},
  {Lico}, {Lindahl}, {Lindqvist}, {Lisakov}, {Liu}, {Liu}, {Liuzzo}, {Lo},
  {Lobanov}, {Loinard}, {Lonsdale}, {Lowitz}, {Lu}, {MacDonald}, {Mao},
  {Marchili}, {Markoff}, {Marrone}, {Marscher}, {Mart{\'\i}-Vidal},
  {Matsushita}, {Matthews}, {Medeiros}, {Menten}, {Mizuno}, {Mizuno},
  {Montgomery}, {Moran}, {Moriyama}, {Moscibrodzka}, {Mulaudzi}, {M{\"u}ller},
  {M{\"u}ller}, {Mus}, {Musoke}, {Myserlis}, {Nagai}, {Nagar}, {Nakamura},
  {Narayanan}, {Natarajan}, {Nathanail}, {Fuentes}, {Neilsen}, {Ni}, {Nowak},
  {Oh}, {Okino}, {Olivares}, {Oyama}, {{\"O}zel}, {Palumbo}, {Paraschos},
  {Park}, {Parsons}, {Patel}, {Pen}, {Pesce}, {Pi{\'e}tu}, {PopStefanija},
  {Porth}, {Prather}, {Psaltis}, {Pu}, {Ramakrishnan}, {Rao}, {Rawlings},
  {Raymond}, {Rezzolla}, {Ricarte}, {Ripperda}, {Roelofs},
  {Romero-Ca{\~n}izales}, {Ros}, {Roshanineshat}, {Rottmann}, {Roy}, {Ruiz},
  {Ruszczyk}, {Rygl}, {S{\'a}nchez}, {S{\'a}nchez-Arg{\"u}elles},
  {S{\'a}nchez-Portal}, {Sasada}, {Satapathy}, {Savolainen}, {Schloerb},
  {Schonfeld}, {Schuster}, {Shao}, {Shen}, {Small}, {Sohn}, {SooHoo}, {Salas},
  {Souccar}, {Stanway}, {Sun}, {Tazaki}, {Tetarenko}, {Tiede}, {Tilanus},
  {Titus}, {Toma}, {Torne}, {Toscano}, {Traianou}, {Trent}, {Trippe}, {Turk},
  {van Bemmel}, {van Langevelde}, {van Rossum}, {Vos}, {Wagner},
  {Ward-Thompson}, {Wardle}, {Washington}, {Weintroub}, {Wharton}, {Wielgus},
  {Wiik}, {Witzel}, {Wondrak}, {Wong}, {Wu}, {Yadlapalli}, {Yamaguchi},
  {Yfantis}, {Yoon}, {Young}, {Younsi}, {Yu}, {Yuan}, {Yuan}, {Anton Zensus},
  {Zhang}, {Zhao}, {Zhao}, {Allardi}, {Chang}, {Chang}, {Chang}, {Chen},
  {Chilson}, {Faber}, {Gale}, {Han}, {Han}, {Hasegawa},
  {Hern{\'a}ndez-Rebollar}, {Huang}, {Jiang}, {Jinchi}, {Kimura}, {Kubo}, {Li},
  {Lin}, {Liu}, {Liu}, {Lu}, {Martin-Cocher}, {Meyer-Zhao}, {Monta{\~n}a},
  {Moraghan}, {Moreno-Nolasco}, {Nishioka}, {Norton}, {Nystrom}, {Ogawa},
  {Oshiro}, {Pradel}, {Principe}, {Raffin}, {Rodr{\'\i}guez-Montoya}, {Shaw},
  {Snow}, {Sridharan}, {Srinivasan}, {Wei}, \& {Yu}}]{eht24}
---. 2024, \aap, 681, A79, \dodoi{10.1051/0004-6361/202347932}

\bibitem[{{Frank} {et~al.}(2002){Frank}, {King}, \&
  {Raine}}]{2002apa..book.....F}
{Frank}, J., {King}, A., \& {Raine}, D.~J. 2002, {Accretion Power in
  Astrophysics: Third Edition} (Cambridge University Press, New York)

\bibitem[{{Gal'tsov} \& {Kobialko}(2019)}]{galtsov19}
{Gal'tsov}, D.~V., \& {Kobialko}, K.~V. 2019, \prd, 99, 084043,
  \dodoi{10.1103/PhysRevD.99.084043}

\bibitem[{{Goldstein}(1974)}]{goldstein74}
{Goldstein}, H. 1974, Zeitschrift fur Physik, 271, 275,
  \dodoi{10.1007/BF01677935}

\bibitem[{{Goldstein} {et~al.}(2002){Goldstein}, {Poole}, {Safko}, \&
  {Addison}}]{goldstein02}
{Goldstein}, H., {Poole}, C., {Safko}, J., \& {Addison}, S.~R. 2002, American
  Journal of Physics, 70, 782, \dodoi{10.1119/1.1484149}

\bibitem[{{Karas} \& {Kop{\'a}{\v{c}}ek}(2021)}]{karas21}
{Karas}, V., \& {Kop{\'a}{\v{c}}ek}, O. 2021, Astronomische Nachrichten, 342,
  357, \dodoi{10.1002/asna.202113934}

\bibitem[{{Karas} {et~al.}(2013){Karas}, {Kop{\'a}{\v{c}}ek}, \&
  {Kunneriath}}]{karas13}
{Karas}, V., {Kop{\'a}{\v{c}}ek}, O., \& {Kunneriath}, D. 2013, International
  Journal of Astronomy and Astrophysics, 3, 18,
  \dodoi{10.4236/ijaa.2013.33A003}

\bibitem[{{Kato} {et~al.}(2008){Kato}, {Fukue}, \&
  {Mineshige}}]{2008bhad.book.....K}
{Kato}, S., {Fukue}, J., \& {Mineshige}, S. 2008, {Black-Hole Accretion Disks
  --- Towards a New Paradigm} (Kyoto University Press, Kyoto)

\bibitem[{{Kerr}(1963)}]{kerr63}
{Kerr}, R.~P. 1963, \prl, 11, 237, \dodoi{10.1103/PhysRevLett.11.237}

\bibitem[{{Kolo{\v{s}}} {et~al.}(2021){Kolo{\v{s}}}, {Tursunov}, \&
  {Stuchl{\'\i}k}}]{kolos21}
{Kolo{\v{s}}}, M., {Tursunov}, A., \& {Stuchl{\'\i}k}, Z. 2021, \prd, 103,
  024021, \dodoi{10.1103/PhysRevD.103.024021}

\bibitem[{{Kop{\'a}{\v{c}}ek} \& {Karas}(2018{\natexlab{a}})}]{kopacek18b}
{Kop{\'a}{\v{c}}ek}, O., \& {Karas}, V. 2018{\natexlab{a}}, in Fourteenth
  Marcel Grossmann Meeting - MG14, ed. M.~{Bianchi}, R.~T. {Jansen}, \&
  R.~{Ruffini}, 1050--1055

\bibitem[{{Kop{\'a}{\v{c}}ek} \& {Karas}(2018{\natexlab{b}})}]{kopacek18}
{Kop{\'a}{\v{c}}ek}, O., \& {Karas}, V. 2018{\natexlab{b}}, \apj, 853, 53,
  \dodoi{10.3847/1538-4357/aaa45f}

\bibitem[{{Kop{\'a}{\v{c}}ek} \& {Karas}(2020)}]{kopacek20}
---. 2020, \apj, 900, 119, \dodoi{10.3847/1538-4357/ababa8}

\bibitem[{{Kop{\'a}{\v{c}}ek} \& {Karas}(2023)}]{kopacek23}
{Kop{\'a}{\v{c}}ek}, O., \& {Karas}, V. 2023, in The Sixteenth Marcel Grossmann
  Meeting. On Recent Developments in Theoretical and Experimental General
  Relativity, Astrophysics, and Relativistic Field Theories, 3999--4009

\bibitem[{{Kop{\'a}{\v{c}}ek} {et~al.}(2014){Kop{\'a}{\v{c}}ek}, {Karas},
  {Kov{\'a}{\v{r}}}, \& {Stuchl{\'\i}k}}]{kopacek14b}
{Kop{\'a}{\v{c}}ek}, O., {Karas}, V., {Kov{\'a}{\v{r}}}, J., \&
  {Stuchl{\'\i}k}, Z. 2014, in Proceedings of RAGtime 10-13: Workshops on black
  holes and neutron stars, 123--132

\bibitem[{{Kov{\'a}{\v{r}}} {et~al.}(2010){Kov{\'a}{\v{r}}},
  {Kop{\'a}{\v{c}}ek}, {Karas}, \& {Stuchl{\'\i}k}}]{kovar10}
{Kov{\'a}{\v{r}}}, J., {Kop{\'a}{\v{c}}ek}, O., {Karas}, V., \&
  {Stuchl{\'\i}k}, Z. 2010, Classical and Quantum Gravity, 27, 135006,
  \dodoi{10.1088/0264-9381/27/13/135006}

\bibitem[{{Kov{\'a}{\v{r}}} {et~al.}(2008){Kov{\'a}{\v{r}}}, {Stuchl{\'\i}k},
  \& {Karas}}]{kovar08}
{Kov{\'a}{\v{r}}}, J., {Stuchl{\'\i}k}, Z., \& {Karas}, V. 2008, Classical and
  Quantum Gravity, 25, 095011, \dodoi{10.1088/0264-9381/25/9/095011}

\bibitem[{{Kumar} \& {Pringle}(1985)}]{1985MNRAS.213..435K}
{Kumar}, S., \& {Pringle}, J.~E. 1985, \mnras, 213, 435,
  \dodoi{10.1093/mnras/213.3.435}

\bibitem[{{Lan{\v{c}}ov{\'a}} {et~al.}(2019){Lan{\v{c}}ov{\'a}}, {Abarca},
  {Klu{\'z}niak}, {Wielgus}, {S{\k{a}}dowski}, {Narayan}, {Schee},
  {T{\"o}r{\"o}k}, \& {Abramowicz}}]{2019ApJ...884L..37L}
{Lan{\v{c}}ov{\'a}}, D., {Abarca}, D., {Klu{\'z}niak}, W., {et~al.} 2019,
  \apjl, 884, L37, \dodoi{10.3847/2041-8213/ab48f5}

\bibitem[{{Levin} \& {Perez-Giz}(2008)}]{2008PhRvD..77j3005L}
{Levin}, J., \& {Perez-Giz}, G. 2008, \prd, 77, 103005,
  \dodoi{10.1103/PhysRevD.77.103005}

\bibitem[{{Levin} \& {Perez-Giz}(2009)}]{levin09}
---. 2009, \prd, 79, 124013, \dodoi{10.1103/PhysRevD.79.124013}

\bibitem[{{Lichtenberg} \& {Lieberman}(1992)}]{lichtenberg92}
{Lichtenberg}, A., \& {Lieberman}, M. 1992, {Regular and Chaotic Dynamics}
  (Springer-Verlag, New York)

\bibitem[{{Liska} {et~al.}(2019){Liska}, {Tchekhovskoy}, {Ingram}, \& {van der
  Klis}}]{2019MNRAS.487..550L}
{Liska}, M., {Tchekhovskoy}, A., {Ingram}, A., \& {van der Klis}, M. 2019,
  \mnras, 487, 550, \dodoi{10.1093/mnras/stz834}

\bibitem[{{Lukes-Gerakopoulos} \& {Kop{\'a}{\v{c}}ek}(2018)}]{glg18}
{Lukes-Gerakopoulos}, G., \& {Kop{\'a}{\v{c}}ek}, O. 2018, International
  Journal of Modern Physics D, 27, 1850010, \dodoi{10.1142/S0218271818500104}

\bibitem[{{Martin} {et~al.}(2007){Martin}, {Pringle}, \&
  {Tout}}]{2007MNRAS.381.1617M}
{Martin}, R.~G., {Pringle}, J.~E., \& {Tout}, C.~A. 2007, \mnras, 381, 1617,
  \dodoi{10.1111/j.1365-2966.2007.12349.x}

\bibitem[{{Marwan} {et~al.}(2007){Marwan}, {Carmen Romano}, {Thiel}, \&
  {Kurths}}]{marwan07}
{Marwan}, N., {Carmen Romano}, M., {Thiel}, M., \& {Kurths}, J. 2007, \physrep,
  438, 237, \dodoi{10.1016/j.physrep.2006.11.001}

\bibitem[{{Meier}(2012)}]{2012bhae.book.....M}
{Meier}, D.~L. 2012, {Black Hole Astrophysics: The Engine Paradigm} (Springer,
  heidelberg, and Praxis Publishing, Chichester),
  \dodoi{10.1007/978-3-642-01936-4}

\bibitem[{{Misner} {et~al.}(2017){Misner}, {Thorne}, \& {Wheeler}}]{mtw}
{Misner}, C.~W., {Thorne}, K.~S., \& {Wheeler}, J.~A. 2017, {Gravitation}
  (Princeton University Press)

\bibitem[{{Mukherjee} {et~al.}(2023){Mukherjee}, {Kop{\'a}{\v{c}}ek}, \&
  {Lukes-Gerakopoulos}}]{mukherjee23}
{Mukherjee}, S., {Kop{\'a}{\v{c}}ek}, O., \& {Lukes-Gerakopoulos}, G. 2023,
  \prd, 107, 064005, \dodoi{10.1103/PhysRevD.107.064005}

\bibitem[{{Mummery} \& {Balbus}(2022)}]{2022PhRvL.129p1101M}
{Mummery}, A., \& {Balbus}, S. 2022, \prl, 129, 161101,
  \dodoi{10.1103/PhysRevLett.129.161101}

\bibitem[{{Narayan} {et~al.}(2003){Narayan}, {Igumenshchev}, \&
  {Abramowicz}}]{2003PASJ...55L..69N}
{Narayan}, R., {Igumenshchev}, I.~V., \& {Abramowicz}, M.~A. 2003, \pasj, 55,
  L69, \dodoi{10.1093/pasj/55.6.L69}

\bibitem[{{Navarro} \& {Pierre Auger Collaboration}(2013)}]{navarro13}
{Navarro}, J.~L., \& {Pierre Auger Collaboration}. 2013, in Journal of Physics
  Conference Series, Vol. 409, Journal of Physics Conference Series, 012115

\bibitem[{{Ott}(1993)}]{ott93}
{Ott}, E. 1993, {Chaos in dynamical systems} (Cambridge University Press,
  Cambridge)

\bibitem[{{Perez-Giz} \& {Levin}(2009)}]{perez-giz09}
{Perez-Giz}, G., \& {Levin}, J. 2009, \prd, 79, 124014,
  \dodoi{10.1103/PhysRevD.79.124014}

\bibitem[{{Perlick} \& {Tsupko}(2022)}]{perlick22}
{Perlick}, V., \& {Tsupko}, O.~Y. 2022, \physrep, 947, 1,
  \dodoi{10.1016/j.physrep.2021.10.004}

\bibitem[{{Rana} \& {Mangalam}(2019)}]{rana19}
{Rana}, P., \& {Mangalam}, A. 2019, Classical and Quantum Gravity, 36, 045009,
  \dodoi{10.1088/1361-6382/ab004c}

\bibitem[{{Rodrigues} {et~al.}(2021){Rodrigues}, {Heinze}, {Palladino}, {van
  Vliet}, \& {Winter}}]{rodrigues21}
{Rodrigues}, X., {Heinze}, J., {Palladino}, A., {van Vliet}, A., \& {Winter},
  W. 2021, \prl, 126, 191101, \dodoi{10.1103/PhysRevLett.126.191101}

\bibitem[{{Shakura}(1987)}]{shakura87}
{Shakura}, N. 1987, Soviet Astronomy Letters, 13, 99

\bibitem[{{Shakura}(2018)}]{2018ASSL..454.....S}
---. 2018, Astrophysics and Space Science Library, Vol. 454, {Accretion Flows
  in Astrophysics} (Springer Cham), 419 pp., \dodoi{10.1007/978-3-319-93009-1}

\bibitem[{{Stein} \& {Warburton}(2020)}]{stein20}
{Stein}, L.~C., \& {Warburton}, N. 2020, \prd, 101, 064007,
  \dodoi{10.1103/PhysRevD.101.064007}

\bibitem[{Strogatz(2019)}]{strogatz19}
Strogatz, S.~H. 2019, Nonlinear Dynamics and Chaos, 2nd edn. (London, England:
  CRC Press)

\bibitem[{{Stuchl{\'\i}k} \& {Kolo{\v{s}}}(2016)}]{stuchlik16}
{Stuchl{\'\i}k}, Z., \& {Kolo{\v{s}}}, M. 2016, European Physical Journal C,
  76, 32, \dodoi{10.1140/epjc/s10052-015-3862-2}

\bibitem[{{Tabor}(1989)}]{tabor89}
{Tabor}, M. 1989, {Chaos and Integrability in Nonlinear Dynamics: An
  Introduction} (Wiley-Interscience, USA)

\bibitem[{{Tavlayan} \& {Tekin}(2021)}]{2021PhRvD.104l4059T}
{Tavlayan}, A., \& {Tekin}, B. 2021, \prd, 104, 124059,
  \dodoi{10.1103/PhysRevD.104.124059}

\bibitem[{{Teo}(2003)}]{teo03}
{Teo}, E. 2003, General Relativity and Gravitation, 35, 1909,
  \dodoi{10.1023/A:1026286607562}

\bibitem[{{Teo}(2021)}]{teo21}
---. 2021, General Relativity and Gravitation, 53, 10,
  \dodoi{10.1007/s10714-020-02782-z}

\bibitem[{{Tursunov} {et~al.}(2016){Tursunov}, {Stuchl{\'\i}k}, \&
  {Kolo{\v{s}}}}]{tursunov16}
{Tursunov}, A., {Stuchl{\'\i}k}, Z., \& {Kolo{\v{s}}}, M. 2016, \prd, 93,
  084012, \dodoi{10.1103/PhysRevD.93.084012}

\bibitem[{{Tursunov} {et~al.}(2020){Tursunov}, {Stuchl{\'\i}k}, {Kolo{\v{s}}},
  {Dadhich}, \& {Ahmedov}}]{tursunov20}
{Tursunov}, A., {Stuchl{\'\i}k}, Z., {Kolo{\v{s}}}, M., {Dadhich}, N., \&
  {Ahmedov}, B. 2020, \apj, 895, 14, \dodoi{10.3847/1538-4357/ab8ae9}

\bibitem[{{Verner}(2010)}]{verner10}
{Verner}, J. 2010, Numerical Algorithms, 53, 383,
  \dodoi{10.1007/s11075-009-9290-3}

\bibitem[{{Wielgus} {et~al.}(2022){Wielgus}, {Lan{\v{c}}ov{\'a}}, {Straub},
  {Klu{\'z}niak}, {Narayan}, {Abarca}, {R{\'o}{\.z}a{\'n}ska}, {Vincent},
  {T{\"o}r{\"o}k}, \& {Abramowicz}}]{2022MNRAS.514..780W}
{Wielgus}, M., {Lan{\v{c}}ov{\'a}}, D., {Straub}, O., {et~al.} 2022, \mnras,
  514, 780, \dodoi{10.1093/mnras/stac1317}

\bibitem[{Wiggins(2006)}]{wiggins06}
Wiggins, S. 2006, Introduction to Applied Nonlinear Dynamical Systems and
  Chaos, Texts in Applied Mathematics (Springer New York).
\newblock \url{https://books.google.cz/books?id=YhXnBwAAQBAJ}

\bibitem[{{Wilkins}(1972)}]{wilkins72}
{Wilkins}, D.~C. 1972, \prd, 5, 814, \dodoi{10.1103/PhysRevD.5.814}

\bibitem[{{Will}(2012)}]{2012CQGra..29u7001W}
{Will}, C.~M. 2012, Classical and Quantum Gravity, 29, 217001,
  \dodoi{10.1088/0264-9381/29/21/217001}

\bibitem[{{Young}(1976)}]{1976PhRvD..14.3281Y}
{Young}, P.~J. 1976, \prd, 14, 3281, \dodoi{10.1103/PhysRevD.14.3281}

\end{thebibliography}

\end{document}